\documentclass[aps,prl,reprint,superscriptaddress,floatfix,nofootinbib]{revtex4-2}
\usepackage{bm}
\usepackage{bbm}
\usepackage{amsmath}
\usepackage{etoolbox}
\usepackage{float}
\usepackage{needspace}
\newcommand{\eg}{{\emph{e.g.~}}}
\newcommand{\ie}{{\emph{i.e.~}}}

\usepackage[utf8]{inputenc}
\usepackage{verbatim}
\usepackage{amsmath,amsfonts,amssymb}
\usepackage[breaklinks,colorlinks]{hyperref}
\usepackage{xr-hyper}
\usepackage{float}
\usepackage{natbib}
\usepackage{tikz}
\usepackage{float}
\usepackage{graphicx}
\usepackage{adjustbox}
\usepackage{tabularx}

\usepackage{amsbsy}
\usepackage{orcidlink}
\hypersetup{
    colorlinks=true,
    allcolors=black,
}

\begin{document}

\preprint{DESY-25-111}
\preprint{CERN-TH-2025-251}

\title{Constraints on light QCD and CP-violating axions from the death line of rotation-powered pulsars}

\author{Samuel J. Witte\orcidlink{0000-0003-4649-3085}}\email{Samuel.Witte@physics.ox.ac.uk}
\affiliation{Rudolf Peierls Centre for Theoretical Physics, University of Oxford, UK}
\affiliation{Deutsches Elektronen-Synchrotron DESY, Notkestra\ss{}e 85, 22607 Hamburg, Germany}
\affiliation{
II. Institute of Theoretical Physics, Universit{\"a}t Hamburg, 22761, Hamburg, Germany}
\author{Andrea Caputo\orcidlink{0000-0003-3516-8332}}\email{Andrea.Caputo@cern.ch} 
\affiliation{Department of Theoretical Physics, CERN, Esplanade des Particules 1, P.O. Box 1211, Geneva 23, Switzerland}
\affiliation{Dipartimento di Fisica, ``Sapienza'' Universit\`a di Roma \& Sezione INFN Roma1, Piazzale Aldo Moro
5, 00185, Roma, Italy}
\affiliation{Department of Particle Physics and Astrophysics, Weizmann Institute of Science, Rehovot 7610001, Israel}

\author{Stefan Stelzl\orcidlink{0000-0001-5964-1054}}\email{SStelzl@ifae.es}
\affiliation{Institute of Physics,Theoretical Particle Physics Laboratory, Ecole Polytechnique Federale de Lausanne,  Switzerland}
\affiliation{Institut de F\'{i}sica d'Altes Energies (IFAE) and
Barcelona Institute of Science and Technology (BIST), 
Campus UAB, 08193 Bellaterra (Barcelona), Spain
}

\author{Alexander Chernoglazov}
\affiliation{Institute for Advanced Study, Princeton, NJ 08540, USA}

\author{Alexander A. Philippov} 
\affiliation{Department of Physics, University of Maryland, College Park, MD 20742, USA}

\author{Surjeet Rajendran}
\affiliation{Department of Physics and Astronomy, Johns Hopkins University, Baltimore, MD 21218, USA}

\begin{abstract}
Dense nuclear matter can modify the effective potential of axions, displacing them from their vacuum minimum, and sourcing large external field gradients (``axion hair"). In the case of neutron stars, axion hair directly modifies the electrodynamic processes operating on the open field-line region, strongly enhancing or suppressing the acceleration experienced by ambient charges. As a result, the point in the neutron star lifetime at which pair-cascades cease -- known as pulsar ``death" -- can be dramatically altered, allowing for much older pulsars to emit observable radio emission. We study the pair discharge process in the presence of axion hair using semi-analytic techniques and particle-in-cell simulations, and use these results alongside pulsar demographics to derive new constraints on light QCD axions with non-negligible axion-photon coupling and CP-violating axion–nucleon interactions. We also illustrate how nearly orthogonal rotators, where emission is observed from both poles (such as in the case of PSR J1906+0746), provides a complementary probe of axion hair.
\end{abstract}

\maketitle

The quantum chromodynamics (QCD) axion is a pseudo-Nambu--Goldstone boson arising from the spontaneous breaking of the Peccei--Quinn symmetry~\cite{Peccei:1977hh, WeinbergAxion, WilczekAxion}, and provides one of the most compelling solutions to the strong CP problem (\ie the question of why the neutron electric dipole moment is so small~\cite{Abel:2020pzs}). In typical UV-complete models, the QCD axion is expected to live in a well-defined region of parameter space; this is because the effective low-energy potential generated by QCD relates the axion mass to the decay constant,
$m_a \sim 5.7\,\mu{\rm eV}\,(10^{12}\,{\rm GeV}/f_a)$, where $f_a$ also sets the characteristic size of the axion couplings. This relation, however, is not fully generic: there exist mechanisms in which the axion remains parametrically lighter than in conventional scenarios~\cite{Hook:2018jle, DiLuzio:2021pxd, Hook:2017psm,Banerjee:2025kov}. Such models are especially interesting from a phenomenological perspective, since localized dense clumps of matter can distort the effective axion potential, displacing the axion from the vacuum minimum and sourcing large field gradients within---and near the boundaries of---these dense objects. Matter-sourcing effects for light axions have been studied in a variety of contexts ranging from the Earth and Sun~\cite{Hook:2017psm} to white dwarfs~\cite{Balkin:2022qer,Bartnick:2025lbg,Bartnick:2025lbv} and neutron stars (NS)~\cite{Hook:2017psm, Balkin:2023xtr, Gomez-Banon:2024oux, Kumamoto:2024wjd, Kahn:2025ytc}. NSs represent one of the most interesting scenarios (albeit also one of the most complicated), as the extreme densities in their interiors allow axion gradients to be sourced across a broader region of parameter space.

\begin{figure}
    \centering
    \includegraphics[width=0.95\linewidth]{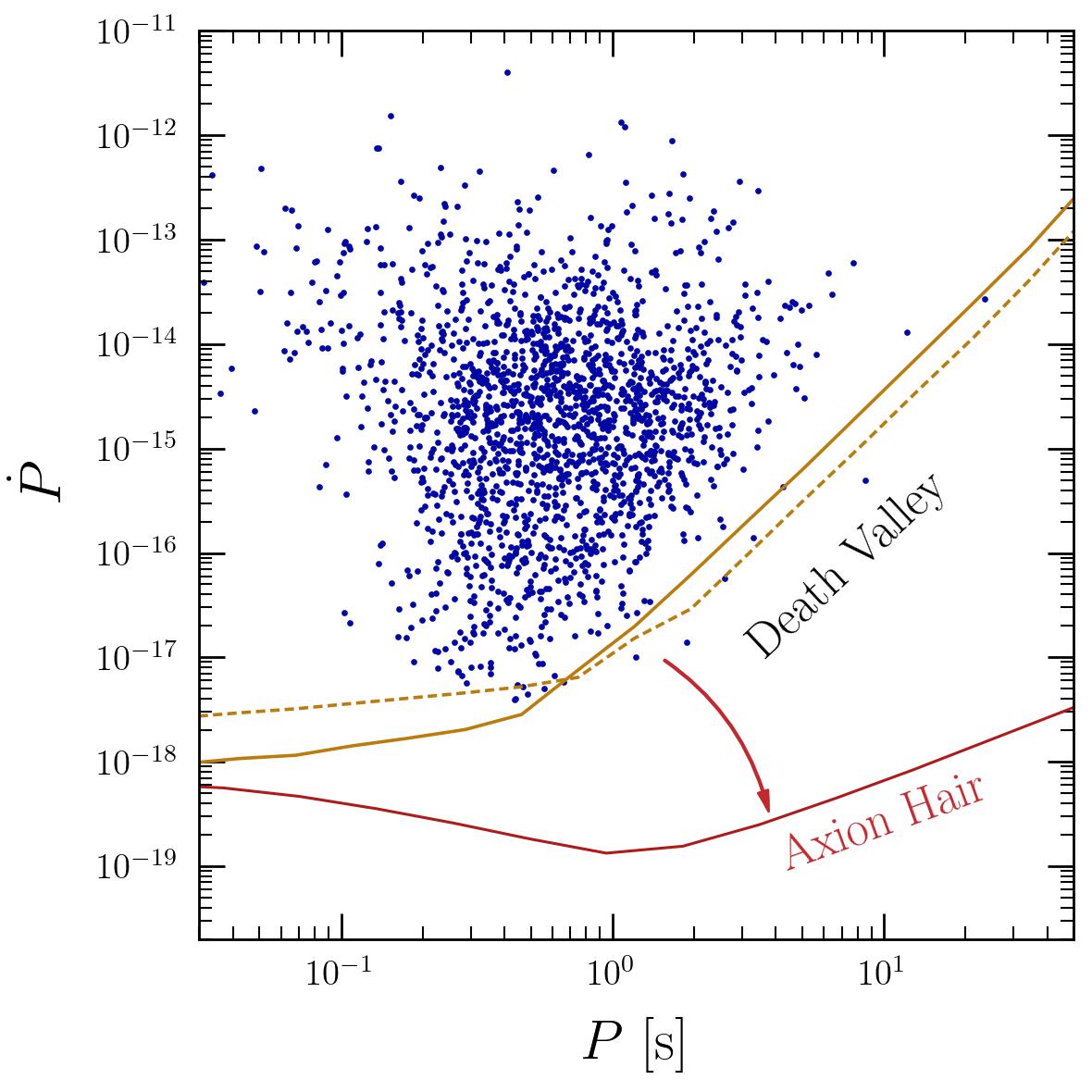}
    \caption{
    Rotational period $P$ and spin-down rate $\dot{P}$ of the rotation-powered pulsar population (blue). Two standard estimates of the pulsar death line, which differentiate the electromagnetically active from the inactive population, are shown in yellow. The red line shows the shift in the death line induced by a light QCD axion with $\epsilon = 10^{-2}$ and $f_a = 10^{16}$~GeV. The construction of the death-line curves and the discharge modeling underlying them are presented in the companion PRD.
    }
    \label{fig:ppdot_std}
\end{figure}

In this work, we propose a novel pulsar signature arising from static axion field gradients, or \emph{axion hair}\footnote{This terminology was also used in \cite{Bai:2023bbg} to describe local axion gradients sourced near NSs.}, near NSs. We demonstrate that axion hair can significantly modify the electrostatic potential drop along the open magnetic field lines of rotation-powered pulsars, allowing pair production and coherent radio emission to occur in much older, more slowly rotating NSs than would otherwise be possible. In effect, axion hair can extend pulsar lifetimes by decoupling the acceleration of primary $e^\pm$ from the rotational period of the pulsar itself, generating a shift in the location of the so-called pulsar ``death line''~\cite{Sturrock1971,RudermanSutherland1975,ChenRuderman93}. We argue that the lack of observation of such an old isolated pulsar population cannot be explained by observational selection effects alone, leading to an apparent incompatibility with the presence of sufficiently long-range axion hair sourced by light axions coupled to nuclear matter. This picture is illustrated in Fig.~\ref{fig:ppdot_std}, where axion hair sourced by a representative light QCD axion induces a large shift in the pulsar death line.

A second, qualitatively different observable follows from the intrinsic \emph{polarity asymmetry} introduced by axion hair: for sufficiently long-range gradients, the two magnetic hemispheres can be driven into different discharge regimes, which can be incompatible with a rare class of objects in which coherent radio emission is observed twice per period (so-called interpulses). We validate our arguments using particle-in-cell (PIC) simulations of the near-surface discharge; the companion article~\cite{Witte:CompanionPRD} provides a complete description of pulsar electrodynamics in the presence of axions hair, including: our numerical setup, and an extended treatment of the discharge physics, global consistency requirements, and systematics.

We also demonstrate how both effects arise in models where axions have, in addition to the coupling to photons, a linear coupling to nucleons (leading to a monopole-dipole interaction)~\cite{Fedderke:2023dwj}; in this case, interpulses observed in PSR J1906+0746 can be used to derive constraints comparable to those obtained using the death line. For this class of axions, both CP and the axion shift symmetry are explicitly broken\footnote{This could also occur for a CP-conserving nucleon coupling should the star be polarized. Here, we have focused the CP-odd coupling to axions, but one could also consider a CP-even coupling, which would induce a shift in the fine structure constant; typically, however, these constraints are much more stringent than their CP-odd counterparts. } (e.g., as in the QCD axion with a nonzero $\theta$ angle~\cite{Pospelov_1998,Moody:1984ba}). It is worth noting that CP-violating nucleon interactions are typically severely constrained by fifth-force experiments~\cite{Gao:2021fyk,Balkin:2023xtr}, highlighting the strength and complementarity of the pulsar-based effects studied here.

\section{Sourcing axions from nuclear matter}

We begin by reviewing how light QCD axions are sourced in nuclear matter, before commenting on the monopole-dipole interaction~\cite{Hook:2017psm, Balkin:2023xtr, Balkin:2020dsr, Balkin:2022qer}. Additional details (including sourcing regimes, phase structure, and stellar-structure/EOS systematics) are discussed in the companion Ref.~\cite{Witte:CompanionPRD}.

Consider a pseudo-scalar field coupled to gluons:
\begin{equation}
\mathcal{L} \supset \frac{1}{2}(\partial a)^2 + \frac{g_s^2}{32\pi^2} \, \frac{a}{f_a} G \tilde{G},
\end{equation}
where $a$ is the axion, $f_a$ the decay constant, $g_s$ the strong coupling, and $G_{\mu\nu}$ the gluon field strength. Below the QCD confinement scale, non-perturbative dynamics generate an effective potential~\cite{DiVecchia:1980yfw, GrillidiCortona:2015jxo}:
\begin{equation}\label{eq:lightQCDpotential}
V(a) = -\epsilon m_\pi^2 f_\pi^2 \left( \sqrt{1 - z_{\rm ud} \sin^2 \left(\frac{a}{2 f_a} \right)}  - 1\right),
\end{equation}
where $z_{\rm ud} = 4 m_u m_d / (m_u + m_d)^2$. The conventional QCD axion corresponds to $\epsilon = 1$, while symmetry-based models producing lighter QCD axions with $\epsilon < 1$ are discussed in~\cite{Hook:2017psm,Hook:2018jle, DiLuzio:2021pxd, DiLuzio:2021gos, Banerjee:2022wzk,Banerjee:2025kov}. Although the mass around the origin remains unchanged, some of these models differ from Eq.~\eqref{eq:lightQCDpotential} in their vacuum structure; we discuss model-building variants in the companion PRD.

The axion mass is given by
\begin{align}
m_a &= \sqrt{\epsilon} \frac{\sqrt{z_{\rm ud}}}{2} \frac{m_\pi f_\pi}{f_a} \nonumber \\
&\simeq 5.8 \times 10^{-7} \, {\rm eV} \left(\frac{\epsilon}{10^{-2}} \right)^{1/2} \left(\frac{10^{12} \, {\rm GeV}}{f_a} \right) \, .
\end{align}
At low energies, a shift-symmetry breaking interaction between axions and nucleons is generated,
\begin{equation}\label{eq:axionnucleoninteraction}
\mathcal{L} \supset  \sigma_{\pi N} \bar{N} N \left( \sqrt{1- z_{\rm ud} \sin^2 \left( \frac{a}{2 f_a}\right)} - 1 \right),
\end{equation}
where $\sigma_{\pi N} \simeq 50$~MeV is the pion-nucleon sigma term. In a background of non-relativistic nucleons with number density $n$ one finds $\langle \bar{N} N \rangle \simeq n \equiv \langle \bar{N} \gamma^0 N \rangle$. The above interaction leads to a new term in the effective axion potential, and if
\begin{equation}
\label{eq:dense_enough}
\sigma_{\pi N} n_b \gtrsim \epsilon m_\pi^2 f_\pi^2 \, , \qquad \Rightarrow \qquad
n_b \gtrsim 0.16\, \text{fm}^{-3} \Big(\frac{\epsilon}{0.4}\Big),
\end{equation}
the minimum of the effective potential shifts from $a = 0$ to $a = \pi f_a$. Equivalently, the interaction in Eq.~\eqref{eq:axionnucleoninteraction} can be interpreted as an axion-dependent nucleon mass.

Eq.~\eqref{eq:dense_enough} is a necessary, but not sufficient condition, for sourcing: the system must also be large enough to accommodate the axion gradient. Sourcing occurs if the system size $R \gg m_a(n)^{-1}$, where $m_a(n)$ is the effective in-density axion mass (note that we reserve the notation of $m_a$ to represent the axion vacuum mass). In NSs this condition is satisfied for $f_a \lesssim 10^{17}~\text{GeV}$~\cite{Balkin:2023xtr}.

Depending on the value of $\epsilon$, there are two different regimes. For $\epsilon\lesssim 0.07$, the axion is sourced all the way until the boundary of the star, while for $\epsilon\gtrsim 0.07$, a phase transition between the axion-sourced phase and the normal nuclear phase happens inside the star. For this Letter, we focus primarily on the first regime (see the companion PRD for a detailed discussion), where the axion profile~\cite{Balkin:2023xtr} is well approximated by
\begin{equation}
a(r) =
\begin{cases}
\pm \pi f_a, & r < R, \\
\pm \pi f_a \, \dfrac{R}{r} \, \exp\left[ -\dfrac{\sqrt{\epsilon z_{\rm ud}} m_\pi f_\pi}{2f_a}(r-R) \right], & r > R,
\end{cases}
\label{eq:axradial}
\end{equation}
with characteristic decay length
\begin{equation}
\lambda \simeq \frac{f_a}{\sqrt{\epsilon}\,m_\pi f_\pi} \simeq 16 \, {\rm cm} \left(\frac{f_a}{10^{12} \, {\rm GeV}} \right) \left(\frac{10^{-2}}{\epsilon} \right)^{1/2}.
\end{equation}

The example above applies explicitly to the light QCD axion, but similar results may arise generically for other light scalars. Consider, for example, a scalar linearly coupled to nucleons,
$\mathcal{L} \supset \frac{1}{2}(\partial a)^2 - \frac{1}{2} m_a^2 a^2 + g_N a \bar{N} N$.
In this case, the minimum of the potential shifts smoothly with ambient density,
\begin{equation}
a_\text{min} = g_N \, \frac{ \langle \bar{N} N \rangle}{ m_a^2}.
\end{equation}
{This continuous shift however only happens on length scales longer than $m_a^{-1}$. On shorter length scales, the axion gradient stops the field from tracking its minimum. Thus there is an axion gradient leaking out of the star.} 
Unlike for the QCD axion, the viable parameter space typically requires $a \ll m_N / g_N$ inside NSs~\cite{Balkin:2023xtr}, implying a negligible back-reaction on the stellar structure, and highlighting the importance of the electrodynamic effects studied here.

\section{Pulsar Electrodynamics and the Implications of Axion Hair}

Having demonstrated how axion hair arises, we turn our attention to the impact of these field gradients on the electromagnetic evolution of NSs.

As a result of the coupling of the axion to quarks and mesons, QCD axions acquire a coupling to electromagnetism via
\begin{equation}
\mathcal{L} \supset -\frac{g_{a\gamma\gamma}}{4} \, a \, F_{\mu\nu} \tilde{F}^{\mu\nu},
\end{equation}
where $g_{a\gamma\gamma} = \alpha_{\rm EM} C / (2\pi f_a)$ and $C$ is an $\mathcal{O}(1)$ model-dependent coefficient. This interaction modifies Gauss' and Amp\`{e}re's laws:
\begin{align}
\nabla \cdot \vec{E} &= \rho - g_{a \gamma \gamma} \vec{B} \cdot \nabla a \, \equiv \rho - \rho_a \, , \label{eq:Gauss} \\
\nabla \times \vec{B} - \partial_t \vec{E} &= \vec{j} + g_{a \gamma \gamma} \dot{a} \vec{B} + g_{a \gamma \gamma} \nabla a \times \vec{E} \, . \label{eq:curlB}
\end{align}
Axion hair sourced by stellar objects is static, $\partial_t a \simeq 0$, and the rotationally induced electric field near the surface of the star is always much less than the magnetic field, \ie $|\vec{E}| / |\vec{B}| \ll 1$, implying the leading correction comes from $\vec{B} \cdot \nabla a$. Comparing the effective axion charge density near the stellar surface, $\rho_a \equiv g_{a\gamma\gamma} \vec{B} \cdot \nabla a$, with the co-rotating (Goldreich--Julian) charge density $\rho_{\rm GJ} \simeq - 2 \vec{\Omega} \cdot \vec{B}$, one sees that for pulsars near the conventional death line,
$|\rho_a / \rho_{\rm GJ}| \sim (1 + m_a R)\,\alpha_{\rm EM}/(R\Omega)\gg 1$.
Thus, on distances $r \lesssim m_a^{-1}$, axion hair rather than rotation controls the local charge density requirement and can dominate the near-field electrodynamics.

\subsection{Pulsar Death}

We briefly review the physics behind the pulsar ``death line,'' which differentiates electromagnetically active from inactive pulsars~\cite{Sturrock1971,RudermanSutherland1975, ChenRuderman93,zhang2000radio,Faucher-Giguere:2005dxp,Konar:2019vhj,beskin2022pulsar,Beskin:2022dbs}.

For typical rotation-powered pulsars, the dominant coherent radio emission is produced near the star on the open field line bundle. The precise emission mechanism has been investigated for decades, with recent progress indicating that inhomogeneous dynamical screening of electric fields by newly produced plasma is likely central to the observed emission~\cite{Philippov:2020jxu,Chernoglazov:2024rvo}. While important details remain open, one aspect is robust: coherent radio emission from the inner magnetosphere requires sustained $e^\pm$ pair production~\cite{SashaReview}.

In standard isolated radio pulsars, pair production is sustained via single-photon magnetic conversion,
$\gamma + B \rightarrow e^- + e^+$, where the high-energy photon is produced as curvature radiation by locally accelerating charges. The optical depth $\tau$ is given by~\cite{daugherty1982electromagnetic,daugherty1983pair}:
\begin{equation}\label{Eq:OpticalDepth}
    \tau \simeq 0.23 \, \alpha_{\rm EM} \, m_e \, \int dx \, \sin\psi \, \left( \frac{B}{B_Q} \right)
    e^{-\frac{8 B_Q m_e}{3 \omega B \sin\psi} } \, ,
\end{equation}
where $B_Q \simeq 4.4 \times 10^{13}$\,G, $B$ is the magnetic field strength, $\omega$ the photon energy, and $\psi$ the angle between the photon momentum and the magnetic field (with $\psi \sim \gamma^{-1} \ll 1$). The optical depth is exponentially suppressed unless $\omega$ is sufficiently large. Using the characteristic energy of curvature radiation $\langle \omega\rangle \sim 3\gamma^3/(2\rho_c)$, one finds that primary electrons must be accelerated to Lorentz factors of order
$\gamma_T \sim 6 \times 10^7 \times (B_Q/B)^{1/3} (\rho_c / 10^8 \, {\rm cm})^{1/3}$
\footnote{Here we fixed $\sin \psi \sim 10^{-3}$ and identified when the exponential suppression becomes $\sim 0.1$~\cite{TimokhinHarding2015}.}
to efficiently trigger pair production~\cite{TimokhinHarding2015,Timokhin:2018vdn}.

The maximal voltage drop along a field line for a NS with polar-cap radius $r_{\rm pc}$ is roughly~\cite{Goldreich:1969sb}\footnote{Pair production typically limits the realized voltage drop to be much smaller than this upper bound.}:
\begin{equation}\label{eq:delta_v}
    \Delta V \simeq \frac{B\, \Omega \, r_{\rm pc}^2}{2} \, .
\end{equation}
This corresponds to a maximal Lorentz factor
\begin{eqnarray}
\gamma_{\rm max} \sim \frac{\Delta V}{e \, m_e} \sim 10^7
\left(\frac{B}{10^{12} \, {\rm G}}\right) \left(\frac{1 \,{\rm s}}{P} \right)^2  \, .
\end{eqnarray}
For rapidly rotating NSs, $\gamma_{\rm max} \gg \gamma_T$, while as pulsars age and spin down, they can eventually reach a point where acceleration becomes too inefficient to ignite pair production. Since pair production is central to radio emission, pulsars become radio quiet and ``die'' once pair cascades can no longer be sustained\footnote{We restrict attention to curvature-radiation-induced pair production relevant for rotation-powered pulsars. Magnetars and millisecond pulsars may instead be supported by inverse Compton or two-photon pair production~\cite{Timokhin:2018vdn,harding2002regimes}, and are excluded from our analysis.}.

Axion hair modifies this picture by producing an additional potential drop that can be much larger than implied by Eq.~\eqref{eq:delta_v} and, crucially, is not tied to the rotational frequency of the pulsar. As a result, sufficiently slowly rotating pulsars can sustain pair production with comparable efficiency to much younger objects. The non-observation of such slowly rotating radio pulsars can therefore be used as evidence against the presence of long-range axion hair.

\subsection{Identifying the Death Valley}

The discussion above provides a qualitative understanding of pulsar death. Here we outline a quantitative calculation capturing the impact of axion-induced electric fields on the pulsar population.

For pulsars with active pair production, electric fields are largely screened and the current flow is set by the large-scale magnetic structure, $\vec{j} \simeq \nabla \times \vec{B}$, approaching a force-free configuration. Deviations arise when the supplied charges cannot simultaneously support the required current and screen the accelerating field: an electric-field gradient forms, and if the voltage drop is sufficiently large, a pair discharge occurs (temporarily screening the field). As pulsars age, pair production eventually ceases, driving the system toward an inactive state with vanishing current and no coherent radio emission. The key question is whether axion hair allows sustained discharge in old pulsars that would otherwise be below the conventional death line.

Using a one-dimensional approximation, one can combine the Lorentz force equation, Gauss' law, and Amp\`{e}re's law (momentarily neglecting axion gradients and radiative losses) into a differential equation governing the evolution of the Lorentz factor along a field line~\cite{TimokhinArons2013}:
\begin{eqnarray}\label{eq:dgam_3}
    \left(\frac{d\gamma}{ds} \right)^2 = 2 \left[\alpha_0 \sqrt{\gamma^2 - 1}  - \gamma + 1 \right] \, .
\end{eqnarray}
Here $s$ is the distance along the field line (normalized to the GJ skin depth), $\vec{j}_{\rm m}\equiv\nabla\times\vec{B}$ is the current density fixed by the large-scale solution, and $\alpha_0 \equiv j_{\rm m}/\rho_{\rm GJ}$ is the discharge parameter. For $\alpha_0 \ge 1$, $E_\parallel$ grows with altitude and particles are accelerated until pair production occurs or geometric corrections suppress acceleration. For $0 \le \alpha_0 < 1$, there exists a maximum Lorentz factor $\gamma^{\rm max}=(1+\alpha_0^2)/(1-\alpha_0^2)$ beyond which particles cannot be accelerated and discharges do not occur. In force-free solutions, $\alpha_0$ is a spatially varying $\mathcal{O}(1)$ function across open field lines; we give explicit examples and discuss field-line dependence in the companion PRD.

Generalizing to include axion hair and curvature losses\footnote{We work with the second-order form rather than the square of the first-order equation.}, one finds~\cite{Caputo:2023cpv}
\begin{eqnarray}\label{eq:dgam_2}
    \frac{d^2\gamma}{ds^2} = \frac{j_{\rm m}}{\rho_{\rm GJ}} \frac{\gamma}{\sqrt{\gamma^2 - 1}} - 1 - \frac{\rho_a}{\rho_{\rm GJ}} - \mathcal{R}_\gamma \, ,
\end{eqnarray}
where $\mathcal{R}_\gamma$ accounts for curvature-radiation losses~\cite{Jackson, Arons1983},
\begin{eqnarray} \label{eq:Rgamma}
\mathcal{R}_\gamma = \frac{8}{3} \frac{e^2}{m_e \rho_c^2} \gamma^3 \frac{d \gamma}{dt} \, ,
\end{eqnarray}
with $\rho_c$ the radius of curvature. For middle-aged and old pulsars, $|\rho_a/\rho_{\rm GJ}|\gg 1$ near the star, motivating the intuition that the near-surface behavior resembles Eq.~\eqref{eq:dgam_3} with $\rho_{\rm GJ}\to\rho_a$, yielding an effective discharge parameter $\alpha^{\rm eff}=\pm j_{\rm m}/\rho_a \sim 0^\pm$ (the sign depends on the magnetic pole). Because the axion profile is exponentially suppressed beyond $r\sim m_a^{-1}$, the system transitions back to the ordinary $\alpha_0$ regime at larger radii, and new null surfaces defined by $\rho_{\rm GJ}+\rho_a=0$ can arise. The full discharge phenomenology across poles, including null-surface formation and return-current consistency, is established and validated with PIC simulations in the companion PRD.

Let us summarize the qualitative outcomes relevant for pulsar demographics, focusing first on regions of the polar cap with $\alpha_0>0$ (return currents $\alpha_0<0$ are addressed below). Without loss of generality take $\vec{\Omega}\cdot\vec{B}>0$ so that $\rho_{\rm GJ}<0$:
\begin{itemize}
    \item Northern pole: $\rho_a>0$. A null surface forms where $\rho_a=-\rho_{\rm GJ}$. The current is carried by electrons, whereas screening $\rho_a$ would require positive charges. As a result, an accelerating field develops and, for sufficiently large $\rho_a$, pair discharge is triggered. This persists even for pulsars that would be below the conventional death line.
    \item Southern pole: $\rho_a<0$. The axion-induced effective charge density can be screened by slowly moving electrons, and beyond $r\gtrsim m_a^{-1}$ particles are accelerated primarily by the rotationally induced field. In this regime, long-range axion hair can effectively displace the location of the near-surface accelerating region, and for sufficiently light axions can reduce the available voltage in one hemisphere.
\end{itemize}

To shift the death line in a globally self-consistent way, pair production must also remain active on field lines carrying return currents, $\alpha_0<0$; otherwise the global solution can collapse to an inactive disk--dome state. In simulations, gaps often open at altitudes $\mathcal{O}(R_{\rm NS})$ and ignite once local conditions allow discharge. Distinguishing the two poles:
\begin{itemize}
    \item Northern pole: $\rho_a>0$. The return-current region can supply the required current while screening the axion-induced field near the surface; particles are then accelerated mainly by the rotationally induced field at $r\gtrsim m_a^{-1}$, and discharge can fail for pulsars below the standard death line.
    \item Southern pole: $\rho_a<0$. The rotationally induced and axion-induced fields can act constructively, enabling return-current particles to reach discharge threshold even when rotation alone would be insufficient.
\end{itemize}
Collectively, this implies that pulsars conventionally considered ``dead'' can remain radio active out to much larger periods, yielding the demographic tension emphasized above. A notable consequence of the polarity asymmetry is a generic tension between long-range axion hair and the existence of interpulses in near-orthogonal rotators; PSR J1906+0746~\cite{Kramer:2008iw,2019Sci...365.1013D} provides an important example, and the detailed interpulse analysis is presented in the companion PRD.

To identify the corresponding ``death valley''\footnote{The term death valley, rather than death line, is often used since there is no unique mapping between $\dot{P}$ and the existence of $e^\pm$ production.}, we solve Eq.~\eqref{eq:dgam_2} over a range of pulsar parameters. At each step we compute the highest-energy curvature photon emitted by the primary particle, propagate it outward (including Schwarzschild effects for a $M=1.4\,M_\odot$ star~\cite{Witte2021,McDonald:2023shx}), and evaluate the optical depth using Eq.~\eqref{Eq:OpticalDepth}. We define $\ell_{e^\pm}$ as the distance traveled by the photon when the optical depth reaches unity and define the pair-production distance as
$d_{\rm pp}\equiv \ell_{\rm acc}+\ell_{\rm rad}+\ell_{\rm mfp}$,
where $\ell_{\rm acc}$ is the acceleration distance, $\ell_{\rm mfp}$ satisfies $\tau(\ell_{\rm mfp})=1$, and
\begin{eqnarray}\label{eq:ellrad}
    \ell_{\rm rad} \sim \sqrt{\frac{8\pi}{3}} \sqrt{\frac{\omega}{\omega_c}} \, \frac{\rho_c \, e^{\omega / \omega_c}}{e^2 \gamma} \, .
\end{eqnarray}
The gap height is determined by minimizing $d_{\rm pp}$ along the trajectory; pulsars with non-finite gap heights are classified as dead. The full numerical implementation, as well as a simple analytical recipe, can be found in the Appendices. 

We map the results into the $P$--$\dot{P}$ plane assuming $\dot{P} \simeq \beta B^2 / P$, with $\beta = \pi^2 R_{\rm NS}^6 / I_{\rm NS} \simeq 6 \times 10^{-40} \, {\rm s/G^2}$ for canonical NS parameters. We compare in Fig.~\ref{fig:ppdot_std} death lines without (yellow) and with (red) axion hair for a light QCD axion with $\epsilon=10^{-2}$ and $f_a=10^{16}$~GeV. The figure highlights the main conclusion: axion hair induces a large, observationally incompatible shift in pulsar death.

\section{Results and Conclusions}

\begin{figure*}
    \centering
    \includegraphics[width=0.475\textwidth]{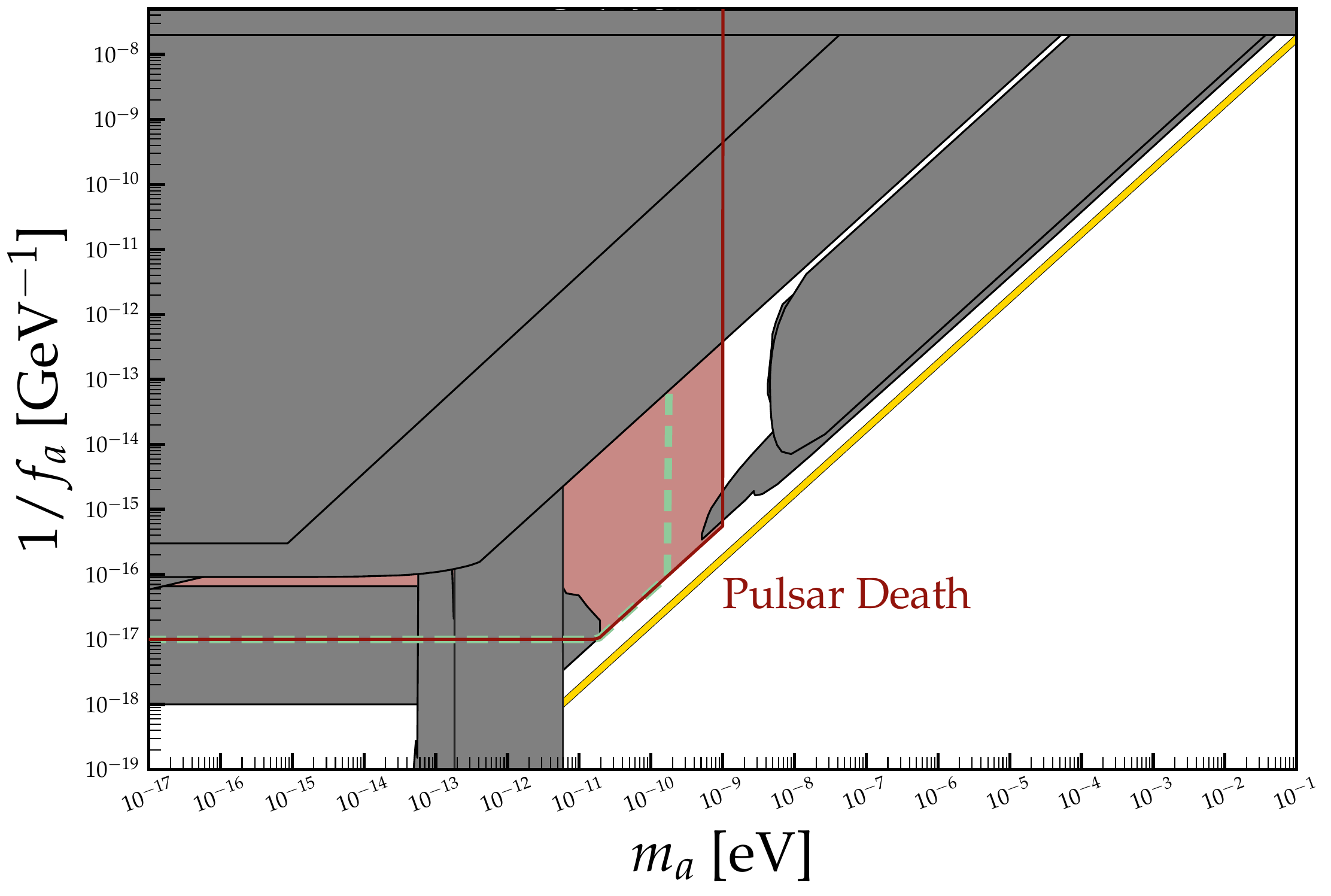}
    \includegraphics[width=0.49\textwidth]{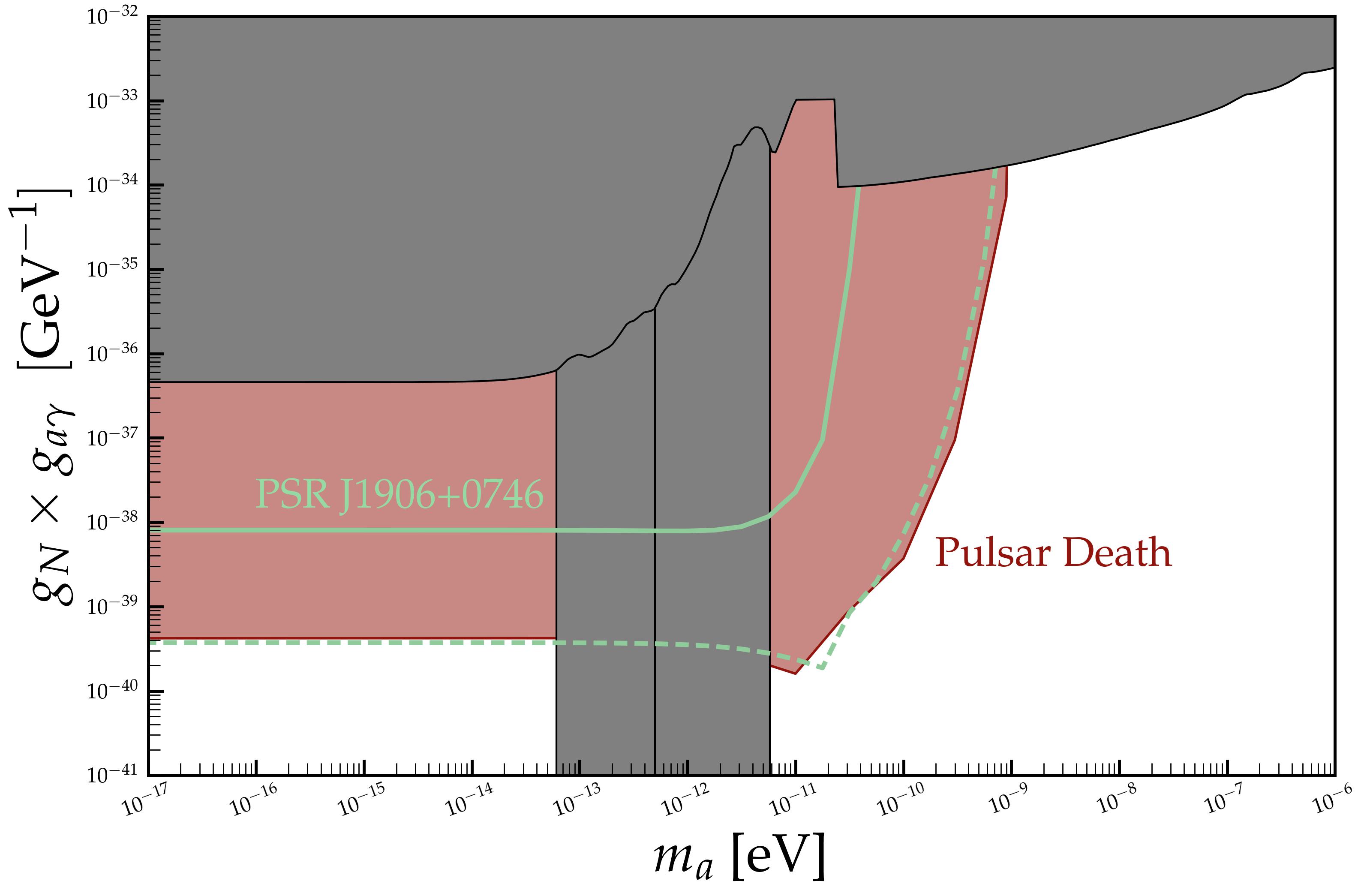}
    \caption{
    Constraints on the light QCD axion (left), set at a confidence level of greater than  $95\%$ (see text for details), and the linearly coupled axion (right) , set at the $95\%$ confidence level.  Grey regions are excluded by existing constraints~\cite{Hook:2017psm,Springmann:2024ret,Balkin:2022qer,Zhang:2021mks,Gomez-Banon:2024oux,Kumamoto:2024wjd,Witte:2024drg,Caputo:2025oap,Smith:1999cr,Berge:2017ovy,Noordhuis:2022ljw,Benabou:2025jcv,Ning:2024eky,Reynes:2021bpe,Haxton:2025xqz,Ning:2025kyu,Fiorillo:2025gnd}. Note that for the QCD axion we assumed an untuned axion-photon coupling $C \gtrsim 10^{-2}$. Note further that for the linearly coupled axion, we take the strongest constraints in this mass range of the axion-photon coupling, which are set by a combination of pulsar polar caps~\cite{Noordhuis:2022ljw}, magnetic white dwarf polarization~\cite{Benabou:2025jcv}, NuSTAR Observations of M82 and M87~\cite{Ning:2024eky}, and Chandra observations of H1821+643~\cite{Reynes:2021bpe}. Red regions highlight the pulsar-death analysis presented here. We truncate the death-line analysis at $m_a = 10^{-9}\,{\rm eV}$; for larger masses the axion-induced voltage drop becomes comparable to the transverse scale of the return-current field-line bundle and the global magnetospheric response may be altered (see discussion below). Green lines in the right panel show constraints (solid) from interpulses in PSR J1906+0746; the interpulse modeling and the sign dependence are presented in the companion PRD. The yellow line in the left panel corresponds to the conventional QCD axion relation.
    }
    \label{fig:limits}
\end{figure*}

The preceding section allows one to estimate, for a fixed set of pulsar parameters, whether a pulsar produces near-field radio emission. The remaining question is whether one can observationally distinguish a pulsar population whose lifetimes have been extended by axion hair from a population evolving under standard electrodynamics.

Fig.~\ref{fig:ppdot_std} suggests that the observed pulsar population broadly aligns with standard expectations for pulsar death; however one must confirm that the absence of high-period radio pulsars is not dominated by observational biases. To test this, we perform a population synthesis analysis (following \eg~\cite{lyne1985galactic,faucher2006birth,Bates:2013uma,Gullon:2014dva,Ronchi:2021arl,rea2024long}) by:
$(i)$ sampling pulsar birth locations,
$(ii)$ drawing pulsar parameters at birth from parameterized distributions,
$(iii)$ determining present-day locations by sampling ages and evolving motion in the Galactic potential,
$(iv)$ evolving magneto-rotational spin down to obtain present-day $P$, $\dot{P}$ and magnetic fields,
$(v)$ applying observational selection cuts (radio flux thresholds, beaming geometry, pulse width, survey sensitivity, etc.),
$(vi)$ comparing simulated and observed populations via a test statistic, and
$(vii)$ optimizing over free parameters.
We test multiple distributions and assumptions to (1) show that minimal models can fit observations, (2) demonstrate that models without a death line are disfavored, and (3) assess the importance of each assumption. \emph{All technical details of the population synthesis, selection modeling, and statistical procedure are provided in the Appendices.}

To infer axion-compatible parameter space, we repeat the population synthesis using death lines specific to each model. We apply a conservative $50\%$ visibility cut for pulsars below the line to account for hemispheric asymmetries in pair production. For each parameter set, we generate up to 10 mock populations, compute the mean and standard deviation of pulsars below the death line, and identify where a $2\sigma$ downward statistical fluctuation exceeds 14---corresponding to a $2\sigma$ excess of ATNF pulsars below the fiducial line (limits on the light QCD axion are highly insensitive to this threshold). The resulting exclusions are shown in Fig.~\ref{fig:limits}\footnote{We have verified that constraints on the axion-photon interaction for the light QCD axion are sub-dominant to those presented here.}. Recall that the region of parameter space being probed for the light QCD axion is bounded to $f_a \leq 10^{17}$GeV, $m_a \lesssim 10^{-9}$eV, and $\epsilon \lesssim 0.07$ -- within this range, all parameter space is excluded at greater than $95\%$ confidence level (note that this is not exactly at $95\%$, but greater than this value -- this is because one cannot recover the conventional death line by smoothly vary e.g. $\epsilon$ at fixed axion mass). Note that for our fiducial analysis, we have chosen to not consider axion masses $m_a \gtrsim 10^{-9}$ eV. This is because our one-dimensional treatment is only valid when the transverse scale of the pair producing field lines can be neglected -- for return currents, this implies pair production should occur at distances away from the surface $L_{\rm pp }\gtrsim r_{\rm pc} / \zeta$, with $\zeta \gtrsim 2$ characterizing the fractional size of the polar cap spanned by the return current field lines. For low mass axions, the potential drop is localized at distances larger than this threshold. For heavy axions, however, this is not the case. In order to determine where this characteristic threshold cut should be, we adopt the analytic formalism described in the Appendices to compute death lines for a variety of axion masses, but inducing an artificial shift in the surface of the neutron star by a distance $r_{\rm pc} / 2$ -- this effectively serves to remove the near-field axion potential drop from the analysis, and allows us to assess only the impact of the exponentially suppressed tail. 

Let us highlight that the constraints derived on the light QCD axion have been obtained using an axion-photon coupling $g_{a\gamma\gamma} = C \alpha_{\rm EM} / (2\pi f_a)$, with $C=1$. More generally, $C \equiv E/N - 1.92$, with $E/N$ being the ratio of electromagnetic to color anomaly coefficients. For the KSVZ, DSVZ-I axions, this would imply $C\sim-1.92$ and $C\sim 0.75$ respectively. In general, $\mathcal{O}(1)$ variations in $C$ induce very small shifts in the death lines, but have no impact on the derived constraints (this is because the region of the $P-\dot{P}$ diagram driving the constraints is typically quite far from the derived death lines). By recomputing the death lines with suppressed values of $C$, and comparing with synthesized pulsar populations, we estimate that our constraints are expected to be valid for values of $C \gtrsim 10^{-2}$; for smaller values, which require a fine-tuned cancellation between the anomaly coefficients and the contribution from the pion, our constraints would not apply.

We have shown that axion gradients sourced from dense nuclear matter can decouple the electric field responsible for acceleration and pair production in the inner magnetosphere from the rotational period of the pulsar itself, leading to a population of slowly rotating radio-active stars and, in some cases, an incompatibility with interpulses. The non-observation of such a population leads to competitive constraints on the light QCD axion and on linear axion--nucleon interactions, as summarized in Fig.~\ref{fig:limits}. The companion PRD provides the complete electrodynamic derivations, PIC simulation details, and extended systematics underlying these results, while the Appendices contain the population synthesis and statistical inference procedure, and additionally draw on Refs.~\cite{Galishnikova:2020gmu,Akmal:1998cf,Gralla:2017nbw,beloborodov2008polar,malov2011geometry,zhang1997three,Harding:1998ma,Harding:2001at,johnston2020galactic,Kiziltan:2009rx,Hobbs:2005yx,lorimer2006parkes,Cieslar:2018jzd,Graber:2023jgz,du2024initial,Philippov:2013aha,Passamonti:2016nmf,Pons:2007vf,Pons:2019zyc,Cumming:2004mf,Aguilera:2007xk,Vigano:2021olr,Mereghetti:2015asa,Igoshev:2018fcp,yao2017new,berger2014kolmogorov,rizzo2016energy,peacock1983two,fasano1987multidimensional,rozwadowska2021rate,SKAOPulsarScienceWorkingGroup:2025zsy,padmanabh2023mpifr}.

\vspace{.2cm}
\section{Acknowledgments}%
We thank Anson Hook, Konstantin Springmann and Kai Bartnick for enlightening discussions, and Ani Prabhu for their useful comments on the manuscript. SJW acknowledges support from a Royal Society University Research Fellowship (URF-R1-231065). This work is also supported by the Deutsche Forschungsgemeinschaft under Germany's Excellence Strategy EXC 2121 Quantum Universe 390833306. AC is supported by an ERC STG grant (``AstroDarkLS'', grant No. 101117510). AC acknowledges the Weizmann Institute of Science for hospitality at different stages of this project and the support from the Benoziyo Endowment Fund for the Advancement of
Science. 
SS acknowledges financial support from the Spanish Ministry of Science and Innovation (MICINN) through the Spanish State Research Agency, under Severo Ochoa Centres of Excellence Programme 2025-2029 (CEX2024001442-S).
This work is also part of the R\&D\&i project PID2023-146686NB-C31, funded by MICIU/AEI/10.13039/501100011033/ and by ERDF/EU. IFAE is partially funded by the CERCA program of the Generalitat de Catalunya.
A.Ch. is supported by Martin A. and Helen Chooljian Member Fund and the Fund for Memberships in Natural Sciences.  This article/publication is based upon work from COST Action COSMIC WISPers CA21106, supported by COST (European Cooperation in Science and Technology). This work was supported by a grant from the Simons Foundation (MP-SCMPS-00001470), as well as an Alfred P. Sloan Fellowship and a Packard Foundation Fellowship in Science and Engineering to AP. This work was supported by the U.S.~Department of Energy~(DOE), Office of Science, National Quantum Information Science Research Centers, Superconducting Quantum Materials and Systems Center~(SQMS) under Contract No.~DE-AC02-07CH11359.  S.R.\ is supported in part by the U.S.~National Science Foundation~(NSF) under Grant No.~PHY-2412361.The work of S.R.\  was also supported by the Simons Investigator Award No.~827042.

\bibliographystyle{bibi}
\bibliography{biblio}

\providecommand{\href}[2]{#2}\begingroup\raggedright\begin{thebibliography}{100}

\bibitem{Peccei:1977hh}
R.~D. Peccei and H.~R. Quinn, \emph{{CP Conservation in the Presence of
  Instantons}}, \href{https://doi.org/10.1103/PhysRevLett.38.1440}{\emph{Phys.
  Rev. Lett.} {\bfseries 38} (1977) 1440}. [,328(1977)].

\bibitem{WeinbergAxion}
S.~Weinberg, \emph{A new light boson?},
  \href{https://doi.org/10.1103/PhysRevLett.40.223}{\emph{Phys. Rev. Lett.}
  {\bfseries 40} (1978) 223}.

\bibitem{WilczekAxion}
F.~Wilczek, \emph{Problem of strong {$P$} and {$T$} invariance in the presence
  of instantons}, \href{https://doi.org/10.1103/PhysRevLett.40.279}{\emph{Phys.
  Rev. Lett.} {\bfseries 40} (1978) 279}.

\bibitem{Abel:2020pzs}
C.~Abel et~al., \emph{{Measurement of the Permanent Electric Dipole Moment of
  the Neutron}},
  \href{https://doi.org/10.1103/PhysRevLett.124.081803}{\emph{Phys. Rev. Lett.}
  {\bfseries 124} (2020) 081803}
  [\href{https://arxiv.org/abs/2001.11966}{{\ttfamily 2001.11966}}].

\bibitem{Hook:2018jle}
A.~Hook, \emph{{Solving the Hierarchy Problem Discretely}},
  \href{https://doi.org/10.1103/PhysRevLett.120.261802}{\emph{Phys. Rev. Lett.}
  {\bfseries 120} (2018) 261802}
  [\href{https://arxiv.org/abs/1802.10093}{{\ttfamily 1802.10093}}].

\bibitem{DiLuzio:2021pxd}
L.~Di~Luzio, B.~Gavela, P.~Quilez and A.~Ringwald, \emph{{An even lighter QCD
  axion}}, \href{https://doi.org/10.1007/JHEP05(2021)184}{\emph{JHEP}
  {\bfseries 05} (2021) 184}
  [\href{https://arxiv.org/abs/2102.00012}{{\ttfamily 2102.00012}}].

\bibitem{Hook:2017psm}
A.~Hook and J.~Huang, \emph{{Probing axions with neutron star inspirals and
  other stellar processes}},
  \href{https://doi.org/10.1007/JHEP06(2018)036}{\emph{JHEP} {\bfseries 06}
  (2018) 036} [\href{https://arxiv.org/abs/1708.08464}{{\ttfamily
  1708.08464}}].

\bibitem{Banerjee:2025kov}
A.~Banerjee, M.~A. Buen-Abad and A.~Hook, \emph{{Stacking the Deck: Gambling on
  a Light QCD Axion}},  \href{https://arxiv.org/abs/2507.02049}{{\ttfamily
  2507.02049}}.

\bibitem{Balkin:2022qer}
R.~Balkin, J.~Serra, K.~Springmann, S.~Stelzl and A.~Weiler, \emph{{White
  dwarfs as a probe of exceptionally light QCD axions}},
  \href{https://doi.org/10.1103/PhysRevD.109.095032}{\emph{Phys. Rev. D}
  {\bfseries 109} (2024) 095032}
  [\href{https://arxiv.org/abs/2211.02661}{{\ttfamily 2211.02661}}].

\bibitem{Bartnick:2025lbg}
K.~Bartnick, K.~Springmann, S.~Stelzl and A.~Weiler, \emph{{$\phi$-Dwarfs:
  White Dwarfs probe Quadratically Coupled Scalars}},
  \href{https://arxiv.org/abs/2509.25305}{{\ttfamily 2509.25305}}.

\bibitem{Bartnick:2025lbv}
K.~Bartnick, D.~Koester, R.-P. Kudritzki, K.~Springmann, S.~Stelzl and
  A.~Weiler, \emph{{New ultralight scalar particles and the mass-radius
  relation of white dwarfs - the important role of Sirius B}},
  \href{https://arxiv.org/abs/2510.06312}{{\ttfamily 2510.06312}}.

\bibitem{Balkin:2023xtr}
R.~Balkin, J.~Serra, K.~Springmann, S.~Stelzl and A.~Weiler, \emph{{Heavy
  neutron stars from light scalars}},
  \href{https://doi.org/10.1007/JHEP02(2025)141}{\emph{JHEP} {\bfseries 02}
  (2025) 141} [\href{https://arxiv.org/abs/2307.14418}{{\ttfamily
  2307.14418}}].

\bibitem{Gomez-Banon:2024oux}
A.~G\'omez-Ba\~n\'on, K.~Bartnick, K.~Springmann and J.~A. Pons,
  \emph{{Constraining Light QCD Axions with Isolated Neutron Star Cooling}},
  \href{https://doi.org/10.1103/PhysRevLett.133.251002}{\emph{Phys. Rev. Lett.}
  {\bfseries 133} (2024) 251002}
  [\href{https://arxiv.org/abs/2408.07740}{{\ttfamily 2408.07740}}].

\bibitem{Kumamoto:2024wjd}
M.~Kumamoto, J.~Huang, C.~Drischler, M.~Baryakhtar and S.~Reddy, \emph{{Pi in
  the Sky: Neutron Stars with Exceptionally Light QCD Axions}},
  \href{https://arxiv.org/abs/2410.21590}{{\ttfamily 2410.21590}}.

\bibitem{Kahn:2025ytc}
Y.~Kahn, M.~Wentzel and N.~Yunes, \emph{{Effects of Lighter-than-QCD Axions on
  Neutron Star Tidal Deformability}},
  \href{https://arxiv.org/abs/2511.13812}{{\ttfamily 2511.13812}}.

\bibitem{Bai:2023bbg}
Y.~Bai and C.~H. de~Lima, \emph{{Electrobaryonic axion: hair of neutron
  stars}}, \href{https://doi.org/10.1007/JHEP05(2024)312}{\emph{JHEP}
  {\bfseries 05} (2024) 312}
  [\href{https://arxiv.org/abs/2311.18794}{{\ttfamily 2311.18794}}].

\bibitem{Sturrock1971}
P.~A. {Sturrock}, \emph{{A Model of Pulsars}},
  \href{https://doi.org/10.1086/150865}{\emph{\apj} {\bfseries 164} (1971)
  529}.

\bibitem{RudermanSutherland1975}
M.~A. {Ruderman} and P.~G. {Sutherland}, \emph{{Theory of pulsars: polar gaps,
  sparks, and coherent microwave radiation.}},
  \href{https://doi.org/10.1086/153393}{\emph{\apj} {\bfseries 196} (1975) 51}.

\bibitem{ChenRuderman93}
K.~{Chen} and M.~{Ruderman}, \emph{{Pulsar Death Lines and Death Valley}},
  \href{https://doi.org/10.1086/172129}{\emph{\apj} {\bfseries 402} (1993)
  264}.

\bibitem{Witte:CompanionPRD}
S.~J. Witte, A.~Caputo, S.~Stelzl, A.~Chernoglazov, A.~A. Philippov and
  S.~Rajendran, \emph{Axion hair and pulsar electrodynamics: modeling,
  discharge dynamics, and particle-in-cell simulations},  2026.
\newblock in preparation.

\bibitem{Fedderke:2023dwj}
M.~A. Fedderke, J.~O. Thompson, R.~Cervantes, B.~Giaccone, R.~Harnik, D.~E.
  Kaplan, S.~Posen and S.~Rajendran, \emph{{Measuring axion gradients with
  photon interferometry}},
  \href{https://doi.org/10.1103/PhysRevD.109.015025}{\emph{Phys. Rev. D}
  {\bfseries 109} (2024) 015025}
  [\href{https://arxiv.org/abs/2304.11261}{{\ttfamily 2304.11261}}].

\bibitem{Pospelov_1998}
M.~Pospelov, \emph{Odd interaction of axion with matter},
  \href{https://doi.org/10.1103/physrevd.58.097703}{\emph{Physical Review D}
  {\bfseries 58} (1998) }.

\bibitem{Moody:1984ba}
J.~E. Moody and F.~Wilczek, \emph{{NEW MACROSCOPIC FORCES?}},
  \href{https://doi.org/10.1103/PhysRevD.30.130}{\emph{Phys. Rev. D} {\bfseries
  30} (1984) 130}.

\bibitem{Gao:2021fyk}
C.~Gao and A.~Stebbins, \emph{{Structure of stellar remnants with coupling to a
  light scalar}},
  \href{https://doi.org/10.1088/1475-7516/2022/07/025}{\emph{JCAP} {\bfseries
  07} (2022) 025} [\href{https://arxiv.org/abs/2110.07012}{{\ttfamily
  2110.07012}}].

\bibitem{Balkin:2020dsr}
R.~Balkin, J.~Serra, K.~Springmann and A.~Weiler, \emph{{The QCD axion at
  finite density}}, \href{https://doi.org/10.1007/JHEP07(2020)221}{\emph{JHEP}
  {\bfseries 07} (2020) 221}
  [\href{https://arxiv.org/abs/2003.04903}{{\ttfamily 2003.04903}}].

\bibitem{DiVecchia:1980yfw}
P.~Di~Vecchia and G.~Veneziano, \emph{{Chiral Dynamics in the Large n Limit}},
  \href{https://doi.org/10.1016/0550-3213(80)90370-3}{\emph{Nucl. Phys. B}
  {\bfseries 171} (1980) 253}.

\bibitem{GrillidiCortona:2015jxo}
G.~Grilli~di Cortona, E.~Hardy, J.~Pardo~Vega and G.~Villadoro, \emph{{The QCD
  axion, precisely}},
  \href{https://doi.org/10.1007/JHEP01(2016)034}{\emph{JHEP} {\bfseries 01}
  (2016) 034} [\href{https://arxiv.org/abs/1511.02867}{{\ttfamily
  1511.02867}}].

\bibitem{DiLuzio:2021gos}
L.~Di~Luzio, B.~Gavela, P.~Quilez and A.~Ringwald, \emph{{Dark matter from an
  even lighter QCD axion: trapped misalignment}},
  \href{https://doi.org/10.1088/1475-7516/2021/10/001}{\emph{JCAP} {\bfseries
  10} (2021) 001} [\href{https://arxiv.org/abs/2102.01082}{{\ttfamily
  2102.01082}}].

\bibitem{Banerjee:2022wzk}
A.~Banerjee, J.~Eby and G.~Perez, \emph{{From axion quality and naturalness
  problems to a high-quality $\mathbb{Z}_{4\mathcal{N}}$ QCD relaxion}},
  \href{https://arxiv.org/abs/2210.05690}{{\ttfamily 2210.05690}}.

\bibitem{zhang2000radio}
B.~Zhang, A.~K. Harding and A.~G. Muslimov, \emph{Radio pulsar death line
  revisited: Is psr j2144--3933anomalous?}, {\emph{The Astrophysical Journal}
  {\bfseries 531} (2000) L135}.

\bibitem{Faucher-Giguere:2005dxp}
C.-A. Faucher-Giguere and V.~M. Kaspi, \emph{{Birth and evolution of isolated
  radio pulsars}}, \href{https://doi.org/10.1086/501516}{\emph{Astrophys. J.}
  {\bfseries 643} (2006) 332}
  [\href{https://arxiv.org/abs/astro-ph/0512585}{{\ttfamily
  astro-ph/0512585}}].

\bibitem{Konar:2019vhj}
S.~Konar and U.~Deka, \emph{{Radio Pulsar Sub-Populations (I) : The Curious
  Case of Nulling Pulsars}},
  \href{https://doi.org/10.1007/s12036-019-9608-z}{\emph{J. Astrophys. Astron.}
  {\bfseries 40} (2019) 42} [\href{https://arxiv.org/abs/1908.07681}{{\ttfamily
  1908.07681}}].

\bibitem{beskin2022pulsar}
V.~Beskin and P.~Litvinov, \emph{Pulsar death line revisited--i. almost vacuum
  gap}, {\emph{Monthly Notices of the Royal Astronomical Society} {\bfseries
  510} (2022) 2572}.

\bibitem{Beskin:2022dbs}
V.~S. Beskin and A.~Y. Istomin, \emph{{Pulsar death line revisited
  \textendash{} II. \textquoteleft{}The death valley\textquoteright{}}},
  \href{https://doi.org/10.1093/mnras/stac2423}{\emph{Mon. Not. Roy. Astron.
  Soc.} {\bfseries 516} (2022) 5084}
  [\href{https://arxiv.org/abs/2207.04723}{{\ttfamily 2207.04723}}].

\bibitem{Philippov:2020jxu}
A.~Philippov, A.~Timokhin and A.~Spitkovsky, \emph{{Origin of Pulsar Radio
  Emission}}, \href{https://doi.org/10.1103/PhysRevLett.124.245101}{\emph{Phys.
  Rev. Lett.} {\bfseries 124} (2020) 245101}
  [\href{https://arxiv.org/abs/2001.02236}{{\ttfamily 2001.02236}}].

\bibitem{Chernoglazov:2024rvo}
A.~Chernoglazov, A.~Philippov and A.~Timokhin, \emph{{Coherence of
  Multidimensional Pair Production Discharges in Polar Caps of Pulsars}},
  \href{https://doi.org/10.3847/2041-8213/ad7e24}{\emph{Astrophys. J. Lett.}
  {\bfseries 974} (2024) L32}
  [\href{https://arxiv.org/abs/2409.15409}{{\ttfamily 2409.15409}}].

\bibitem{SashaReview}
A.~{Philippov} and M.~{Kramer}, \emph{{Pulsar Magnetospheres and Their
  Radiation}},
  \href{https://doi.org/10.1146/annurev-astro-052920-112338}{\emph{Annual
  Review of Astronomy and Astrophysics} {\bfseries 60} (2022) 495}.

\bibitem{daugherty1982electromagnetic}
J.~Daugherty and A.~K. Harding, \emph{Electromagnetic cascades in pulsars},
  {\emph{Astrophysical Journal, Part 1, vol. 252, Jan. 1, 1982, p. 337-347.}
  {\bfseries 252} (1982) 337}.

\bibitem{daugherty1983pair}
J.~K. Daugherty and A.~K. Harding, \emph{Pair production in superstrong
  magnetic fields}, .

\bibitem{TimokhinHarding2015}
A.~N. {Timokhin} and A.~K. {Harding}, \emph{{On the Polar Cap Cascade Pair
  Multiplicity of Young Pulsars}},
  \href{https://doi.org/10.1088/0004-637X/810/2/144}{\emph{\apj} {\bfseries
  810} (2015) 144} [\href{https://arxiv.org/abs/1504.02194}{{\ttfamily
  1504.02194}}].

\bibitem{Timokhin:2018vdn}
A.~N. Timokhin and A.~K. Harding, \emph{{On the maximum pair multiplicity of
  pulsar cascades}},
  \href{https://doi.org/10.3847/1538-4357/aaf050}{\emph{Astrophys. J.}
  {\bfseries 871} (2019) 12}
  [\href{https://arxiv.org/abs/1803.08924}{{\ttfamily 1803.08924}}].

\bibitem{Goldreich:1969sb}
P.~Goldreich and W.~H. Julian, \emph{{Pulsar electrodynamics}},
  \href{https://doi.org/10.1086/150119}{\emph{Astrophys. J.} {\bfseries 157}
  (1969) 869}.

\bibitem{harding2002regimes}
A.~K. Harding, A.~G. Muslimov and B.~Zhang, \emph{Regimes of pulsar pair
  formation and particle energetics}, {\emph{The Astrophysical Journal}
  {\bfseries 576} (2002) 366}.

\bibitem{TimokhinArons2013}
A.~N. Timokhin and J.~Arons, \emph{{Current Flow and Pair Creation at Low
  Altitude in Rotation Powered Pulsars' Force-Free Magnetospheres: Space-Charge
  Limited Flow}}, \href{https://doi.org/10.1093/mnras/sts298}{\emph{Mon. Not.
  Roy. Astron. Soc.} {\bfseries 429} (2013) 20}
  [\href{https://arxiv.org/abs/1206.5819}{{\ttfamily 1206.5819}}].

\bibitem{Caputo:2023cpv}
A.~Caputo, S.~J. Witte, A.~A. Philippov and T.~Jacobson, \emph{{Pulsar Nulling
  and Vacuum Radio Emission from Axion Clouds}},
  \href{https://doi.org/10.1103/PhysRevLett.133.161001}{\emph{Phys. Rev. Lett.}
  {\bfseries 133} (2024) 161001}
  [\href{https://arxiv.org/abs/2311.14795}{{\ttfamily 2311.14795}}].

\bibitem{Jackson}
J.~D. Jackson, \emph{{Classical Electrodynamics}}. Wiley, 1998.

\bibitem{Arons1983}
J.~{Arons}, \emph{{Pair creation above pulsar polar caps : geometrical
  structure and energetics of slot gaps.}},
  \href{https://doi.org/10.1086/160771}{\emph{\apj} {\bfseries 266} (1983)
  215}.

\bibitem{Kramer:2008iw}
M.~Kramer and S.~Johnston, \emph{{High-precision geometry of a double-pole
  pulsar}}, \href{https://doi.org/10.1111/j.1365-2966.2008.13780.x}{\emph{Mon.
  Not. Roy. Astron. Soc.} {\bfseries 390} (2008) 87}
  [\href{https://arxiv.org/abs/0807.5013}{{\ttfamily 0807.5013}}].

\bibitem{2019Sci...365.1013D}
G.~{Desvignes}, M.~{Kramer}, K.~{Lee}, J.~{van Leeuwen}, I.~{Stairs},
  A.~{Jessner}, I.~{Cognard}, L.~{Kasian}, A.~{Lyne} and B.~W. {Stappers},
  \emph{{Radio emission from a pulsar{\textquoteright}s magnetic pole revealed
  by general relativity}},
  \href{https://doi.org/10.1126/science.aav7272}{\emph{Science} {\bfseries 365}
  (2019) 1013} [\href{https://arxiv.org/abs/1909.06212}{{\ttfamily
  1909.06212}}].

\bibitem{Witte2021}
S.~J. Witte, D.~Noordhuis, T.~D. Edwards and C.~Weniger, \emph{Axion-photon
  conversion in neutron star magnetospheres: The role of the plasma in the
  {Goldreich-Julian} model},
  \href{https://doi.org/10.1103/physrevd.104.103030}{\emph{Physical Review D}
  {\bfseries 104} (2021) }.

\bibitem{McDonald:2023shx}
J.~I. McDonald and S.~J. Witte, \emph{{Generalized ray tracing for axions in
  astrophysical plasmas}},
  \href{https://doi.org/10.1103/PhysRevD.108.103021}{\emph{Phys. Rev. D}
  {\bfseries 108} (2023) 103021}
  [\href{https://arxiv.org/abs/2309.08655}{{\ttfamily 2309.08655}}].

\bibitem{Springmann:2024ret}
K.~Springmann, M.~Stadlbauer, S.~Stelzl and A.~Weiler, \emph{{A Universal Bound
  on QCD Axions from Supernovae}},
  \href{https://arxiv.org/abs/2410.19902}{{\ttfamily 2410.19902}}.

\bibitem{Zhang:2021mks}
J.~Zhang, Z.~Lyu, J.~Huang, M.~C. Johnson, L.~Sagunski, M.~Sakellariadou and
  H.~Yang, \emph{{First Constraints on Nuclear Coupling of Axionlike Particles
  from the Binary Neutron Star Gravitational Wave Event GW170817}},
  \href{https://doi.org/10.1103/PhysRevLett.127.161101}{\emph{Phys. Rev. Lett.}
  {\bfseries 127} (2021) 161101}
  [\href{https://arxiv.org/abs/2105.13963}{{\ttfamily 2105.13963}}].

\bibitem{Witte:2024drg}
S.~J. Witte and A.~Mummery, \emph{{Stepping up superradiance constraints on
  axions}}, \href{https://doi.org/10.1103/PhysRevD.111.083044}{\emph{Phys. Rev.
  D} {\bfseries 111} (2025) 083044}
  [\href{https://arxiv.org/abs/2412.03655}{{\ttfamily 2412.03655}}].

\bibitem{Caputo:2025oap}
A.~Caputo, G.~Franciolini and S.~J. Witte, \emph{{Superradiance Constraints
  from GW231123}},  \href{https://arxiv.org/abs/2507.21788}{{\ttfamily
  2507.21788}}.

\bibitem{Smith:1999cr}
G.~L. Smith, C.~D. Hoyle, J.~H. Gundlach, E.~G. Adelberger, B.~R. Heckel and
  H.~E. Swanson, \emph{{Short range tests of the equivalence principle}},
  \href{https://doi.org/10.1103/PhysRevD.61.022001}{\emph{Phys. Rev. D}
  {\bfseries 61} (2000) 022001}
  [\href{https://arxiv.org/abs/2405.10982}{{\ttfamily 2405.10982}}].

\bibitem{Berge:2017ovy}
J.~Berg\'e, P.~Brax, G.~M\'etris, M.~Pernot-Borr\`as, P.~Touboul and J.-P.
  Uzan, \emph{{MICROSCOPE Mission: First Constraints on the Violation of the
  Weak Equivalence Principle by a Light Scalar Dilaton}},
  \href{https://doi.org/10.1103/PhysRevLett.120.141101}{\emph{Phys. Rev. Lett.}
  {\bfseries 120} (2018) 141101}
  [\href{https://arxiv.org/abs/1712.00483}{{\ttfamily 1712.00483}}].

\bibitem{Noordhuis:2022ljw}
D.~Noordhuis, A.~Prabhu, S.~J. Witte, A.~Y. Chen, F.~Cruz and C.~Weniger,
  \emph{{Novel Constraints on Axions Produced in Pulsar Polar Cap Cascades}},
  \href{https://arxiv.org/abs/2209.09917}{{\ttfamily 2209.09917}}.

\bibitem{Benabou:2025jcv}
J.~N. Benabou, C.~Dessert, K.~C. Patra, T.~G. Brink, W.~Zheng, A.~V. Filippenko
  and B.~R. Safdi, \emph{{Search for Axions in Magnetic White Dwarf
  Polarization at Lick and Keck Observatories}},
  \href{https://arxiv.org/abs/2504.12377}{{\ttfamily 2504.12377}}.

\bibitem{Ning:2024eky}
O.~Ning and B.~R. Safdi, \emph{{Leading Axion-Photon Sensitivity with NuSTAR
  Observations of M82 and M87}},
  \href{https://doi.org/10.1103/PhysRevLett.134.171003}{\emph{Phys. Rev. Lett.}
  {\bfseries 134} (2025) 171003}
  [\href{https://arxiv.org/abs/2404.14476}{{\ttfamily 2404.14476}}].

\bibitem{Reynes:2021bpe}
J.~S. Reyn\'es, J.~H. Matthews, C.~S. Reynolds, H.~R. Russell, R.~N. Smith and
  M.~C.~D. Marsh, \emph{{New constraints on light axion-like particles using
  Chandra transmission grating spectroscopy of the powerful cluster-hosted
  quasar H1821+643}}, \href{https://doi.org/10.1093/mnras/stab3464}{\emph{Mon.
  Not. Roy. Astron. Soc.} {\bfseries 510} (2021) 1264}
  [\href{https://arxiv.org/abs/2109.03261}{{\ttfamily 2109.03261}}].

\bibitem{Haxton:2025xqz}
W.~C. Haxton, X.~Liu, A.~McCutcheon and A.~Ray, \emph{{A Continuous Galactic
  Line Source of Axions: The Remarkable Case of Na23}},
  \href{https://doi.org/10.1103/n3v9-2wml}{\emph{Phys. Rev. Lett.} {\bfseries
  135} (2025) 222701} [\href{https://arxiv.org/abs/2505.03038}{{\ttfamily
  2505.03038}}].

\bibitem{Ning:2025kyu}
O.~Ning, A.~Ray and B.~R. Safdi, \emph{{Axion lines from nuclear de-excitations
  in galactic stellar populations}},
  \href{https://doi.org/10.1103/xyjw-9bw7}{\emph{Phys. Rev. D} {\bfseries 113}
  (2026) 035010} [\href{https://arxiv.org/abs/2509.03569}{{\ttfamily
  2509.03569}}].

\bibitem{Fiorillo:2025gnd}
D.~F.~G. Fiorillo, {\'A}.~Gil~Muyor, H.-T. Janka, G.~G. Raffelt and
  E.~Vitagliano, \emph{{Axion-photon conversion in transient compact stars:
  Systematics, constraints, and opportunities}},
  \href{https://doi.org/10.1088/1475-7516/2026/03/053}{\emph{JCAP} {\bfseries
  03} (2026) 053} [\href{https://arxiv.org/abs/2509.13322}{{\ttfamily
  2509.13322}}].

\bibitem{lyne1985galactic}
A.~Lyne, R.~Manchester and J.~Taylor, \emph{The galactic population of
  pulsars}, {\emph{Monthly Notices of the Royal Astronomical Society}
  {\bfseries 213} (1985) 613}.

\bibitem{faucher2006birth}
C.-A. Faucher-Giguere and V.~M. Kaspi, \emph{Birth and evolution of isolated
  radio pulsars}, {\emph{The Astrophysical Journal} {\bfseries 643} (2006)
  332}.

\bibitem{Bates:2013uma}
S.~Bates, D.~Lorimer, A.~Rane and J.~Swiggum, \emph{{PsrPopPy: An open-source
  package for pulsar population simulations}},
  \href{https://doi.org/10.1093/mnras/stu157}{\emph{Mon. Not. Roy. Astron.
  Soc.} {\bfseries 439} (2014) 2893}
  [\href{https://arxiv.org/abs/1311.3427}{{\ttfamily 1311.3427}}].

\bibitem{Gullon:2014dva}
M.~Gull\'on, J.~A. Miralles, D.~Vigan\`o and J.~A. Pons, \emph{{Population
  synthesis of isolated Neutron Stars with magneto-rotational evolution}},
  \href{https://doi.org/10.1093/mnras/stu1253}{\emph{Mon. Not. Roy. Astron.
  Soc.} {\bfseries 443} (2014) 1891}
  [\href{https://arxiv.org/abs/1406.6794}{{\ttfamily 1406.6794}}].

\bibitem{Ronchi:2021arl}
M.~Ronchi, V.~Graber, A.~Garcia-Garcia, J.~A. Pons and N.~Rea, \emph{{Analyzing
  the Galactic Pulsar Distribution with Machine Learning}},
  \href{https://doi.org/10.3847/1538-4357/ac05bd}{\emph{Astrophys. J.}
  {\bfseries 916} (2021) 100}
  [\href{https://arxiv.org/abs/2101.06145}{{\ttfamily 2101.06145}}].

\bibitem{rea2024long}
N.~Rea, N.~Hurley-Walker, C.~Pardo-Araujo, M.~Ronchi, V.~Graber, F.~C. Zelati,
  D.~De~Martino, A.~Bahramian, S.~J. McSweeney, T.~J. Galvin et~al.,
  \emph{Long-period radio pulsars: population study in the neutron star and
  white dwarf rotating dipole scenarios}, {\emph{The Astrophysical Journal}
  {\bfseries 961} (2024) 214}.

\bibitem{Galishnikova:2020gmu}
A.~K. Galishnikova, A.~A. Philippov and V.~S. Beskin, \emph{{Simulations of the
  radio polarization of a precessing pulsar PSR J1906+0746}},
  \href{https://doi.org/10.1093/mnras/staa2070}{\emph{Mon. Not. Roy. Astron.
  Soc.} {\bfseries 497} (2020) 2831}
  [\href{https://arxiv.org/abs/2008.13750}{{\ttfamily 2008.13750}}].

\bibitem{Akmal:1998cf}
A.~Akmal, V.~R. Pandharipande and D.~G. Ravenhall, \emph{{The Equation of state
  of nucleon matter and neutron star structure}},
  \href{https://doi.org/10.1103/PhysRevC.58.1804}{\emph{Phys. Rev. C}
  {\bfseries 58} (1998) 1804}
  [\href{https://arxiv.org/abs/nucl-th/9804027}{{\ttfamily nucl-th/9804027}}].

\bibitem{Gralla:2017nbw}
S.~E. Gralla, A.~Lupsasca and A.~Philippov, \emph{{Inclined Pulsar
  Magnetospheres in General Relativity: Polar Caps for the Dipole, Quadrudipole
  and Beyond}},
  \href{https://doi.org/10.3847/1538-4357/aa978d}{\emph{Astrophys. J.}
  {\bfseries 851} (2017) 137}
  [\href{https://arxiv.org/abs/1704.05062}{{\ttfamily 1704.05062}}].

\bibitem{beloborodov2008polar}
A.~M. Beloborodov, \emph{Polar-cap accelerator and radio emission from
  pulsars}, {\emph{The Astrophysical Journal} {\bfseries 683} (2008) L41}.

\bibitem{malov2011geometry}
I.~Malov and E.~Nikitina, \emph{The geometry of radio pulsar magnetospheres},
  {\emph{Astronomy reports} {\bfseries 55} (2011) 878}.

\bibitem{zhang1997three}
B.~Zhang, G.~Qiao, W.~Lin and J.~Han, \emph{Three modes of pulsar inner gap},
  {\emph{The Astrophysical Journal} {\bfseries 478} (1997) 313}.

\bibitem{Harding:1998ma}
A.~K. Harding and A.~G. Muslimov, \emph{{Particle acceleration zones above
  pulsar polar caps: electron and positron pair formation fronts}},
  \href{https://doi.org/10.1086/306394}{\emph{Astrophys. J.} {\bfseries 508}
  (1998) 328} [\href{https://arxiv.org/abs/astro-ph/9805132}{{\ttfamily
  astro-ph/9805132}}].

\bibitem{Harding:2001at}
A.~K. Harding and A.~Muslimov, \emph{{Pulsar polar cap heating and surface
  thermal x-ray emission. 2. Inverse Compton radiation pair fronts}},
  \href{https://doi.org/10.1086/338985}{\emph{Astrophys. J.} {\bfseries 568}
  (2002) 862} [\href{https://arxiv.org/abs/astro-ph/0112392}{{\ttfamily
  astro-ph/0112392}}].

\bibitem{johnston2020galactic}
S.~Johnston, D.~Smith, A.~Karastergiou and M.~Kramer, \emph{The galactic
  population and properties of young, highly energetic pulsars}, {\emph{Monthly
  Notices of the Royal Astronomical Society} {\bfseries 497} (2020) 1957}.

\bibitem{Kiziltan:2009rx}
B.~Kiziltan and S.~E. Thorsett, \emph{{Millisecond Pulsar Ages: Implications of
  Binary Evolution and a Maximum Spin Limit}},
  \href{https://doi.org/10.1088/0004-637X/715/1/335}{\emph{Astrophys. J.}
  {\bfseries 715} (2010) 335}
  [\href{https://arxiv.org/abs/0909.1562}{{\ttfamily 0909.1562}}].

\bibitem{Hobbs:2005yx}
G.~Hobbs, D.~R. Lorimer, A.~G. Lyne and M.~Kramer, \emph{{A Statistical study
  of 233 pulsar proper motions}},
  \href{https://doi.org/10.1111/j.1365-2966.2005.09087.x}{\emph{Mon. Not. Roy.
  Astron. Soc.} {\bfseries 360} (2005) 974}
  [\href{https://arxiv.org/abs/astro-ph/0504584}{{\ttfamily
  astro-ph/0504584}}].

\bibitem{lorimer2006parkes}
D.~R. Lorimer, A.~Faulkner, A.~Lyne, R.~N. Manchester, M.~Kramer,
  M.~McLaughlin, G.~Hobbs, A.~Possenti, I.~Stairs, F.~Camilo et~al., \emph{The
  parkes multibeam pulsar survey--vi. discovery and timing of 142 pulsars and a
  galactic population analysis}, {\emph{Monthly Notices of the Royal
  Astronomical Society} {\bfseries 372} (2006) 777}.

\bibitem{Cieslar:2018jzd}
M.~Cie\'slar, T.~Bulik and S.~Os\l{}owski, \emph{{Markov chain Monte Carlo
  population synthesis of single radio pulsars in the Galaxy}},
  \href{https://doi.org/10.1093/mnras/staa073}{\emph{Mon. Not. Roy. Astron.
  Soc.} {\bfseries 492} (2020) 4043}
  [\href{https://arxiv.org/abs/1803.02397}{{\ttfamily 1803.02397}}].

\bibitem{Graber:2023jgz}
V.~Graber, M.~Ronchi, C.~Pardo-Araujo and N.~Rea, \emph{{Isolated Pulsar
  Population Synthesis with Simulation-based Inference}},
  \href{https://doi.org/10.3847/1538-4357/ad3e78}{\emph{Astrophys. J.}
  {\bfseries 968} (2024) 16}
  [\href{https://arxiv.org/abs/2312.14848}{{\ttfamily 2312.14848}}].

\bibitem{du2024initial}
S.-S. Du, X.-J. Liu, Z.-C. Chen, Z.-Q. You, X.-J. Zhu and Z.-H. Zhu, \emph{On
  the initial spin period distribution of neutron stars}, {\emph{The
  Astrophysical Journal} {\bfseries 968} (2024) 105}.

\bibitem{Philippov:2013aha}
A.~Philippov, A.~Tchekhovskoy and J.~G. Li, \emph{{Time evolution of pulsar
  obliquity angle from 3D simulations of magnetospheres}},
  \href{https://doi.org/10.1093/mnras/stu591}{\emph{Mon. Not. Roy. Astron.
  Soc.} {\bfseries 441} (2014) 1879}
  [\href{https://arxiv.org/abs/1311.1513}{{\ttfamily 1311.1513}}].

\bibitem{Passamonti:2016nmf}
A.~Passamonti, T.~Akg\"un, J.~A. Pons and J.~A. Miralles, \emph{{The relevance
  of ambipolar diffusion for neutron star evolution}},
  \href{https://doi.org/10.1093/mnras/stw2936}{\emph{Mon. Not. Roy. Astron.
  Soc.} {\bfseries 465} (2017) 3416}
  [\href{https://arxiv.org/abs/1608.00001}{{\ttfamily 1608.00001}}].

\bibitem{Pons:2007vf}
J.~A. Pons and U.~Geppert, \emph{{Magnetic field dissipation in neutron star
  crusts: From magnetars to isolated neutron stars}},
  \href{https://doi.org/10.1051/0004-6361:20077456}{\emph{Astron. Astrophys.}
  {\bfseries 470} (2007) 303}
  [\href{https://arxiv.org/abs/astro-ph/0703267}{{\ttfamily
  astro-ph/0703267}}].

\bibitem{Pons:2019zyc}
J.~A. Pons and D.~Vigan\`o, \emph{{Magnetic, thermal and rotational evolution
  of isolated neutron stars}},
  \href{https://doi.org/10.1007/s41115-019-0006-7}{\emph{Liv. Rev. Comput.
  Astrophys.} {\bfseries 5} (2019) 3}
  [\href{https://arxiv.org/abs/1911.03095}{{\ttfamily 1911.03095}}].

\bibitem{Cumming:2004mf}
A.~Cumming, P.~Arras and E.~G. Zweibel, \emph{{Magnetic field evolution in
  neutron star crusts due to the Hall effect and Ohmic decay}},
  \href{https://doi.org/10.1086/421324}{\emph{Astrophys. J.} {\bfseries 609}
  (2004) 999} [\href{https://arxiv.org/abs/astro-ph/0402392}{{\ttfamily
  astro-ph/0402392}}].

\bibitem{Aguilera:2007xk}
D.~N. Aguilera, J.~A. Pons and J.~A. Miralles, \emph{{2D Cooling of Magnetized
  Neutron Stars}},
  \href{https://doi.org/10.1051/0004-6361:20078786}{\emph{Astron. Astrophys.}
  {\bfseries 486} (2008) 255}
  [\href{https://arxiv.org/abs/0710.0854}{{\ttfamily 0710.0854}}].

\bibitem{Vigano:2021olr}
D.~Vigan\`o, A.~Garc\'\i{}a-Garc\'\i{}a, J.~A. Pons, C.~Dehman and V.~Graber,
  \emph{{Magneto-thermal evolution of neutron stars with coupled Ohmic, Hall
  and ambipolar effects via accurate finite-volume simulations}},
  \href{https://doi.org/10.1016/j.cpc.2021.108001}{\emph{Comput. Phys. Commun.}
  {\bfseries 265} (2021) 108001}
  [\href{https://arxiv.org/abs/2104.08001}{{\ttfamily 2104.08001}}].

\bibitem{Mereghetti:2015asa}
S.~Mereghetti, J.~Pons and A.~Melatos, \emph{{Magnetars: Properties, Origin and
  Evolution}}, \href{https://doi.org/10.1007/s11214-015-0146-y}{\emph{Space
  Sci. Rev.} {\bfseries 191} (2015) 315}
  [\href{https://arxiv.org/abs/1503.06313}{{\ttfamily 1503.06313}}].

\bibitem{Igoshev:2018fcp}
A.~P. Igoshev, \emph{{Ages of radio pulsar: long-term magnetic field
  evolution}}, \href{https://doi.org/10.1093/mnras/sty2945}{\emph{Mon. Not.
  Roy. Astron. Soc.} {\bfseries 482} (2019) 3415}
  [\href{https://arxiv.org/abs/1810.12922}{{\ttfamily 1810.12922}}].

\bibitem{yao2017new}
J.~Yao, R.~Manchester and N.~Wang, \emph{A new electron-density model for
  estimation of pulsar and frb distances}, {\emph{The Astrophysical Journal}
  {\bfseries 835} (2017) 29}.

\bibitem{berger2014kolmogorov}
V.~W. Berger and Y.~Zhou, \emph{Kolmogorov--smirnov test: Overview},
  {\emph{Wiley statsref: Statistics reference online} (2014) }.

\bibitem{rizzo2016energy}
M.~L. Rizzo and G.~J. Sz{\'e}kely, \emph{Energy distance}, {\emph{wiley
  interdisciplinary reviews: Computational statistics} {\bfseries 8} (2016)
  27}.

\bibitem{peacock1983two}
J.~A. Peacock, \emph{Two-dimensional goodness-of-fit testing in astronomy},
  {\emph{Monthly Notices of the Royal Astronomical Society} {\bfseries 202}
  (1983) 615}.

\bibitem{fasano1987multidimensional}
G.~Fasano and A.~Franceschini, \emph{A multidimensional version of the
  kolmogorov--smirnov test}, {\emph{Monthly Notices of the Royal Astronomical
  Society} {\bfseries 225} (1987) 155}.

\bibitem{rozwadowska2021rate}
K.~Rozwadowska, F.~Vissani and E.~Cappellaro, \emph{On the rate of core
  collapse supernovae in the milky way}, {\emph{New Astronomy} {\bfseries 83}
  (2021) 101498}.

\bibitem{SKAOPulsarScienceWorkingGroup:2025zsy}
{\scshape SKAO Pulsar Science Working Group} Collaboration, L.~Levin et~al.,
  \emph{{Understanding the Neutron Star Population with the SKAO telescopes}},
  \href{https://arxiv.org/abs/2512.16156}{{\ttfamily 2512.16156}}.

\bibitem{padmanabh2023mpifr}
P.~Padmanabh, E.~Barr, S.~Sridhar, M.~Rugel, A.~Damas-Segovia, A.~Jacob,
  V.~Balakrishnan, M.~Berezina, M.~Bernadich, A.~Brunthaler et~al., \emph{The
  mpifr--meerkat galactic plane survey--i. system set-up and early results},
  {\emph{Monthly Notices of the Royal Astronomical Society} {\bfseries 524}
  (2023) 1291}.

\end{thebibliography}\endgroup

\appendix

\clearpage
\twocolumngrid

\begin{figure}
    \includegraphics[width=0.45\textwidth]{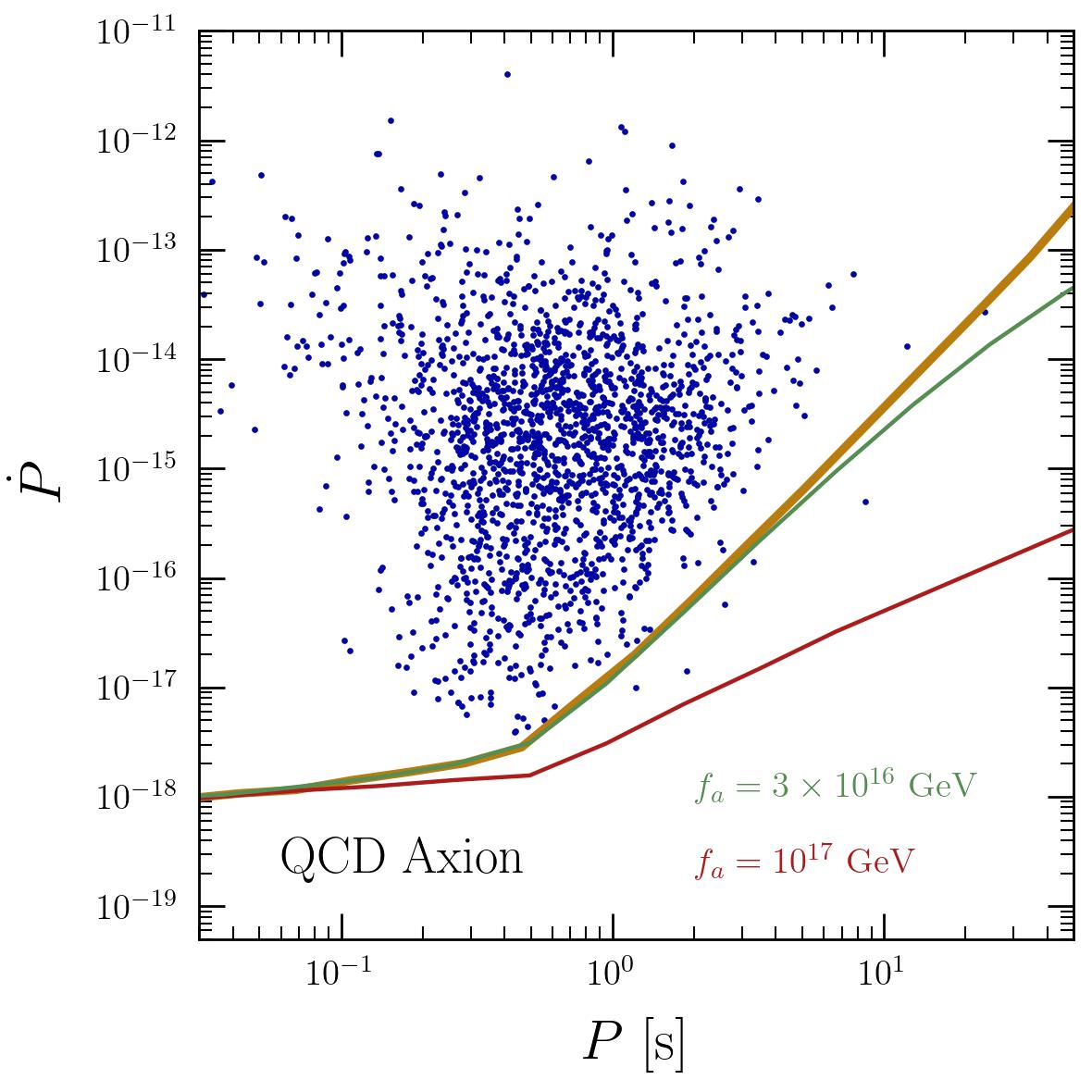}
    \caption{Shift in the death line that would arise should the core of neutron star be sufficiently dense to support the sourcing of the QCD axion (\ie, $\epsilon = 1$). Here, we have assumed a core radius of $R_c = 5$ km. }\label{fig:ppdot_realqcd}
\end{figure}

\section{Appendix A: Speculations about the vanilla QCD axion}
It remains an open question as to whether the vanilla QCD axion is sourced within the very high density matter in the core of the neutron star. If this is the case, probes of the neutron star equation of state may be one of the most promising ways to directly probe the QCD axion. However, the modification of the structure of neutron stars predominantly is confined to the core, making surface probes like cooling insensitive to sourcing of the QCD axion deep inside the star.
Since the probe discussed here is sensitive to the axion gradient at the surface, and the QCD axion would be sourced by the core of the neutron star, it is unclear whether there would be sufficient leakage outside of the core to shift the pulsar death line. In order to directly address this question, we assume the QCD axion is sourced within a core of radius $R_c \sim 5$ km, and adopt a radial axion profile consistent with Eq.~\ref{eq:axradial}, but taking $R \rightarrow R_c = 5$ km. The shift in the death line produced from this profile is shown for two values of $f_a$ in Fig.~\ref{fig:ppdot_realqcd}. Here, one can see that a fraction of parameter space near $f_a \sim 10^{17}$ GeV may still be accessible, despite the strong suppression of the axion field at the surface of the star. It is worth highlighting that we have chosen to truncate our analysis conservatively at $f_a \lesssim 10^{17}$ GeV, however, should QCD axions be sourced with slightly higher decay constants, these axions will leave discernible imprints on the pulsar population.

\section{Appendix B: The One-Sided Lighthouse}
One of the interesting features that is produced by these axion field gradients is an asymmetry in the sign of the axion-induced electric field across each of the poles. %
In particular, for the case of sufficiently light axion hair around active pulsars, screening can effectively turn off pair production along a large part of one pole, while leaving it unchanged in the other. A natural question to ask is whether there exist clear observational consequences that could arise from this asymmetry. 

One possible avenue is to exploit double-pole pulsars --  a rare class of nearly orthogonal rotating pulsars that allow for radio emission to be observed from both polar caps~\cite{Kramer:2008iw,2019Sci...365.1013D}. Depending on the relative alignment, the line of sight of Earth may pass through one pole where pair production is enhanced, and another where it is suppressed. In the event that the the line of sight passes both return currents, or both out-flowing currents, one is guaranteed to observe at least one pole where pair production is screened (this follows from the fact that for fixed $\rho_{\rm GJ}$ and $\alpha_0$, the axion charge density has an asymmetry). For nearly maximally mis-aligned rotators, there is also a possibility that one observes radio emission from out-flowing current on one pole and return current on another; depending on the axion and field configuration (and which pole contains the out-flowing and return currents), one may expect axion hair to either screen both emission zones, or neither. Thus, the geometry of the magnetic field and the sign of the axion hair must be uniquely identified in order to derive constraints.

There exists a remarkable pulsar, PSR J1906+0746, for which interpulses are observed, and which is in binary that induces a small precession such that different parts of the polar cap are observed over the timescales of years. The specific geometry of this configuration with respect to two polar caps is clearly illustrated in Fig. 1 of \cite{Galishnikova:2020gmu}, where one can see the line of sight  is {\emph{typically}}\footnote{Precision of the orbit causes the field line bundle being probed to shift.} traversing two return current regions, both with the same sign of $\rho_{\rm GJ}$, but with opposite signs of $\rho_a$. For sufficiently small axion masses, and sufficiently large values of $\rho_a$, one can ensure that axion hair suppresses pair production on one of these return currents. In addition, depending on the sign of $\rho_a$, one may also expect the precession to allow one to test the screening induced by axion hair along field lines with out-flowing currents.

Let us derive an approximate condition for the axion mass and coupling at which this effect should arise, focusing momentarily on the effect on the discharge in the return current zone (since these lead to an unambiguous signal, irrespective of the sign of $\rho_a$). PSR J1906+0746 has a rotational period of $\sim 0.144$ seconds, a spin down rate of $\dot{P} = 2.0 \times 10^{-14}$, and an inferred surface magnetic field strength of $1.7 \times 10^{12}$ G, and an inclination of $\alpha=99.41^\circ$~\cite{2019Sci...365.1013D}. For a rotational period $P \sim 0.144$ s, the conventional death line is sitting near $\dot{P} \sim 10^{-18}$; since $\dot{P}\propto B^2$, one can infer that an identical pulsar with a magnetic field $B \lesssim 10^{10}$ G would be inactive. For dipolar fields, the magnetic field scales as $B \propto (R_{\rm NS} / r)^3$, implying that the mean free path of curvature photons in PSR J1906+0746 becomes finite only at a distance of $\sim 4$ stellar radii above the surface of the star. This implies that axion hair will suppress emission along one pole in PSR J1906+0746 if $|\rho_a(r) / \rho_{\rm GJ}(r)| \gg 1$ for $r \lesssim 5 \, R_{\rm NS}$. For the light QCD axion, and assuming $g_{a\gamma\gamma} \sim \alpha_{\rm EM} / (2\pi f_a)$, this amounts to the condition $m_a \gtrsim 8 \pi \cos{\alpha} / (\alpha_{\rm EM} P)$, which cannot be satisfied while ensuring $m_a \times 4 \times R_{\rm NS} \ll 1 $ (this statement, however, is not necessarily true for other pulsars). For the CP-violating axion, one finds instead in the low-mass limit
\begin{eqnarray}  \label{eq:linear_light}
    m_a & \lesssim & 10^{-12} \,{\rm eV} \\
   g_N \times g_{a\gamma}&  \gtrsim &  8 \times 10^{-39} \, {\rm GeV}^{-1} \, .
\end{eqnarray}
This constraint is only marginally weaker than the constraint derived using the death line, and is included (computing the constraint using the full field profile across all axion masses, not just in the massless limit) in Fig.~\ref{fig:limits} for comparison (solid line).

It is worth highlighting that this constraint becomes significantly stronger if axions also turn off the discharge along the field lines with out-flowing currents. This is because acceleration necessarily takes place close to the surface, and is limited by the traverse size of the polar cap region; here, the condition that axion screening is sufficient to suppress emission is merely that $\rho_a$ is large on distances scales $ \gg 2 r_{\rm pc}$ from the stellar surface. Repeating the procedure above, but imposing  $\rho_a(r) / \rho_{\rm GJ}(r) \gg 1$ for $r \lesssim R_{
\rm NS} + 2 \, r_{\rm pc}$ instead leads to a sensitivity in the massless axion limit at the level of 
\begin{eqnarray}  \label{eq:linear_light_2}
    m_a & \lesssim & 10^{-12} \,{\rm eV} \\[6pt]
   g_N \times g_{a\gamma}&  \gtrsim &  2 \times 10^{-39} \, {\rm GeV}^{-1} \, .
\end{eqnarray}
There is a subtlety, however; in order to adopt this stronger constraint, one must confidently probe out-flowing field lines with ${\rm{sign}}(\rho_a) = {\rm{sign}}(\rho_{\rm GJ})$. Observationally, is no clear observational signature which allows one to differentiate the sign of $\rho_{\rm GJ}$; moreover, both $g_{a\gamma\gamma}$ and $g_N$ can take either sign, meaning $\rho_a$ is also undetermined. Therefore, the most promising route is to hope for the detection of a similar object that exclusively probes the out-flowing field lines on both poles. In order to merely demonstrate the potential power of such a probe, we include a dashed line in Fig.~\ref{fig:limits} to demarcate the region of parameter space for which efficient screening along the out-flowing field line should alter the emission (this is done for both models).

\section{Appendix C: Field profiles for the linear interaction case}
Here we provide the fitting formulas for the axion profile in the linear model, analogous to Eq.~\ref{eq:axradial}. For $m_a \lesssim 10^{-11} \, \rm eV$ the profile is
\begin{equation}
a \simeq  1.5\times 10^{13} \, \rm{GeV}\,\left(\frac{g_N}{10^{-23}}\right) \frac{R_\text{star}}{ \, r} \exp[-m_a (r - R_\text{star})].
\label{eq:lin1}
\end{equation}

In the opposite limit, when the axion mass is large, $1/m_a \lesssim \rm km$, a better fit to the numeric solution outside the star is
\begin{eqnarray}
a &\simeq& 10^{11} \, \rm{GeV} \, \left(\frac{g_N}{10^{-23}}\right)\left(\frac{2 \times 10^{-10}\rm eV}{m_a}\right)^2 \nonumber \\ &\times & \frac{R_c}{r}\exp[-m_a (r - R_c)],
\label{eq:lin2}
\end{eqnarray}
with $R_c \simeq 10.5 \, \rm km $. We numerically verified this fit for masses up to $10^{-9} \, \rm eV$, where the field value outside the star is already quite small.
For such heavy axion masses, the axion field profile has non-negligible dependence on the density profile of the star, in particular on the density within a distance of $1/m_a$ from the edge of the star. The axion profile above was found adopting the APR equation of state \cite{Akmal:1998cf} and a NS with mass $1 M_\odot$.
When the inverse of the scalar mass is comparable to the size of the entire star, that is to say $m_a \sim \mathcal{O}(\text{few})10^{-11}-10^{-10} \, \rm eV$, than the the solution outside the star is well approximated by 
\begin{eqnarray}\label{Eq:SolIntermediate}
    a &\simeq& 2.5 \times 10^{11} \, \rm{GeV} \, \left(\frac{g_N}{10^{-23}}\right)\left(\frac{10^{-10} \, \rm eV}{m_a}\right)^2  \nonumber \\ &\times &  \frac{R_\text{star}}{r} \exp[-m_a (r - R_\text{star})],
\label{eq:lin3}
\end{eqnarray}
where $R_\text{star} = 12 \, \rm km$ and the numerical prefactor have been fixed using the APR density profile. Details about the derivation of these formulas are provided in the companion PRD.  

\clearpage
In the following we provide details about the construction of the pulsar death valley, both numerically and with a simple analytical recipe, as well as our neutron star population modeling and the statistical analysis on which our bounds are based.  

\section{Appendix D: Constructing the pulsar death valley}
\label{sec:pulsardeath}

In this section we describe in detail how, for a fixed set of pulsar parameters, we determine if the pulsar is active. We start with our full, numerical solution and then we also provide an easy analytical solution which agrees extremely well with the numerics for the parameters of interest. 

\subsection{Full numerical solution}

We start by fixing the surface magnetic field strength $B_0$, the rotational frequency $\Omega$, and the curvature of the field lines $\rho_c$ of a given neutron star. Our fiducial analysis adopts a radius of curvature consistent with a dipolar field configuration, 
\begin{eqnarray}
    \rho_c \simeq 9 \times 10^7 \, \left(\frac{\theta_0}{\theta} \right) \sqrt{\frac{P}{1 \,{\rm sec}}} \, {\rm cm} \, ,
\end{eqnarray}
$\theta_0$ is the angular opening of the polar cap and $\theta$ defines the field line of interest (note that we also run analyses fixing the value of $\rho_c$, and varying this fixed value across a wide range of characteristic scales consistent with both dipolar and quadrupolar fields). We then fix the current demanded by the twist on the local field line bundle under consideration $j_m \equiv |\nabla \times \vec{B}| = \alpha_0 \rho_{\rm GJ}$; we adopt a pre-factor $\alpha_0 = 2$ which is roughly consistent with characteristic values obtained in the force-free simulations of inclined pulsars \citep{Gralla:2017nbw}. 

We then evolve Eq.~\ref{eq:dgam_2} to determine the acceleration of the primary particles along the field lines. We stop the acceleration at a distance scale $r \sim 2 r_{\rm pc}$, as this is where the one-dimensional approximation breaks down~\cite{beloborodov2008polar}. At each time step, one can compute the energy radiated per unit frequency per unit distance, which is given by~\cite{beskin2022pulsar}
\begin{eqnarray}
    dI = \frac{\sqrt{3}}{2 \pi} \frac{e^2}{\rho_c} \, \gamma \, F(\omega / \omega_c) \, d\omega \, d\ell \, 
\end{eqnarray}
where 
\begin{eqnarray}
    F(x) \equiv x \int_x^{\infty} \, K_{5/3}(x) \, dx ,
\end{eqnarray}
and $K_{5/3}$ is the modified Bessel function of the second kind. The characteristic photon energy emitted is $\omega_c \sim 3 \gamma^3 / (2 \rho_c)$, but pair cascades are initiated by the highest energy photons, which are typically a factor of a few higher than this value. Taking the limit $\omega \gg \omega_c$,
we can estimate
\begin{eqnarray}
    F(x) \sim \sqrt{\frac{\pi x}{2}} \, e^{-x} \left( 1 + \frac{55}{72}\frac{1}{x} + \cdots \right) \, .
\end{eqnarray}
We only expect a gamma-ray of energy $\omega$ to be radiated over a distance scale $\ell_{\rm rad}$ when the total energy radiated exceeds the photon energy, i.e.
\begin{eqnarray}
    \omega \leq \int_{\ell_0}^{\ell_0 + \ell_{\rm rad}} \, \int_{\omega}^{\infty} \, dI(\omega^\prime) \, d\omega^\prime \, d\ell \, .
\end{eqnarray}
Assuming that the radiation length is sufficiently small that $\gamma$ can be treated as a constant, one can define an implicit relation between $\ell_{\rm rad}$ and the maximum photon energy radiated given in Eq.~\ref{eq:ellrad}. Notice that as $\omega / \omega_c \rightarrow \infty$ the radiation length $\ell_{\rm rad} \rightarrow \infty$, but the mean free path of the photon $\ell_{e^\pm}$, obtained by inverting the optical depth given in Eq.~\ref{Eq:OpticalDepth} goes to zero. 
Conversely, in the limit $\omega / \omega_c \rightarrow 1$, the radiation length shrinks, but $\ell_{e^\pm}$ becomes large. This implies an implicit trade-off between production of the photon and pair production, as one might expect. 

At each point in the evolution of the primary particle, we compute the total distance traveled, and then project onto two dimensional location of a specific field line of interest. A photon of fixed energy $\omega$ is emitted from that location parallel to the magnetic field line itself, and the photon trajectory is traced by solving
\begin{eqnarray}
    \frac{d x^\mu}{d\lambda} = \frac{\partial \mathcal{H}}{\partial k_\mu} \\
    \frac{d k_\mu}{d\lambda} = -\frac{\partial \mathcal{H}}{\partial x^\mu}
\end{eqnarray}
with $\mathcal{H} = g^{\mu\nu} k_\mu k_\nu$. Here, we adopt a Schwarzschild metric for a $M = 1.4 \, M_\odot$ star with  a radius of 12 km. For the sake of illustration, we also consider in an additional variant considered here the case of an aligned quadrupolar field -- the procedure here is the same, however the structure and strength of the magnetic field have been altered accordingly. Note that for a purely quadrupolar field, the magnetic field lines follow the solution 
\begin{eqnarray}
    \frac{dr}{r} = \frac{d\theta(3 \cos^2\theta - 1)}{\sin 2\theta} \, ,
\end{eqnarray}
which implies that open field lines will be confined to $\theta \lesssim 0.62 \times (R_{\rm NS} \times \Omega)$~\cite{malov2011geometry}. Since the open field bundle is notably reduced with respect to a dipolar pulsar, this tends to imply a larger characteristic radius of curvature of the field line. For this reason, we take $\rho_c \sim 10^9$ cm in our fiducial quadrupole analysis (although quadrupolar field can also have characteristic values of $\rho_c$ much smaller than dipolar fields in certain regimes). 

Since the gap height is effectively the sum of the acceleration length $\ell_{\rm acc}$, the radiation length, and the mean free path, one can ascertain the energy of pair producing photons, and the explicit gap height, by minimizing the sum of these three quantities over the path of the primary particle. The transition between a small but finite gap height and an infinite gap height (implying pair production never occurs) is extremely rapid, and thus in practice we define pulsar death as the point where the gap height, computed using the procedure outlined above, exceeds two kilometers. 

One can gain a bit of analytic intuition for this problem  by noting that optical depth is exponentially suppressed until $\chi \equiv (\omega / m_e) \times (B / B_q) \times \sin\psi  / 2 \gtrsim \mathcal{O}(1/7)$, and $\psi$ always remains small near the neutron star; collectively, these conditions suggests that if pair production occurs, the mean free path should be approximately given by
\begin{eqnarray}\label{eq:lguess}
    \ell_{e^\pm} \sim \frac{2 \rho_b}{7} \left(\frac{m_e}{\omega} \right) \left( \frac{B_q}{B}\right) \, .
\end{eqnarray}
For a fixed value of $\gamma$, Eqns.~\ref{eq:ellrad} and~\ref{eq:lguess} suggest that the minimum of $(\ell_{\rm rad} + \ell_{e^\pm})$ is roughly obtained by solving
\begin{eqnarray}\label{eq:higherg}
    \left(\frac{\omega}{\omega_c} \right)^{5/2} \, e^{\omega / \omega_c} \left(1 - \frac{55}{72}\frac{\omega_c}{\omega} + \cdots \right) \simeq  \\[10pt] 40 \times \left(\frac{\rho_c}{10^7 \, {\rm cm}} \right) \left(\frac{10^{12} \, {\rm G}}{B} \right) \, \left(\frac{10^7}{\gamma} \right)^2 \, . \nonumber
\end{eqnarray}
For pulsars near the death line, this typically yields an energy $\omega \sim \mathcal{O}(5-10) \times \omega_c$. One can then use the optical depth to turn this into a condition on the boost factor of the primary particle. This estimate works reasonably well across much of the relevant parameter space, but does fail in certain regimes.

The discussion outlined above assumes that the primary pair production channel is via single photon $e^\pm$ pair production, with the incident high energy gamma ray being produced by curvature radiation. This is a valid assumption across a wide range of the pulsar parameter space (including for standard radio pulsars), however it is not necessarily valid for certain sub-populations. Below, we briefly discuss two alternative pair production channels.

For high magnetic field pulsars and magnetars, resonant inverse Compton scattering (often referred to as `RICS') can provide an alternative plasma loading mechanism~\cite{zhang1997three,Harding:1998ma,Harding:2001at,Timokhin:2018vdn}. Here, hot thermal photons emitted from the stellar surface (in particular, photons emitted from the so-called `hot spots', which are generated as highly energetic particles are accelerated into the surface of the star) can up-scatter off of primary charges accelerated from the stellar surface, or off of secondary plasma particles initiated by curvature radiation -- the final state photon sits at pair-producing energies, and will thus contribute to the generation of plasma via the same single-photon pair production process discussed above. Resonant scattering off of secondary plasma can play an important role in establishing the maximum multiplicity of pair cascades in strong field pulsars~\cite{Timokhin:2018vdn}, but is less interesting from the prospective of pulsar death, as pair cascades must be initiated from curvature photons. Resonant scattering off primary particles, on the other hand, typically have a multiplicity that is too low to efficiently screen the electric field~\cite{Harding:2001at,Timokhin:2018vdn}, implying they are not relevant from the perspective of establishing pulsar death.

\begin{figure*}
    \centering
    \includegraphics[width=0.49\linewidth]{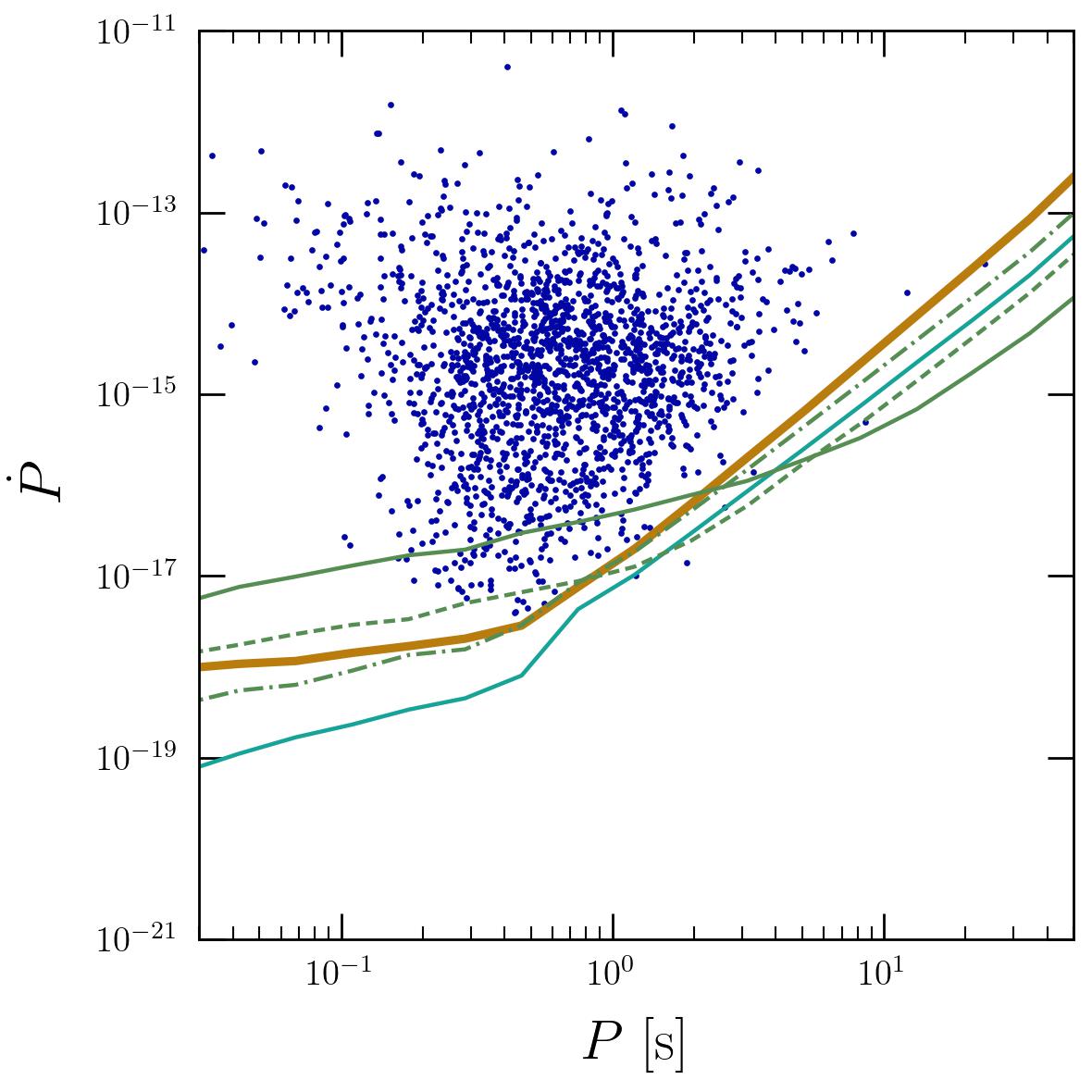}
    \includegraphics[width=0.49\linewidth]{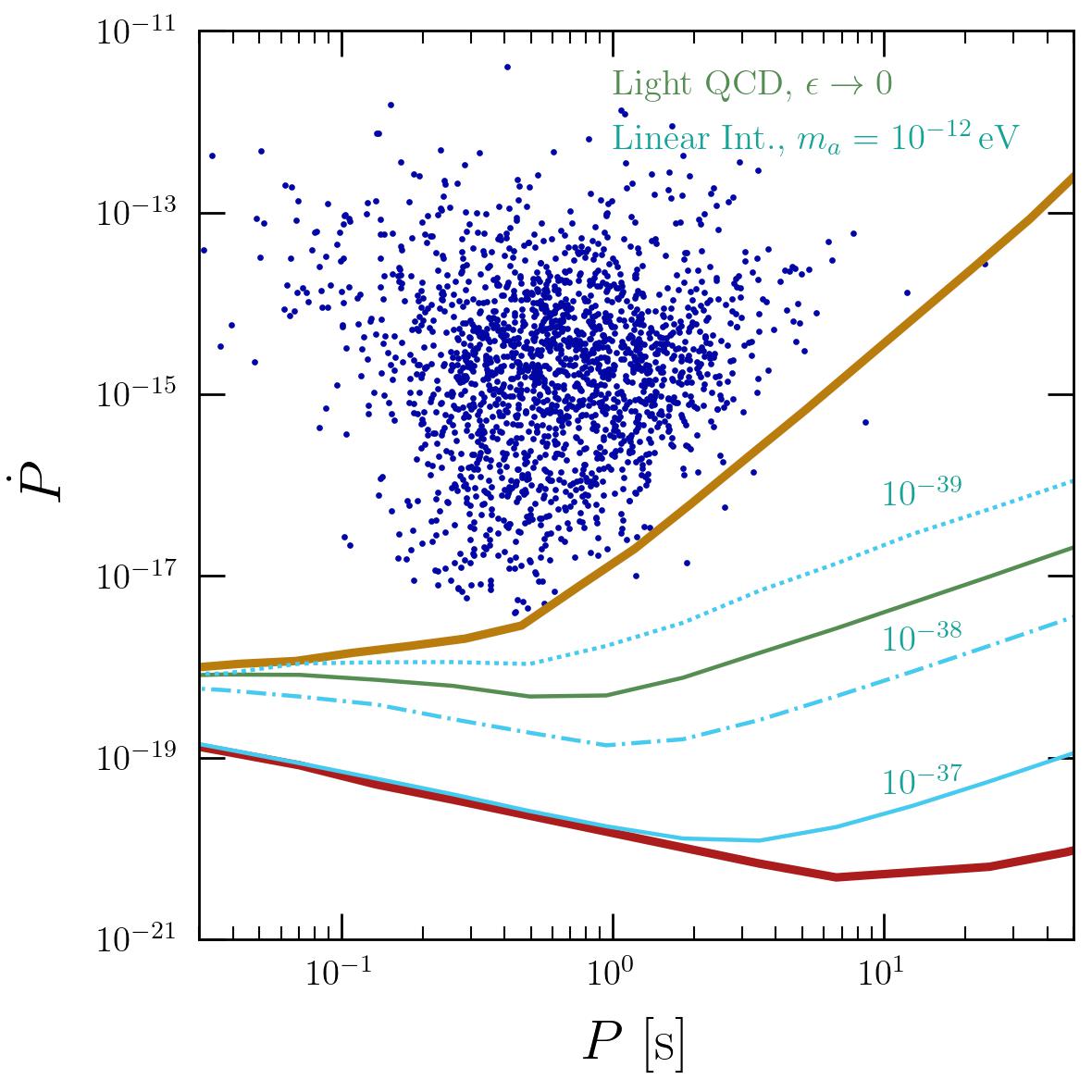}
    \caption{ Left: pulsar death line computed using:  a dipolar field, and a radius of curvature fixed to the last open field line (yellow),  a dipolar field, but imposing a radius of curvature of $\rho_c = 10^6, 10^7, $ or $10^8$ cm (green; solid, dashed, and dot-dashed), or an order one quadrupolar-dipolar field with a fixed radius of curvature of $\rho_c = 10^9$ cm (blue). Right: pulsar death line in axion models. Red curve (bold) shows the light QCD axion with $\epsilon = 10^{-2}$ and $f_a = 10^{14}$ GeV, while the green curve shows the death line when $m_a \gg R_{\rm NS}$, corresponding to the limit when $\epsilon \rightarrow 0$. In the case of the linear axion-nucleon coupling (labeled `linear int.'), we fix the axion mass to $m_a = 10^{-12}$ eV, and vary the product of the couplings $g_{a\gamma} \times g_N$ (see small number next to each line, shown in units of ${\rm GeV}^{-1}$). In all cases the radius of curvature if fixed to that of the last open field line. }
    \label{fig:pdot_extra}
\end{figure*}

In Fig.~\ref{fig:pdot_extra}, we demonstrate how the death line shifts in response to changes in the underlying assumptions. In particular, the left panel of Fig.~\ref{fig:pdot_extra} shows the impact of fixing the radius of curvature to $\rho_c = 10^6$ cm (green, solid), $\rho_c = 10^7$ cm (green, dashed), and $\rho_c = 10^8$ cm (green, dot-dashed), or including a quardupole moment aligned with the dipole, and with a characteristic field strength equal to the dipolar field strength at the surface. Since a quadrupole inherently alters the curvature and structure of the open field line bundle, we fix the radius of curvature in this example to $\rho_c =10^9 $ cm (blue).

In the right hand panel of Fig.~\ref{fig:pdot_extra}, we we show a variety of different death lines computed for both the light QCD axion and the linearly-coupled axion. For the light QCD axion, we note that the effect of the axion gradient has two terms
\begin{eqnarray}
    g_{a \gamma }\partial_r a \sim \frac{\alpha_0}{2 \, R_{\rm NS}} \left(1 + m_a R_{\rm NS} \right) \, .
\end{eqnarray}
For small axion masses (or equivalently when $\epsilon \rightarrow 0$), the axion induced voltage drop tends toward a fixed value, while for large axion masses the effect scales proportionally to the mass itself. Since there exists a maximal value of $m_a$ for a fixed $f_a$, we expect the death lines for the light QCD axion to asymptote to a fixed value when $m_a \times R_{\rm NS} \ll 1$, and be maximally deviating when $\epsilon \sim 0.07$ (the maximal value allowed in our analysis). Therefore, we only plot these two limits.

In the case of the linearly coupled axion (blue lines), we instead fix the axion mass and vary the product of the axion-nucleon and axion-photon coupling. Here, we can see the shift in the death line scales directly with this product, as one might expect.

\subsection{A simple analytical solution}
\label{sec:ansol}

Before delving into the matter of neutron star populations and pulsar data, we find it useful to provide also a simple, analytical solution for the problem of pair discharge and the determination of a pulsar death line.

As we reported in the main text, the equation to be solved reads
\begin{equation}\label{eq:dgamm}
    \frac{d^2\gamma}{ds^2}
    = \frac{e}{m_e}\left(
        \frac{\gamma\,\alpha_0\,\rho_{\rm GJ}}{\sqrt{\gamma^2 - 1}}
        - \rho_{\rm GJ}
        - g_{a\gamma}\,B\,\partial_r a
    \right)
    - \mathcal{R}_\gamma,
\end{equation}
where \(s\) is the distance from the NS surface, \(\gamma\) is the gamma-factor for the accelerated electrons and \(\mathcal{R}_\gamma\) accounts for radiative losses, see Eq. \ref{eq:Rgamma}. The boundary condition is that the electrons start at rest at $s=0$. We start considering the case \textit{without} axions (\(g_{a\gamma}=0\)), fix \(\alpha_0=2\), and approximate
\(\gamma/\sqrt{\gamma^2-1}\approx1\), which is a good approximation given that charges will be accelerated to large values of $\gamma$ pretty quickly. Hence the equation simplifies to
\begin{equation}\label{eq:dgammSimp}
    \frac{d^2\gamma}{ds^2}
    = \frac{e}{m_e}\,\rho_{\rm GJ}
    - \mathcal{R}_\gamma.
\end{equation}

\begin{figure}
    \centering
    \includegraphics[width=\linewidth]{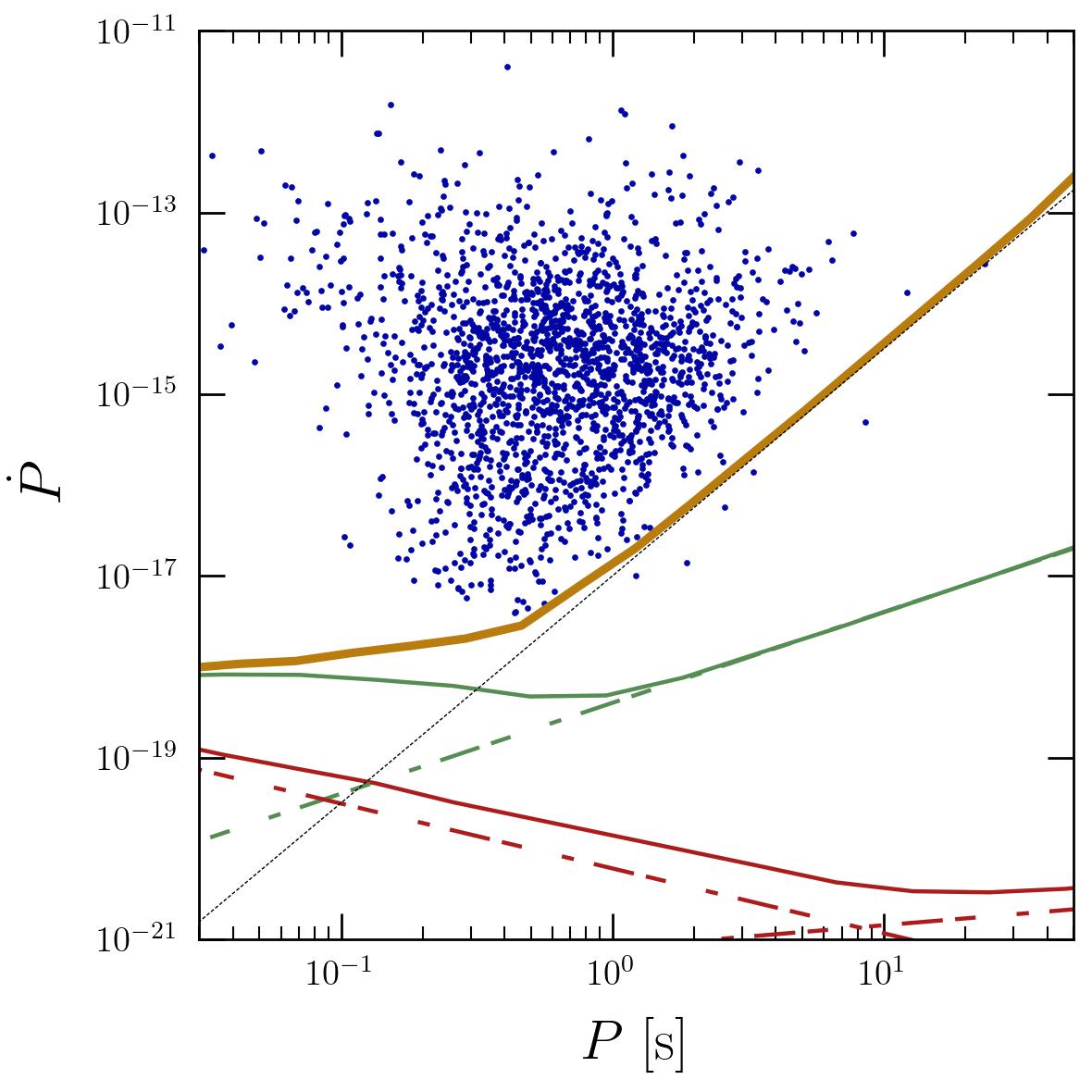}
    \caption{$P-\dot{P}$ diagram comparing the numeric results for the standard fiducial analysis (yellow), and light QCD axion in the small (green) and large mass limit (red) with our analytic approximations (dashed lines). For the small mass limit of the light QCD axion we only display the large $P$ derivation, as the transition observed below $P \lesssim 1$s has multiple competing scales which make analytic estimates less tractable.  }
    \label{fig:analytic_comp}
\end{figure}

One can then identify two different regimes: the regime where the GJ charge density \textit{always} dominates over the curvature radiation back-reaction and the electrons keep accelerating up to $2 \, r_{\rm cap}$; the regime where the back-reaction is large enough such that at some point $\frac{d^2\gamma_e}{ds^2} \sim 0$ and therefore $d\gamma_e/ds \sim \rm const.$  (this corresponds to the radiation-reaction limited regime, mostly expected in millisecond pulsars). Here we consider only the first regime, which, as we will show, describes well pulsar with large periods. The equation becomes
\begin{equation}\label{eq:GJacc}
    \frac{d^2\gamma}{ds^2}
    = \frac{e \, \rho_{\rm GJ}}{m_e},
\end{equation}
whose solution yields an analytical estimate for the maximum Lorentz factor at \(2\,r_{\rm pc}\)
\begin{equation}\label{eq:gamma_max}
    \gamma_{\rm max}
    \simeq
    \frac{8\pi\,e}{P \, m_e}\,B_{\rm ref}
    \biggl(\frac{r_{\rm pc}^2}{1+2\,r_{\rm pc}/R_{\rm NS}}\biggr)
    \sqrt{\frac{P}{s}\,\Bigl(\frac{\dot{P}}{10^{-15}}\Bigr)},
\end{equation}
where we used the standard scaling
\begin{equation}
    B = B_{\rm ref}\,
    \Bigl(\frac{P}{\rm s}\,\frac{\dot{P}}{10^{-15}}\Bigr)^{1/2}.
\end{equation}

Then, we can impose an approximate condition for pair production as
\begin{equation}\label{Eq:PairCondition}
\chi \equiv \frac{3 \, \xi_\gamma \gamma_{\rm max}^3}{2 \, m_e\, \rho_c} \frac{s_{\rm min} \, B(s_{\rm min})/\rho_c}{B_Q} \simeq \chi_{\rm min} \simeq 1/7,
\end{equation}
where $\chi_{\rm min} \simeq 1/7$ has been tuned to the numerical solution of $\tau = 1$, $\xi_\gamma$ is taken to be an order-one coefficient which accounts for the correction between the pair producing photon energy and the typical photon energy ($\omega \equiv \xi_\gamma \omega_c$), and where $s_{\rm min}$ is the maximum of the function $s B(s)$. For a dipolar magnetic field this reads $s_{\rm min} = R_{\rm NS}/2$. Plugging Eq.~\ref{eq:gamma_max} into \ref{Eq:PairCondition} we find the simple scaling
\begin{eqnarray}\label{eq:pdot_std_analytic}
 \dot{P} &\sim 8 \times 10^{-18} \left(\frac{P}{s}\right)^{5/2}\Big(\frac{1.3\cdot10^{12}\rm G}{B_{\rm ref}}\Big)^2 \times \\ & \times\Big(\frac{\chi_{\rm min}}{1/7}\frac{5}{\xi_\gamma}\Big)^{1/2}\Big(\frac{12\,\rm km}{R_{\rm NS}}\Big)^5,\nonumber
\end{eqnarray}
which we checked to reproduce very well the numerical results for $P \gtrsim 0.5 \, \rm s$.

Let us now add the axions into the picture, starting with the case of very small axion masses. In the region of parameter space where axions are sourced, the axion effective charge density always dominates over the GJ charge density; therefore, the differential equation to solve is simply
\begin{equation}\label{eq:dgammAxionLimit}
    \frac{d^2\gamma_e}{ds^2}
    = -\frac{e \, g_{a\gamma}}{m_e}\,B(s)\,\partial_r a \simeq \frac{e\, B(s)}{2\, m_e} \frac{\alpha_{\rm EM}R_{\rm NS}}{(R_{\rm NS}+s)^2},
\end{equation}
where we neglected the exponential factor in the axion field (which is approximately one for small axion masses), and we used $g_{a\gamma} \sim \frac{\alpha_{\rm EM}}{2\pi f_a}$. Using the same procedure and reference values as in the case without axions, we find
\begin{eqnarray}\label{eq:pdotlowmass_highp}
\dot{P} &=  4\cdot 10^{-19} \Big(\frac{P}{s}\Big) \Big(\frac{1.3 \cdot 10^{12}\rm G}{B_{\rm ref}}\Big)^2 \times \\ & \times\Big(\frac{\chi_{\rm min}}{1/7}\frac{5}{\xi_\gamma}\Big)^{1/2}\Big(\frac{12\,\rm km}{R_{\rm NS}}\Big)^{7/2}, \nonumber
\end{eqnarray}
which again agrees very well with the full numerical results for $P \gtrsim 1 \, \rm s$.
One can also do the same computation for the massive axion case, solving the simplified equation (valid for $m_a R_{\rm NS} \gg 1$)
\begin{equation}\label{eq:dgammAxionLimitMss}
\frac{d^2\gamma_e}{ds^2}
    \simeq \frac{\alpha_{\rm EM}e\, B(s)}{2\, m_e} e^{-m_a s}m_a.
\end{equation}
In this case we find
\begin{eqnarray}\label{eq:pdothighmass}
\dot{P} &=  8\cdot 10^{-22} \Big(\frac{P}{s}\Big)^{1/4} \Big(\frac{1.3 \cdot 10^{12}\rm G}{B_{\rm ref}}\Big)^2 \\ & \times\Big(\frac{\chi_{\rm min}}{1/7}\frac{5}{\xi_\gamma}\Big)^{1/2}\Big(\frac{12\,\rm km}{R_{\rm NS}}\Big)^{11/4} \nonumber.
\end{eqnarray}

In all three cases (without axions, and with either massless or massive axions), the analytical formulas reproduce the numerical results extremely well, provided that radiation losses are negligible. This condition holds for large periods, i.e., on the right-hand side of the ``kink'' feature in the full numerical curves.

For smaller periods, the friction term during the acceleration cannot be neglected, and in general it becomes more difficult to derive simple analytical results. However, in the massive axion case, owing to the abrupt fall of the axion profile and the associated acceleration, one can still obtain a reasonably good analytical approximation. To this end, we solve Eq.~\ref{eq:dgammAxionLimitMss} and the equation of motion neglecting friction, up to the point at which friction becomes dominant and $\gamma_e$ saturates. To estimate where this transition occurs, we adopt a simple energetic argument, imposing
\begin{equation}
\int_0^{s_1} ds \Bigg(\frac{\alpha_{\rm EM} e B(s)}{2 \, m_e} e^{-m_a s} m_a\Bigg) = \int_0^{s_1} ds \, \mathcal{R}_\gamma,
\end{equation}
where $\mathcal{R}_\gamma$ is evaluated on the solution of Eq.~\ref{eq:dgammAxionLimitMss}. Following this procedure we get
\begin{eqnarray}\label{eq:pdotlowmass}
\dot{P} &=  6\cdot 10^{-21} \Big(\frac{P}{s}\Big)^{-5/7} \Big(\frac{\rho_c}{9\times10^7 \, \rm cm}\Big)^{4/7}\Big(\frac{1.3 \cdot 10^{12}\rm G}{B_{\rm ref}}\Big)^2 \nonumber\\ & \times\Big(\frac{\chi_{\rm min}}{1/7}\frac{5}{\xi_\gamma}\Big)^{8/7}\Big(\frac{12\,\rm km}{R_{\rm NS}}\Big)^{8/7} .
\end{eqnarray}

In Fig.~\ref{fig:analytic_comp} we compare the analytic approximations obtained in Eqns.\ref{eq:pdot_std_analytic},\ref{eq:pdotlowmass_highp}, \ref{eq:pdothighmass} and \ref{eq:pdotlowmass}, with the numerical results outlined in the preceding sections. We show results for the standard death line, computed without an axion (yellow), and two death lines for the light QCD axion, one obtained in the low mass $m_a R_{\rm NS} \ll 1$ (green) limit and one obtained in the high mass limit $m_a R_{\rm NS} \gg 1$ (red). Analytic estimates are shown in dashed lines. In both the standard scenario and the massless axion limit, analytic approximations are only applicable above the break in the power law, while for the heavy axion we plot the limiting approximations at $P\rightarrow \infty$ and $P \rightarrow 0$. All analytic results appear to be in excellent agreement with numerical calculations.

\begin{figure}
    \centering
    \includegraphics[width=\linewidth]{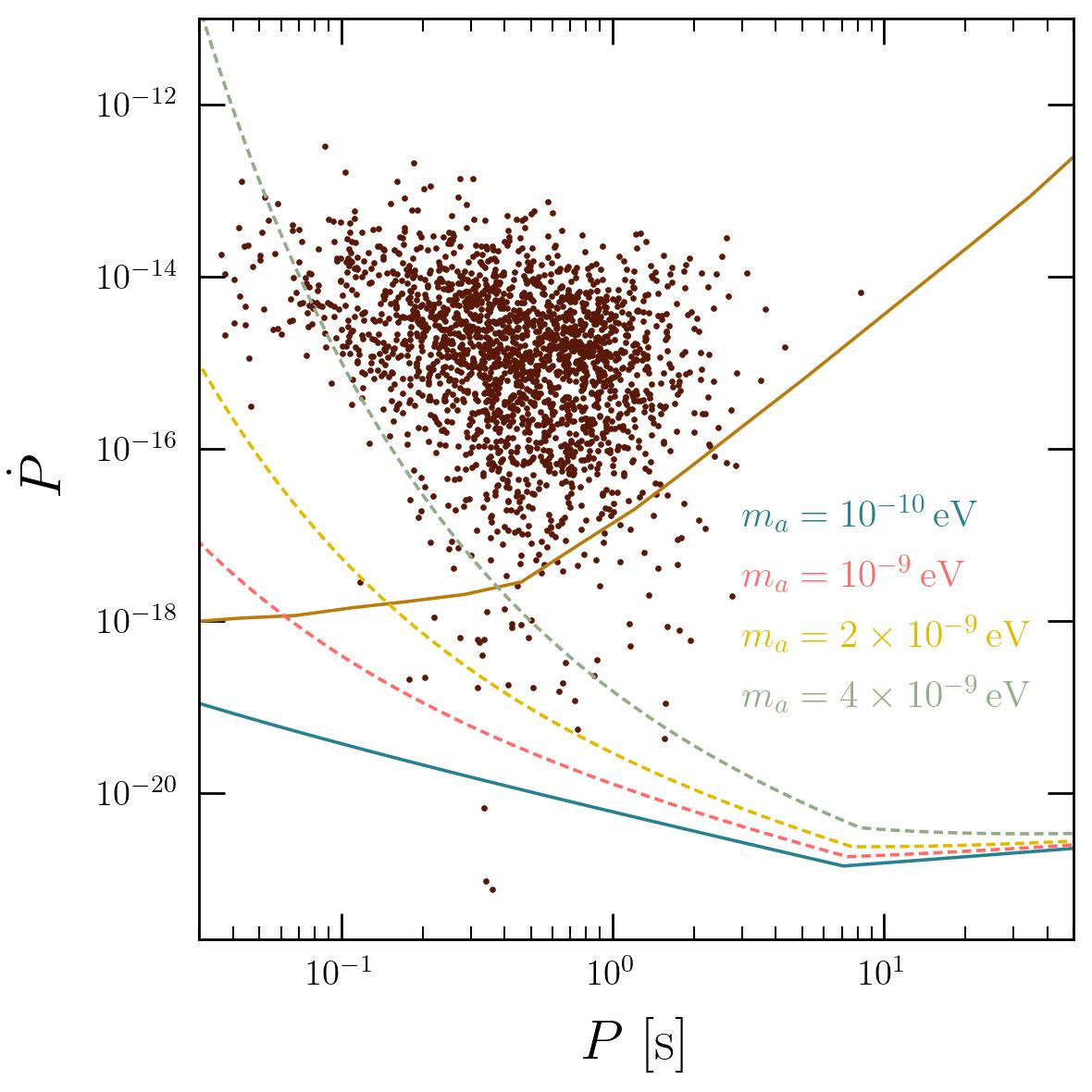}
    \caption{$P-\dot{P}$ diagram showing analytic derivation of death line produced under the assumption that the axion induced voltage drop on distances $r \leq r_{\rm pc}/2$ is zero; this analysis is performed in order to ensure that the one-dimensional treatment of pair production on the return current remains valid, and justifies the cut-off scale of $m_a \leq 10^{-9}$ eV. The neutron star population corresponds to one mock population realization of Model 10 (which is the best-fitting population that does not impose a death line cutoff).    }
    \label{fig:returncut}
\end{figure}

As mentioned in the proceeding section, we impose a cut-off threshold on the axion mass, requiring $m_a \leq 10^{-9}$ eV. This comes from the fact that the longitudinal extent of the axion induced voltage drop becomes smaller than the cross-sectional field line bundle of the return current. In order to assess this cut-off threshold more precisely, we use the analytic approach here to re-derive approximate death lines under the assumption that the axion induces no additional electric field on distances $r \leq r_{\rm pc} / 2$ from the neutron star surface. We plot these revised death lines in Fig.~\ref{fig:returncut} for various axion masses. Instead of showing the ATNF pulsar population, we plot instead one mock population realization produced using Model 10 of Appendix~\ref{secapp:pop}, which corresponds to the best-fitting model that does include a hard cut-off in the death line. At low axion masses ($m_a \lesssim 10^{-10}$ eV), the new cut-off has no effect, while at higher axion masses the death line weakens for lower-period pulsars. The adopted threshold cut of $m_a \leq 10^{-9}$ eV can be seen by eye to be conservative, as it is still in strong tension with nearly all of the pulsars sitting below the standard death line.

\section{Appendix E: Is there Evidence for a Death Line?}
\label{sec:popsyn}

In this section the question we seek to address is whether there is clear evidence for a death line in the pulsar distribution (at least in some rough definition). Only after having established the need of such a line, one may then study if the presence of axion hair unambiguously shift where such a death line would lie.

\begin{figure*}
    \centering
    \includegraphics[width=0.32\textwidth]{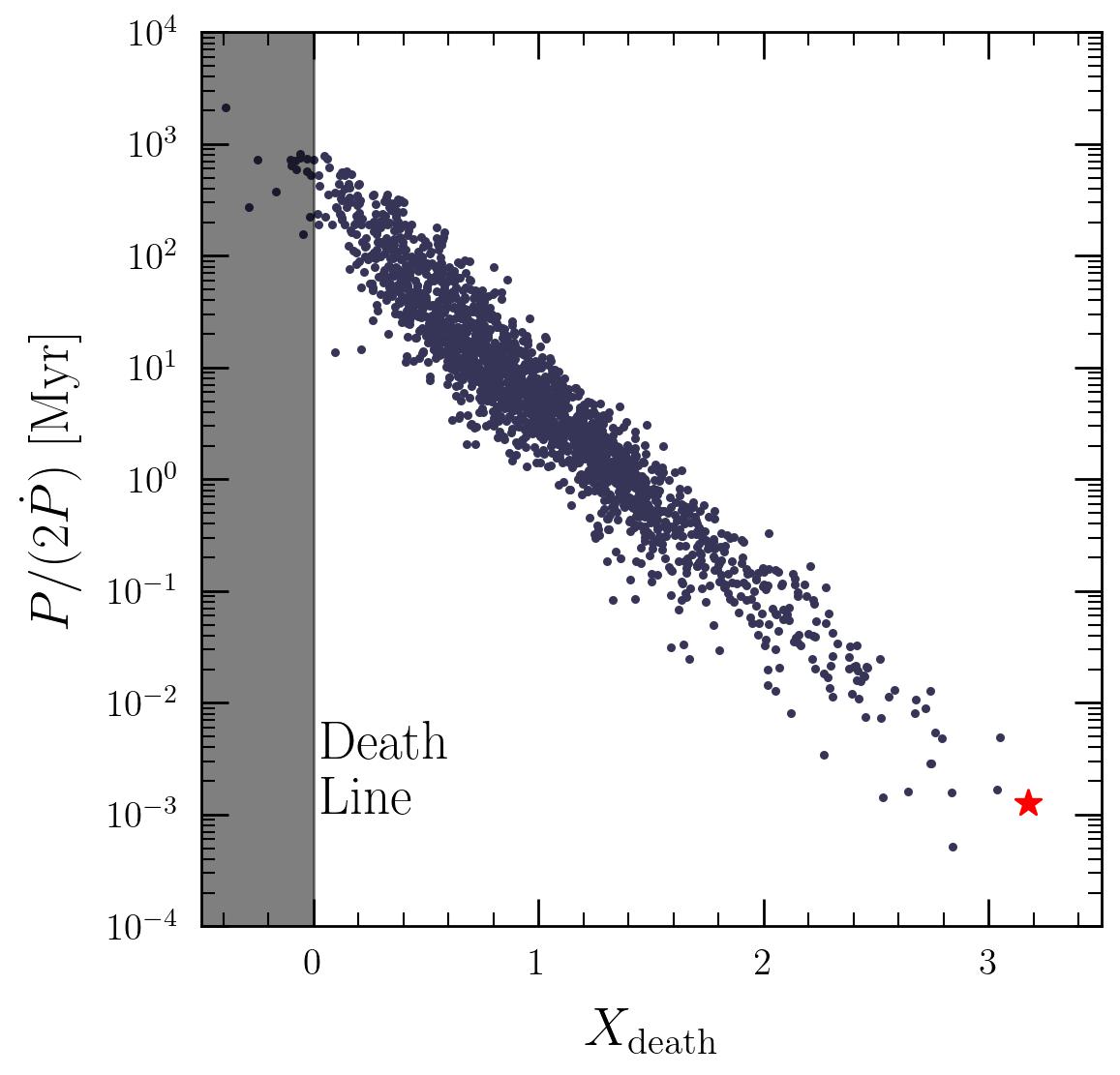}
    \includegraphics[width=0.32\textwidth]{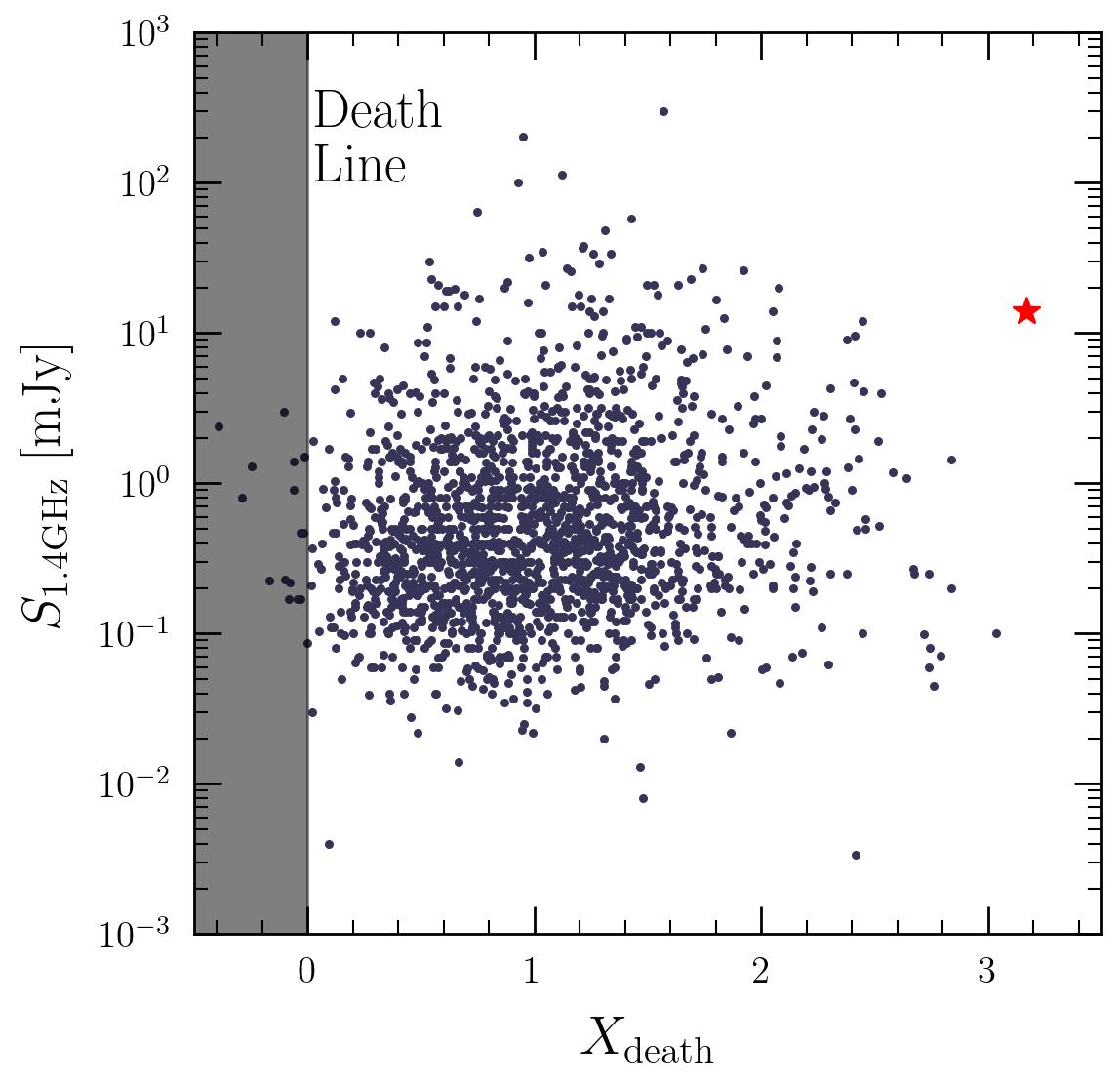}
    \includegraphics[width=0.32\textwidth]{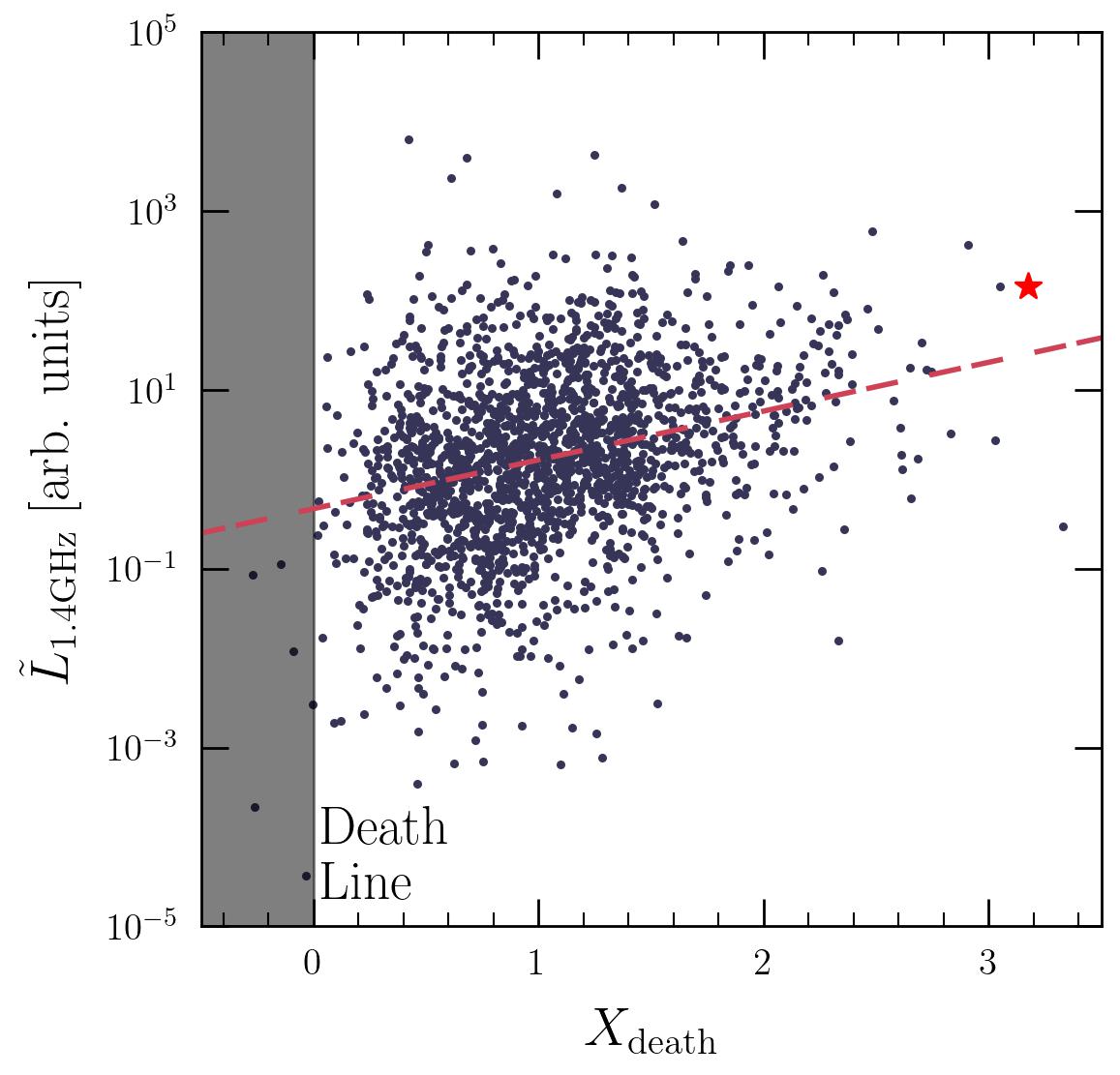}
    \caption{Proxy for the pulsar age $\tau_{\rm age} \sim P / (2 \dot{P})$ (left), flux density (center), and inferred pulsar luminosity (right) at $\nu = 1.4$ GHz as a function of effective distance from the death line $X_{\rm death} = \pm {\rm Min}\left[\sqrt{ (\log P - \log P_{\rm death})^2 + (\log \dot{P} - \log \dot{P}_{\rm death})^2} \right]$, where the $+/-$ corresponds to pulsar above/below the death line (\ie, the death valley has been highlighted in gray), and we have used the fiducial death line shown in Fig.~\ref{fig:ppdot_std}. Magnetars and millisecond pulsars have been removed from the dataset, and the Crab pulsar has been highlighted with a red star to serve as a reference value. In the right panel, a power law has been fit to the data to illustrate the general trend. }
    \label{fig:lumin}
\end{figure*}

Let us recall that from a theoretical perspective, the existence of a pulsar death line (or, more appropriately, a death valley) is apparent: pair cascades are a crucial ingredient in the generation of near-field coherent radio emission, and the generation of pair cascades require a sufficient number density of very high energy radiation to be present near the star, something which can only be sustained for pulsars with sufficiently high magnetic fields and rotational frequencies. Pulsars lose their rotational energy and their magnetic field strength as they age, and thus one expects old pulsars to become radio quiet. This correlation can be seen by noting that a rough proxy for the pulsars age is given by $\tau_{\rm age} \sim P / (2 \dot{P})$, which follows from integrating $\dot{P}$ assuming negligible initial spin period and a constant value of $B$, and plotting $\tau_{\rm age}$ as a function of some distance modulus $X_{\rm death} = \pm {\rm Min}\left[\sqrt{ (\log P - \log P_{\rm death})^2 + (\log \dot{P} - \log \dot{P}_{\rm death})^2} \right]$ which allows one to define some notion of how close a pulsar is to the death line. Here, the $+/-$ sign corresponding to pulsars above/below the death line, and we have used the fiducial death line shown in Fig.~\ref{fig:ppdot_std}. A scatter plot for the pulsar distribution (neglecting MSPs and magnetars) is shown in the left panel of Fig.~\ref{fig:lumin}. Now, assuming the birth rate of pulsars at times $t \gtrsim \mathcal{O}(100)$Myr is not significantly lower than it is today (which would be extremely odd, given that the star formation rate was much more active in the past), one would expect more than an order of magnitude more pulsars with ages above 100 Myr then with less than 100 Myr. Fig.~\ref{fig:lumin}, however, shows that this population is minuscule, suggesting that the old pulsar population is simply not observed.

Now the next question that immediately arises is whether the lack of observation of the old pulsar population could simply be a limit of current telescopes observational sensitivity -- or said equivalently, are old pulsars simply too faint to be observed? The short answer is that there exists no observational evidence to suggest that this is the case. If one plots the observed radio flux density (at, say, $\nu = 1.4$ GHz) as a function of the death distance modulus $X_{\rm death}$, one sees only a very mild correlation, with the general scaling only a small effect relative to the scatter in the population -- this is shown explicitly in the central panel of Fig.~\ref{fig:lumin}\footnote{When flux density measurements at this frequency are not available, we extrapolate from observations at 400 MHz or 2 GHz using the typical frequency scaling relation $S \propto \nu^{-1.4}$ to provide an approximate estimate. This re-scaling has no impact on the inferred trend in this figure.}. At this point, one may try to argue that older pulsars should be preferentially close to Earth, and perhaps the choice of plotting the flux density rather than the intrinsic luminosity has somehow disguised a more prominent feature that arises near the death line. The radio luminosity for each pulsar can be inferred by inverting the equation for the flux density 
\begin{eqnarray}
    S \simeq \frac{L}{d^2 d\Omega} \, ,
\end{eqnarray}
where $d$ is the inferred distance of each pulsar and $d\Omega = 4\pi (1- \cos\rho_b)$ is the fractional sky coverage of the radio beams. Here, $\rho_b$ is the angular opening of each beam, which from geometric considerations is roughly given by
\begin{eqnarray}\label{eq:rhob}
    \rho_b \simeq \sqrt{\frac{9\pi h_{\rm em}}{2P} }\, ,
\end{eqnarray}
with $h_{\rm em}$ being so-called `emission height' (the emission height being a proxy for the height above the surface at which the radio emission last scattered\footnote{The name here is something of a misnomer, and should {\emph{not}} be interpreted as the height at which radio emission is produced.}). For typical pulsars, the inferred emission heights tends to take on values $h_{\rm em} \sim 300$ km, with no apparent dependence on the pulsar properties~\cite{johnston2020galactic}. The scaling trend between luminosity and $X_{\rm death}$, plotted in the right panel of Fig.~\ref{fig:lumin} (a power-law fit to the data has been shown with a red dashed line in order to highlight the trend), is stronger than that of the flux density, but is still minor with respect to the general scatter in the population. As such, one is forced to conclude: (1) that there are many unobserved old pulsars, and (2) the lack of observation does not stem from a limitation of radio telescope sensitivity. Of course neither of these statements are new -- the former is natural consequence of pulsar evolution, and the latter has long been appreciated by simulating mock distributions in the pulsar population~\cite{lyne1985galactic,faucher2006birth,Bates:2013uma,Gullon:2014dva,Ronchi:2021arl,rea2024long}. It is perhaps worth highlighting that the former point could have also been inferred from the mere existence of the MSP population; MSPs are old pulsars (with $\sim \mathcal{O}({\rm Gyr})$ ages~\cite{Kiziltan:2009rx}) which have been, or are being, spun up by a binary companion -- the process of spin up causes these objects to traverse along fixed trajectories in the $P-\dot{P}$ plane, moving from the death valley to the lower left hand side of the diagram (and thus the large number of MSPs observed necessitates a large population of unobserved pulsars in the death valley itself).

The role of selection effects in determining biases in the observed pulsar population has been an active area of investigation for many years. In order to understand the potential impact of such effects, we rely on the ability to forward model the full pulsar population -- the procedure for simulating mock distributions, and the procedure for factoring in observable biases, is the subject of the following section.

\section{Appendix F: Neutron Star Population Synthesis}\label{secapp:pop}

As explained in the main text, in order to determine the importance of a pulsar death line in neutron star evolution, one must account for the wide range of selection effects which bias the observed pulsar distribution. The most natural way to account for these is via population synthesis -- we outline below the standard procedure for generating mock neutron star populations, and highlight the relative importance of each assumption in generating high quality fits to the observed pulsar distribution.

\subsection{Spatial distribution}

\begin{figure*}
    \centering
    \includegraphics[width=0.7\linewidth]{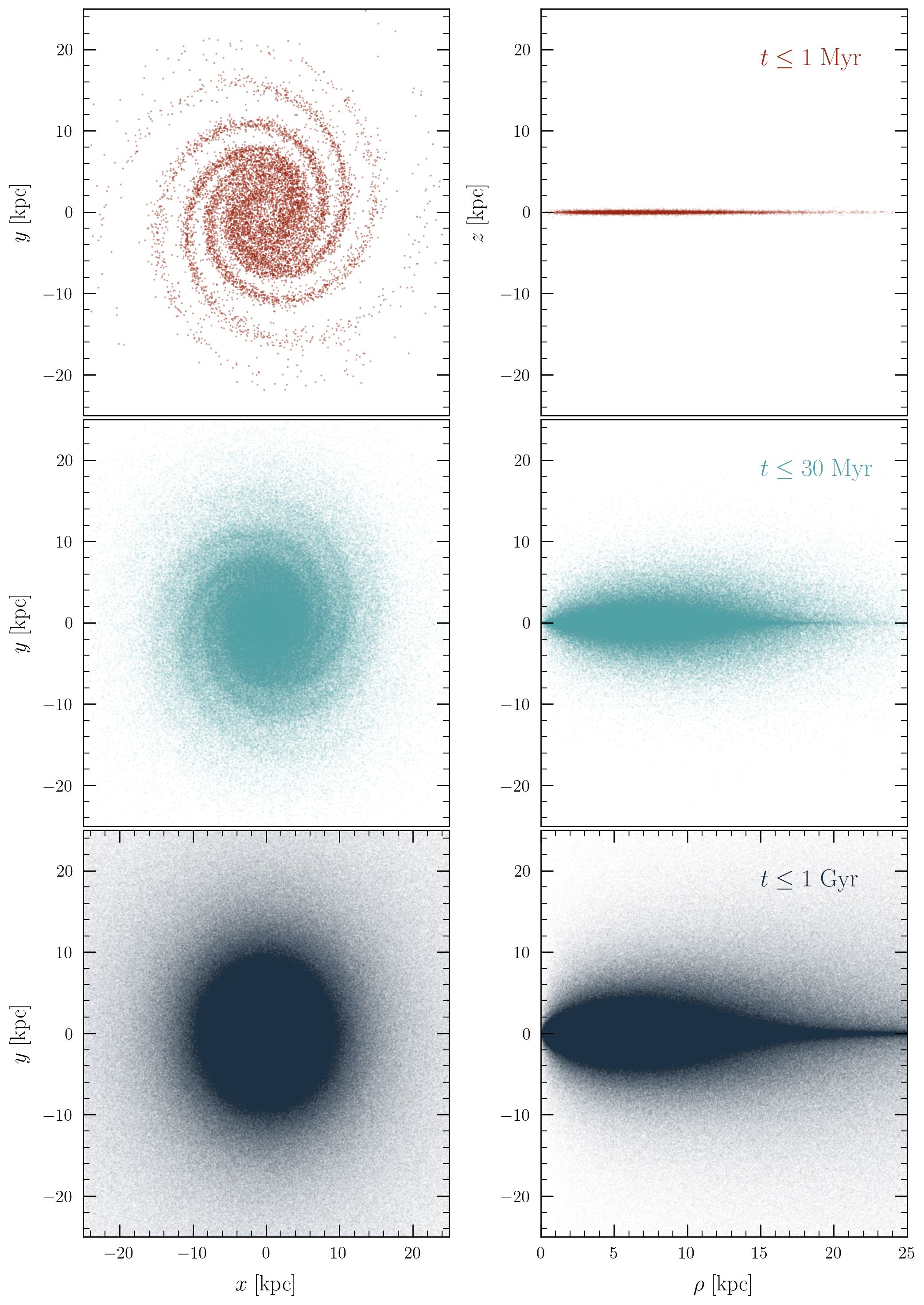}
    \caption{Spatial distribution in the $x-y$ (left) and $\rho-z$ (right) plane of a simulated pulsar population (prior to imposing any selection biases). The various panels highlight the time evolution in the distribution, including only pulsars born within the last $1$ Myr (top), $30$ Myr (center), and $1$ Gyr (bottom). Since there exist strong variations in the number of pulsars satisfying each of these criteria, point size and opacity in each panel has been adjusted in order to make the distributions apparent. }
    \label{fig:spatial}
\end{figure*}

One of the primary ingredients in generating mock pulsar distributions is the determination of their spatial distribution. Neutron stars are born predominantly from core collapse supernova, and thus the distribution of pulsars at birth closely traces that of the stellar distribution in the galactic disk. As a result of the asymmetry of the core collapse supernova, however, neutron stars are imparted with sizable kick velocities -- these initial velocities serve to displace their orbits, leading to an ever-growing discrepancy between the stellar and neutron star distributions. In order to account for this effect, one can sample the location of a neutron star at birth, sample an initial kick velocity, and evolve the neutron star over the corse of its lifetime through the Galactic potential. In this work, we adopt a kick distribution
\begin{eqnarray}
    P_k(v) = \sqrt{\frac{2}{\pi}}\frac{v^2}{\sigma_k^3} e^{-v^2 / \sigma_k^2}
\end{eqnarray}
with $\sigma_k = 260$ km /s, which is broadly consistent with the observed proper motions of pulsars~\cite{Hobbs:2005yx}.
Illustrations of the spatial distributions of a mock neutron star population (produced using the procedure outlined below) are shown in Fig.~\ref{fig:spatial}, where the three panels highlight young neutron stars with ages $t \leq 1$ Myr (top), middle-aged neutron stars with ages $t \leq 30$ Myr (middle), and old neutron stars with ages $t \leq 1$ Gyr (bottom). The two panels in Fig.~\ref{fig:spatial} show a top down, and side on, look of the Milky Way. The spiral arm structure is clearly visible in the initial distribution, and is washed out over time. The point size and opacity in each panel has been adjusted to ease visibility (the old pulsar population has individual objects extending to large radii, beyond the scale of the galaxy, and thus the entire region plotted has some lower point density background). 

In order to generate the spatial distribution of neutron stars at birth, we follow the procedure which has adopted in a majority of population synthesis models (see \eg\cite{lorimer2006parkes,Faucher-Giguere:2005dxp,Bates:2013uma,Cieslar:2018jzd,Graber:2023jgz}), which involves adopting fits to the observed  massive OB stellar distribution (\ie the stellar distribution leading to the production of neutron stars). In particular, the radial distribution is assumed to follow 
\begin{eqnarray}
    p(\rho) &\propto& \left(\frac{\rho}{R_\odot} \right)^{1.9} e^{-5 \left(\frac{\rho - R_\odot}{R_\odot} \right)} \, 
\end{eqnarray}
with $R_\odot = 8.5$ kpc, and the height is narrowly confined to the stellar disk, following an exponential distribution 
\begin{eqnarray}
p_z &\propto& e^{-|z| / h_c} 
\end{eqnarray}
with scale height $h_c = 0.18$ kpc. The angular direction in the $z=0$ plane is obtained by randomly selecting on the spiral arms, whose loci are described by
\begin{eqnarray}
    \theta_{\ell, i }(\rho) = k_i \log(\rho / \rho_{0,i})  + \theta_{0,i}
\end{eqnarray}
with $k_{1/2} = 4.25$ kpc, $k_{3/4} = 4.89$ kpc, $\theta_{0, 1} = 1.57$ rad, $\theta_{0, 2} = 4.71$ rad, $\theta_{0, 3} = 4.09$ rad, $\theta_{0, 4} = 0.95$ rad, $\rho_{0,1/2} = 3.48$ kpc, and $\rho_{0, 3/4} = 4.90$. Scatter along each arm is included by shifting the pulsar from the central part of the arm by a randomly selected distance $r_{\rm raw}$, which is drawn from a normal distribution with mean zero and standard deviation 0.07 kpc. 

 We evolve pulsars from birth through what is assumed to be an approximately static galactic potential. The galactic potential is given by a sum over three components coming from the galactic nucleus, the bulge, and a combined contribution from the disk and halo,
\begin{eqnarray}
    \Phi = \Phi_n + \Phi_b + \Phi_{dh}
\end{eqnarray}
with each contribution given by
\begin{eqnarray}
    \Phi_b &=& \frac{- G M_{b}}{R_{b} + r} \\
    \Phi_n &=& \frac{- G M_{n}}{R_{n} + r} \\
    \Phi_{dh} &=& \frac{- G M_{dh}}{\sqrt{(a_G + \sum_i \beta_i \sqrt{z^2 + h_i^2})^2 + b_{dh} + \rho^2}} 
\end{eqnarray}
Here, $r$ is the spherical radial coordinate and $\rho$ the cylindrical radial coordinate, and the free coefficients have been fit to the observed stellar motion, and are given by: $M_b = 9.3 \times 10^9 M_\odot$, $M_n = 1.0 \times 10^{10} M_\odot$, $M_{dh}= 1.45 \times 10^{11} M_\odot$, $\beta_1 = 0.4$, $\beta_2 = 0.5$, $\beta_3 = 0.1$, $h_1 = 0.325$ kpc, $h_2 = 0.090$ kpc, $h_3 = 0.125$ kpc, $a_G = 2.4$ kpc, $b_{dh} = 5.5$ kpc, $b_b = 0.25$ kpc, $b_{n} = 1.5$ kpc~\cite{Faucher-Giguere:2005dxp}.

\subsection{Magneto-rotational spin down}

As mentioned in the previous sections, the dominant observable for most rotationally powered pulsars are the rotational period and the spin down rate of the pulsar, with the latter depending on the dipolar magnetic field\footnote{Spin down is driven by electromagnetic energy losses occurring at radii near the light cylinder, $r \sim \Omega^{-1} \gg r_{\rm NS}$; as such, the spin down is a probe of the slowest falling multipolar magnetic field component, which is the dipolar component (invalidating this statement would typically require quadrupole field strengths at the surface more than $10^4$ times than of the dipolar field strength). We make no attempt in this work to model large higher order multipolar field components, and note that our results will be broadly consistent so long as such multipoles are not abnormally stronger than the dipolar field.}, the rotational period, and the misalignment angle between the rotational and magnetic axes $\chi$. The distributions that characterize these fundamental properties at birth are not known. As a result, attempts to synthesize mock neutron star populations traditionally adopt a reasonable distributions to characterize each of these (typically assuming that their properties are uncorrelated), and fit the free parameters of each distribution using the observed pulsar population. We will follow the same procedure here.

The misalignment angle is often assumed to be randomly selected at birth (\ie, there is no observational or theoretical evidence at the moment to favor a distribution which is preferentially aligned or anti-aligned), and as such we draw this value from a random distribution:
\begin{eqnarray}
    p(\chi_0) &=& \sin \chi_0 / 2 \, .
\end{eqnarray}
Magnetic field and period distributions of observed radio pulsars span many orders of magnitude, and have no strongly apparent skew, and thus in our baseline analyses we adopt a log normal distributions to characterize both the size of the dipolar field and the rotational period at birth
\begin{eqnarray}\label{eq:fit_LN}
    p(B_0) &=& \frac{1}{\sqrt{2\pi \sigma_{\log B}^2}} \, e^{- (\log B_0 - \mu_{\log B})^2 / (2 \, \sigma_{\log  B}^2)} \\[15pt]
     p(P_0) &=&\frac{1}{\sqrt{2\pi \sigma_{\log P}^2}} \, e^{- (\log P_0 - \mu_{\log P})^2 / (2 \,\sigma_{\log P}^2)} \, .
\end{eqnarray}
These parameterizations are commonly adopted in the literature, and have been shown to provide good fits to the observed data~\cite{lyne1985galactic,faucher2006birth,Bates:2013uma,Gullon:2014dva,Ronchi:2021arl,rea2024long}. Alternative parameterizations of the period distribution have also been proposed (see \eg~\cite{du2024initial}), and so for the sake of comparison we also adopt a truncated Weibull distribution\footnote{The Weibull distribution has a long tail extending to large periods. This tail has a minimal impact on the fit, but leads to a strong bias in the predicted pulsar birth rate, and thus it is natural in some sense to impose a truncation scale on the initial distribution.}
\begin{eqnarray}\label{eq:fit_WB}
    p(P_0) \propto \left(\frac{P_0}{P_\lambda} \right)^{k-1} \, e^{(P_0/P_\lambda)^k}  \theta(P_{\rm max} - P_0) 
\end{eqnarray}
which serves to show the relative importance of these parameterizations on the inferred dead neutron star population. 

The free parameters in Eqns.~\ref{eq:fit_LN} include the central values and standard deviations in the magnetic field and period $(\mu_{\log B}, \sigma_{\log B}, \mu_{\log P}, \sigma_{\log P})$, and in the case of the Weibull distribution the period is instead characterized by $(k, P_\lambda, P_{\rm max})$, which capture the skew, the peak, and the truncation scale of the distribution. We fix $P_{\rm max} = 1$ second, which in practice has a minimal impact on the quality of the fit. 

In order to determine the best fit parameters, one must evolve the population from birth to today. Neutron stars lose their rotational energy (which is dumped into particle acceleration and electromagnetic radiation) at a rate
\begin{eqnarray}
    \dot{P} = \beta \frac{B_0^2}{P} (\kappa_0 + \kappa_1 \sin^2\chi) \\
\end{eqnarray}
where $\chi$ is the angle between the rotational period and magnetic field axis (the so-called misalignment angle), the coefficients $\kappa_0 \simeq \kappa_1 \simeq 1$ have been inferred from numerical simulations\footnote{Note that standard vacuum dipole radiation can be computed exactly, and is proportional to $\sin^2\chi$. The presence of a co-rotating plasma that is dragged by the rotating magnetic field, however, yields additional corrections $\mathcal{O}(1)$ corrections to the spin down rate -- while numerical simulations are required to compute these corrections, we note that the energy loss from a plasma-filled force-free magnetosphere is reasonably well constrained.}\cite{Philippov:2013aha}, and $\beta \equiv \pi^2 R_{\rm NS}^6 ./ I_{\rm NS} \simeq 6 \times 10^{-40} \, {\rm s / G^2}$. The process of spin-down drives the neutron star toward a state of alignment, in which $\chi \sim 0$ at a rate given by
\begin{eqnarray}
    \dot{\chi} = - \beta \kappa_2 \, \frac{B^2}{P^2} \sin\chi \cos \chi \, ,
\end{eqnarray}
where $\kappa_2 \sim 1$~\cite{Philippov:2013aha}. Finally, the magnetic field of the neutron star, which is supported by currents inside the star itself, loses energy as those currents slowly dissipate; in general, there are a variety of physical effects which can drive short-term decay of the magnetic field (this includes ambipolar diffusion, and the Hall effect, both of which tend to drive a mild suppression of ultra-strong magnetic fields of young pulsars on $\mathcal{O}({\rm kyr})$ timescales)~\cite{Passamonti:2016nmf,Pons:2007vf,Pons:2019zyc}, but the long-term evolution is believed to be driven by Ohmic dissipation, which is expected to operate on timescales $\tau \gtrsim \mathcal{O}({\rm Myr})$~\cite{Cumming:2004mf,Aguilera:2007xk,Pons:2019zyc}. The efficiency of each of these processes depend on a variety of different factors, including the location of the currents supporting the magnetic field structure, the temperature and composition of the star and crust, the conductivity, etc. Computing the decay of the magnetic field in a self-consistent manner from first principles is an enormously difficult task and remains an active field of research; as a result, we adopt here multiple phenomenological parameterizations of the magnetic field decay, which functional forms and timescales motivated by numerical simulations and pulsar observations~\cite{Gullon:2014dva,Vigano:2021olr,Mereghetti:2015asa,Igoshev:2018fcp}, and attempt to demonstrate that any reasonable choice of magnetic field decay that fits the data well will lead to the same observational consequences. Since we are interested predominantly in  the late-time evolution of older pulsars, we will neglect the kyr evolution of the  magnetic field and focus solely on the evolutionary behavior taking place on $t \gtrsim $Myr timescales.

One of the most extreme parameterizations one can adopt is the assumption that the magnetic field decays  exponential on timescales $\tau_{\rm ohm}$, \ie
\begin{eqnarray}\label{eq:Bexp}
    B(t) = B_0 \, e^{- t / \tau_{\rm ohm}} \, .
\end{eqnarray}
This behavior is roughly consistent with the scaling behavior seen in numerical simulations which have crust-confined magnetic fields (an assumption which is unlikely to describe the long-term behavior of the field decay)~\cite{Gullon:2014dva,Vigano:2021olr}.  The value of $\tau_{\rm ohm}$ is expected to depend on how the magnetic field penetrates the core, the thickness and temperature of the crust, and the impurity of the crust, with reasonable values of the aforementioned leading to the prediction of Ohmic decay timescales ranging from  $\sim \mathcal{O}(\rm Myrs)$ to $\sim \mathcal{O}(\rm Gyrs)$ (the upper value being the limiting case where magnetic fields don't decay).  Observations inferring the kinematic ages of a small subset of the pulsar population (which are far more reliable than the often quoted spin down timescale, $\tau \sim P / (2\dot{P})$), suggest that long-term exponential decay on timescales $\tau \lesssim 10$ Myr is not consistent with the observable population~\cite{Igoshev:2018fcp}. In what follows we will treat $\tau_{\rm ohm}$ as a free parameter ranging from 10 Myr to 10 Gyr.

An alternative parameterization has recently been invoked in \cite{Graber:2023jgz},  which takes the form
\begin{eqnarray}\label{eq:Bpl}
    B(t) = B_0 \times \begin{cases}
    \left(1 + \frac{t}{\tau_1}\right)^{a_1} \left(1 + \frac{t}{\tau_2}\right)^{a_2-a_1}  \times \\[5pt] \left(1 + \frac{t}{\tau_{\rm late}}\right)^{a_{\rm late} - a_2} \hspace{.31cm} \tau_1 < \tau_2 < \tau_{\rm late} \\[15pt]
    \left(1 + \frac{t}{\tau_1}\right)^{a_1}   \times \\[5pt] \left(1 + \frac{t}{\tau_{\rm late}}\right)^{a_{\rm late} - a_1} \hspace{.4cm} \tau_1 < \tau_{\rm late} < \tau_2  \\[15pt]
    \left(1 + \frac{t}{\tau_{\rm late}}\right)^{a_{\rm late}}  \hspace{.9cm} \tau_{\rm late} < \tau_1 < \tau_2 
    \end{cases}
\end{eqnarray}
with $\tau_1 = A_1 B_0^{b_1}$, $\tau_2 = A_2 B_0^{b_2}$, $\tau_{\rm late} = 2 {\rm Myr}$, $A_1 = 10^{14} {\rm yr} {\rm G}^{-b1}$, $b_1 = -0.8$, $A_2 = 6 \times 10^{8} {\rm yr} \, {\rm G}^{-b2}$, $b_2 = -0.2$, $a_1 = -0.13$, $a_2 = -3.0$, and $a_{\rm late} = -3.0$. Here, the parameters have been adjusted to fit the decay observed in simulations of magnetic field decay~\cite{Vigano:2021olr} (which again make implicit assumptions about the confinement of magnetic fields to the crust).

At late times, both parameterizations predict that the magnetic field will continue to decay to very small values. As mentioned above, this is unlikely to be a realistic description of the late-time evolution. In order to investigate a potential impact of the unknown late-time evolution, we perform two different types of analyses, one in which these fields are allowed to decay indefinitely, and one in which the magnetic field decays until settling into a constant asymptotic value -- we choose a final state value by sampling from a log-normal distribution with mean $\mu_{\log_{10} B, {\rm final} } = 8.5$ and standard deviation $\sigma_{\log_{10} B, {\rm final}} = 0.5$, which is broadly consistent with the properties inferred from millisecond pulsars. These choices are not expected to effect the observable radio pulsar distribution (and thus are expected to fit the data equally well), but have the potential to alter the distribution of dead pulsars (potentially making them relevant when axion hair is present).

\begin{table*}
    \centering
    \scriptsize                      %
    \setlength{\tabcolsep}{5pt}      %
    \begin{adjustbox}{max width=\textwidth}
        \begin{tabular}{c|c|c|c|c|c|c|c|c}
          Model    &  $B_0$ Dist. & $P_0$ Dist. & $B(t)$ & $B_{\rm final} $ & Use $\Gamma_{\rm br}$ &  Death Line & Fit Params.  & $p/p_0$  \\[5pt] \hline
        1    & LN   & LN & Exp. & Yes & Yes & Yes & $(\sigma_{\log{B}}, \log_{10}B, \sigma_{\log{P}}, \log_{10}P)$ &  $0.2$ \\[5pt]
            &   &  & $\tau = 10 $ Myr &  &  &  & $(0.33,12.46,0.73,-0.82)$&   \\[5pt] \hline
        2     & LN   & LN & Exp. & Yes & Yes & Yes & $(\sigma_{\log{B}}, \log_{10}B, \sigma_{\log{P}}, \log_{10}P)$ & $< 4 \times 10^{-4}$ \\[5pt]
           &   &  &  $\tau = 10 $ Gyr &  &  &  & $(0.71,13.26,0.14,-1.32)$ &  \\[5pt] \hline
        3    & LN   & WB & Exp. & Yes & Yes & Yes & $(\sigma_{\log{B}}, \log_{10}B, P_\lambda, k)$ & $0.51$ \\[5pt]
           &   &  & $\tau = 10 $ Myr  &  &  &  & $(0.35,12.41,0.39,58.65)$ &   \\[5pt] \hline
        4    & LN   & WB & Exp. & Yes & Yes & Yes & $(\sigma_{\log{B}}, \log_{10}B, P_\lambda, k)$ & $< 4 \times 10^{-4}$  \\[5pt]
           &   &  & $\tau= 10 $ Gyr &  &  &  & $(0.41,12.60,0.25,32.82)$ &   \\[5pt] \hline
        5    & LN   & LN & Exp. & No & Yes & Yes & $(\sigma_{\log{B}}, \log_{10}B, \sigma_{\log{P}}, \log_{10}P)$ & $0.52$ \\[5pt]
           &   &  & $\tau = 10 $ Myr &  &  &  & $(0.40,12.45,0.66,-0.71)$ &   \\[5pt] \hline
        6    & LN   & LN & Exp. & Yes & No & Yes & $(\sigma_{\log{B}}, \log_{10}B, \sigma_{\log{P}}, \log_{10}P)$ & $0.9$ \\[5pt]
           &   &  & $\tau = 10 $ Myr &  &  &  & $(0.40,12.46,0.52,-0.98)$ & \\[5pt] \hline
        7    & LN   & LN & Exp. & Yes & Yes & No & $(\sigma_{\log{B}}, \log_{10}B, \sigma_{\log{P}}, \log_{10}P)$ & $< 4 \times 10^{-4}$ \\[5pt]
           &   &  & $\tau = 10 $ Myr &  &  &  & $(0.37,12.34,0.58,-1.01)$  & \\[5pt] \hline
        8    & LN   & LN & PL & Yes & Yes & Yes & $(\sigma_{\log{B}}, \log_{10}B, \sigma_{\log{P}}, \log_{10}P)$ & $1.0$ \\[5pt]
           &   &  &  &  &  &  & $(0.36,12.42,0.69,-0.82)$ &  \\[5pt] \hline
        9    & LN   & WB & PL & Yes & Yes & Yes & $(\sigma_{\log{B}}, \log_{10}B, P_\lambda, k)$ &  $0.19$ \\[5pt]
           &   &  &  &  &  &  & $(0.30,12.40,0.37,59.51)$ & \\[5pt] \hline
        10    & LN   & LN & Exp. & No & Yes & No & $(\sigma_{\log{B}}, \log_{10}B, \sigma_{\log{P}}, \log_{10}P)$ & $0.0015$ \\[5pt]
           &   &  & $\tau = 10 $Myr &  &  &  & $(0.30,12.21, 0.40,-0.87)$ &  \\[5pt] \hline
        11    & LN   & WB & Exp. & No & Yes & No & $(\sigma_{\log{B}}, \log_{10}B, P_\lambda, k)$ &  $< 4 \times 10^{-4}$ \\[5pt]
           &   &  & $\tau = 10 $Myr &  &  &  & $(0.20,12.30,0.49,44.42)$  & \\[5pt] \hline
        12    & LN   & LN & PL & No & No & No & $(\sigma_{\log{B}}, \log_{10}B, \sigma_{\log{P}}, \log_{10}P)$ & $< 4 \times 10^{-4}$ \\[5pt]
           &   &  &  &  &  &  & $(0.30,12.34,0.70,-1.00)$   & \\[5pt] \hline
        13    & LN   & WB & PL & No & No & No & $(\sigma_{\log{B}}, \log_{10}B, P_\lambda, k)$ &  $< 4 \times 10^{-4}$ \\[5pt]
           &   &  &  &  &  &  & $(0.61,12.83,0.38,21.97)$ &  \\[5pt] \hline
       14    & LN   & LN & Exp & Yes & No & No & $(\sigma_{\log{B}}, \log_{10}B, \sigma_{\log{P}}, \log_{10}P)$ &  $< 4 \times 10^{-4}$  \\[5pt]
       &   &  & $\tau= 10 $ Myr &  &  &  & $(0.56,12.70,0.53,-0.93)$  & \\[5pt] \hline
       15     & LN   & LN & Exp. & Yes & Yes & No & $(\sigma_{\log{B}}, \log_{10}B, \sigma_{\log{P}}, \log_{10}P)$ & $< 4 \times 10^{-4}$ \\[5pt]
           &   &  &  $\tau = 10 $ Gyr &  &  &  & $(0.55, 12.56, 0.30, -0.98)$ & \\[5pt] \hline
        16    & LN   & WB & Exp. & Yes & Yes & No & $(\sigma_{\log{B}}, \log_{10}B, P_\lambda, k)$ &  $< 4 \times 10^{-4}$ \\[5pt]
           &   &  & $\tau = 10 $ Myr  &  &  &  & $(0.45,12.33,0.38,28.42)$   & \\[5pt] \hline
        17    & LN   & LN & PL & Yes & Yes & No & $(\sigma_{\log{B}}, \log_{10}B, \sigma_{\log{P}}, \log_{10}P)$ &  $4 \times 10^{-4}$ \\[5pt]
           &   &  &  &  &  &  & $(0.35,12.47,0.65,-1.10)$  &  \\[5pt] \hline
        \end{tabular}
    \end{adjustbox}
    \caption{List of parameters used in our neutron star population synthesis. Columns denote: model number (an arbitrary label used for identification), the distribution characterizing the dipolar field strength at birth (`LN' = log-normal), the distribution characterizing the pulsar period at birth (`WB' = Weibull), the magnetic field decay model (with `Exp' corresponding to Eq.~\ref{eq:Bexp}, and `PL' corresponding to Eq.~\ref{eq:Bpl}), whether the late-time evolution of $B(t)$ is truncated at a value determined via the sampling procedure outlined in the text, whether constraints on the pulsar birth rate are applied to the effective likelihood, whether a hard death line is imposed on the simulated population, the best fit parameters of the initial distribution, and the p-value normalized to the best-fit model $p/p_0$.}
    \label{tab:popsyn}
\end{table*}

As the rotational frequency and magnetic field of the pulsar falls, so does the maximum potential drop across the polar cap $(\Delta V)_{\rm max} \propto \Omega^2 B $. Consequently, at some point in the evolutionary history of the neutron star, primary particles pulled from the surface of the neutron star cannot be accelerated to sufficiently large boost factors to ignite pair cascades. Since pair cascades are required for the generation of radio emission, the point at which this occurs defines when the pulsar becomes radio silent, or equivalently, when a pulsar `dies'. Furthermore, without pair cascades the magnetosphere becomes `charge starved', altering the structure of the magnetosphere and the spin down of the pulsar itself (see e.g.~\cite{SashaReview}) -- without a dense plasma surrounding the pulsar, the spin-down coefficients tend toward their vacuum dipole values of $\kappa_0 = 0, \kappa_1 = \kappa_2 = 2/3$. In the following, we will perform two different sets of analyses. In one set of analyses, we will assume pulsar death is abrupt, occurring when pulsars cross our fiducial death line (described in the sections above), leading to an immediate shift in the spin down coefficients $\kappa_i$ and making these pulsars unobservable. In the second analysis, we do not impose any notion of a death line, and we allow pulsars to continue evolving with the $\kappa_i$ values inferred from the force-free limit. The truth likely lies somewhere in between these two analyses, making these well-defined limiting cases.

\subsection{Selection Biases and Statistical Fits}
\label{sec:obsbias}

The procedure outlined in the proceeding section tells one how to sample and forward model each pulsar in the mock population, but in order to compare with the observed population one must fold in all observational biases which determine which small fraction of the total population we observed.

We begin by constructing a sample set of $N_{\rm psr, tot}$ pulsars, where $N_{\rm psr, tot} = \mathcal{P}(\frac{2 \, {\rm psrs}}{\rm cntry} \tau_{\rm max})$, with $\mathcal{P}(x)$ being a Poisson distribution with mean $x$, and $\tau_{\rm max}$ chosen to be a sufficiently large number such that all observable pulsars are expected to have an age $\tau \ll \tau_{\rm max}$ (here we take $\tau_{\rm max} = 0.2$ Gyr). Ages of each pulsar in the sample are drawn from a flat distribution, and then each pulsar is evolved to determine is properties and location today. We then run through each neutron star in our sample and ask the following questions:
\begin{itemize}
    \item Is the pulsar dead? If so, and if the presence of a death line is assumed in the analysis, we remove the pulsar from the population. For a default analyses, we adopt the death line shown in Fig.~\ref{fig:ppdot_std}. For analyses in which a death line is not imposed, pulsars are retrained with 50$\%$ probability -- this selection cut is imposed in order to conservatively avoid the question of how bright the radio emission generated on the return current is.  
    \item Does at least one of the pulsars radio beams point toward Earth? Here, we randomly select a viewing angle on the sky, and use the misalignment angle today $\chi$, alongside the angular opening of the pulsar beam as computed in Eq.~\ref{eq:rhob}, to determine whether part of that beam intersects the angle between the pulsar and Earth. If it does not, the pulsar is removed from the sample. 
    \item Is the flux density above the detection threshold associated to large-scale pulsar surveys? The signal-to-noise ratio of a typical radio survey can be computed using
    \begin{eqnarray}
        \frac{S}{N} = \frac{S_{\rm mean} G \sqrt{n_{\rm pol} \, t_{\rm obs} \, \delta f_{\rm bw}}}{\beta T_{\rm tot}} \sqrt{\frac{P - w_{\rm obs}}{w_{\rm obs}}} \, ,
    \end{eqnarray}
    where $G$ is the receiver gain, $S_{\rm mean} \simeq S_{f,{\rm obs}} w_{\rm obs} / P$ is the mean flux density averaged over a rotational period (with $S_{f,{\rm obs}}$ being the observed flux density at Earth), $t_{\rm obs}$ is the observing time, $\delta f_{\rm bw}$ is the observational bandwidth, $T_{\rm tot} = T_{\rm rcv} + T_{\rm sky}(\theta, \phi)$ is the sum of the receiver temperature and the spatially dependent sky temperature, and $w_{\rm obs}$ is the observed pulse width, which is related to the intrinsic pulse width $w_{\rm int}$ by
    \begin{eqnarray}
        w_{\rm obs} = \sqrt{w_{\rm int}^2 + \tau_{\rm samp}^2 + \tau_{\rm DM}^2 + \tau_{\rm scat}^2} \, .
    \end{eqnarray}
    Here, the observed width receives additional corrections from the scattering of the radio off the intergalactic medium
    \begin{eqnarray}
        \tau_{\rm scat} = 3.6 \times 10^{-9} \, {\rm DM}^{2.2} (1 + 1.94 \times 10^{-3} \, {\rm DM}^2) \, ,
    \end{eqnarray}
    the sampling time of the telescope $\tau_{\rm samp}$, and pulse smearing
    \begin{eqnarray}
        \tau_{\rm DM} = \frac{e^2}{\pi m_e}\frac{\delta f_{\rm ch}}{f^3} \, {\rm DM} \, ,
    \end{eqnarray}
    where $\delta f_{\rm ch}$ is the frequency channel width, and ${\rm DM}$ is the dispersion measure (\ie the column density of free electrons along the line of sight). The dispersion measure of each pulsar in the sample can be computed using the YMW galactic electron model~\cite{yao2017new}. In order to determine the intrinsic luminosity of the pulsar, we note that the distribution of inferred luminosities from the observed pulsar population roughly follows
    \begin{eqnarray}
        L = L_0  \, \dot{P}^\alpha \left(\frac{1 \, {\rm s}}{P}\right)^{3\alpha}
    \end{eqnarray}
    where $\alpha \sim 0.48$ is a fitting parameter~\cite{Faucher-Giguere:2005dxp,Gullon:2014dva,Graber:2023jgz}. In practice, there is scatter about this mean distribution, and thus we sample from a lognormal distribution with mean zero  and standard deviation $\sigma_L =0.8$. The flux at Earth can be directly inferred knowing the distance and the opening angle of each radio beam.

     Here, we adopt characteristic values for the radio telescope parameters consistent with what was performed in the broad sky pulsar searches of the PMPS, SMPS, and HTRU surveys -- this includes a signal to noise threshold of $S/N=9$, $\delta f_{\rm ch} = 3$ MHz, $\tau_{\rm samp} = 64\mu$s, $T_{\rm sky}(\nu \simeq 1.4 GHz) \ll T_{\rm rcv} = 21$K, $t_{\rm obs} = 4300$s, $\delta f_{\rm bw} = 288$MHz, $n_{\rm pol} = 2$, and $G = 0.735$ K/Jy (see \eg compilation of observations in Table 1 of ~\cite{Graber:2023jgz}).

\end{itemize}
The collection of all pulsars passing these selection cuts define the observable pulsar sample.

Given a simulated `observable pulsar population', we are left with the question of how to compare the quality of the mock population with that which the underlying observed population. This is often done by defining a test statistic which characterizes the mismatch between the two populations, and then determining the associated p-value by constructing a null distribution, which operates under the hypothesis that in fact both samples are obtained from the same underlying distribution, via bootstrapping. In one-dimension, common examples of the test statistic include: the Komologorov-Smirnov (KS) test~\cite{berger2014kolmogorov} test, which characterizes the maximum mismatch in the empirically derived cumulative distribution functions (CDFs) of the two data sets (\ie $\mathcal{D} \equiv {\rm Max}\left[ \left|{\rm CDF}_{\rm X}- {\rm CDF}_{\rm Y}\right|\right]$), the Cram\'{e}r-von Mises Test, which captures the integrated squared difference between CDFs, the binned $\chi$-squared test, which bins the data and looks at the summed differences in the PDFs of the two distributions, and the  Energy-Distance (ED) metric~\cite{rizzo2016energy}, which computes the expectation values of the pair-wise distances between, and within, the various sets (\ie $\mathcal{D} \equiv 2 \mathcal{E}||X_i - Y_j|| - \mathcal{E}||X_i - X_j|| - \mathcal{E}||Y_i - Y_j||$ where $\mathcal{E}$ is the expectation value, $|| \cdot ||$ is the Euclidean norm). Each approach has various advantages and disadvantages. For example, the KS and ED test are intrinsically more sensitive to deviations in the bulk, rather than the tails, of the distribution (even if the tails are highly incompatible). The $\chi^2$ test on the other hand requires the imposition of an arbitrary binning scheme. 

\begin{figure*}
    \centering
    \includegraphics[width=\linewidth]{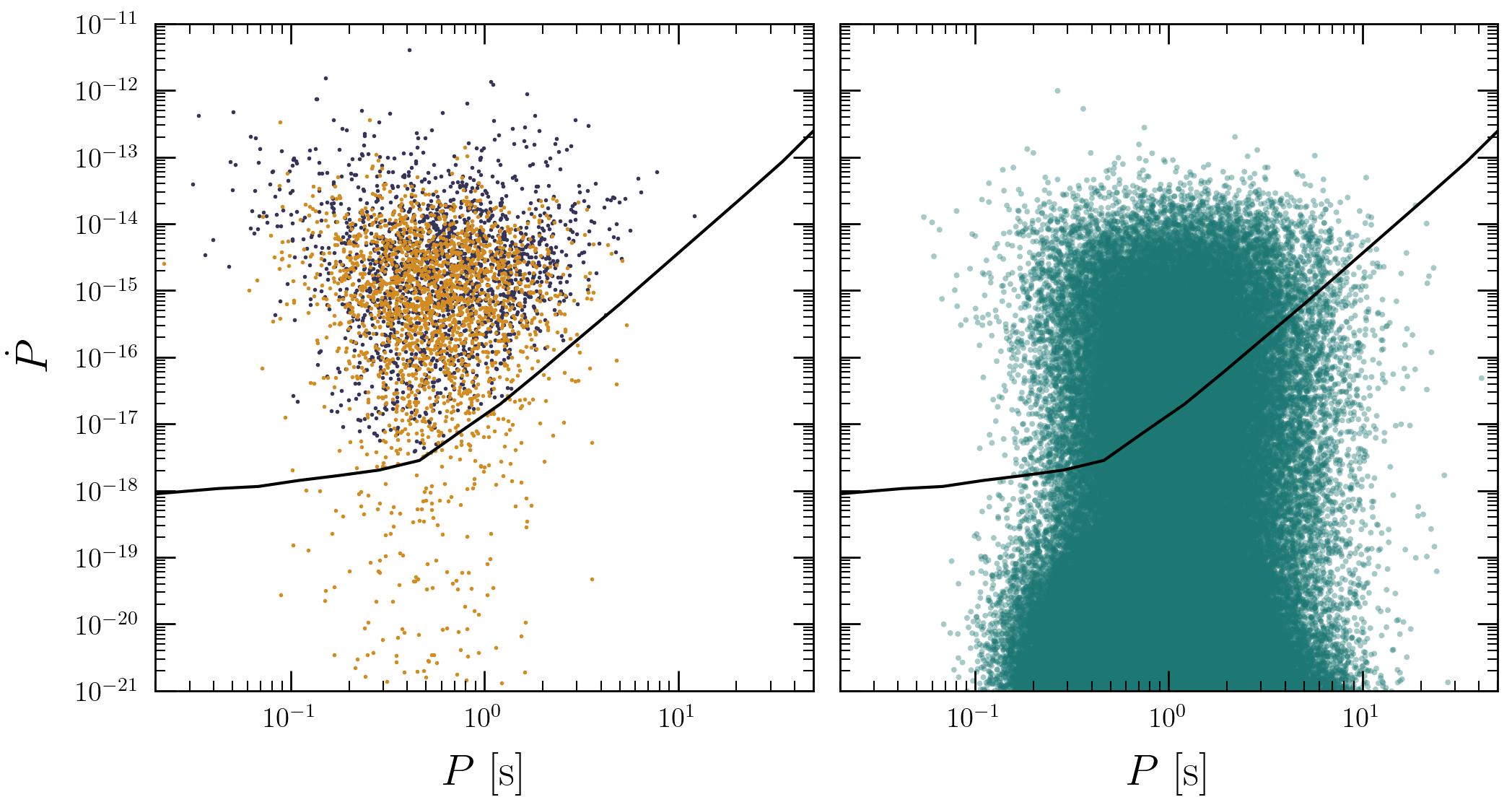}
    \caption{Left: $P-\dot{P}$ diagram showing the comparison between the ATNF catalog (purple) and the best-fit simulated model obtained using an estimated Meerkat flux density threshold and assuming a death line does not exist. In comparison with Fig.~\ref{fig:ppdot_m10}, which only includes older radio surveys, one can see that the number of pulsars beyond the death line has sizably increased. Right: sub-sample of distribution of pulsars which are unobservable.  }
    \label{fig:meerkat}
\end{figure*}

In our case, we hope to quantify the extent to a death line is needed for good fits to the population, and more specifically the extent to which a shifted death line would remain compatible with the observed pulsar distribution. As we will show below, not including a death line in the population models unavoidably leads to the prediction of $\mathcal{O}(100)$ observable pulsars sitting very far below the death line, a number which is largely independent of population modeling assumptions (and a number which grows significantly if one uses a flux density threshold consistent with the currently operating Meerkat telescope, see Fig.~\ref{fig:meerkat}). This is not an exceptionally large number, contributing to the CDF at only the $\mathcal{O}(2\%)$ level, however these pulsars are extremely displaced from the the bulk of the population, and thus would appear as clear outliers would they have been observed. As such,  a CDF based test statistic will not yield sensible results -- statistical fluctuations in the bulk distribution of the population yield test statistics which are consistently larger than the highly extended tail (despite the fact that the bulk distributions look extremely similar -- see figures below), implying the quality if the fit is in most cases dictated by a combination of unaccounted for noise and small systematics that shift pulsars at central $(P, \dot{P})$, rather than what appears to be a robust incompatible feature of the model.

The solution to this issue is that one must incorporate both the fact that the bulk of the pulsar distribution will have a non-negligible intrinsic scatter across different realizations, and the notion that there is some intrinsic cosmic variance (\ie the observed distribution only represents a single realization). These effects can be naturally included using the $\chi^2$ statistic; here, one can include statistical uncertainties using Possonian statistics in each bin, as well as correlated features in the PDF itself (which naturally capture the population-to-population scatter) by including off-diagonal elements in the correlation matrix. The size of the off-diagonal elements are not a priori known, but since the role of these elements are intended to capture small population-to-population scatter, one can assume that they are distance correlated, and introduce a single tunable parameter $\lambda$ such that the off-diagonal elements scale as $C_{ij} = \sigma_i \sigma_j \, e^{-d/\lambda}$ with $d = (i - j)$. As a result, the value of $\lambda$ is  correlated with the adopted binning scheme. In practice, we adopt 35 log-spaced bins across the pulsar parameter space, and take a value of $\lambda = 4$, which appears sufficient to wash out small scale variations without introducing a long-range effects (and yields stable results with respect to order one changes).

Binning and defining PDFs and CDFs in more than one dimension suffers from an intrinsic ambiguity, in that the rescaling of one dimension alters the characteristic value of the distribution function evaluated at the same point. There are often generalized procedures to deal with this issue, see e.g. ~\cite{peacock1983two,fasano1987multidimensional} for an alternative to the KS test, but one simple procedure to avoid this ambiguity is to project the data onto $N_{\rm slc}$ randomly selected one-dimensional slices, analyze each slice independently, and develop a statistic $t_{\rm slc}$ using e.g. the maximum of the set, \eg $t_{\rm slc} \equiv {\rm Max}\left[ \{ t_{i} \} \right]$. This the procedure we adopt here, where in practice we take $10^3$ random projections in the $(\log_{10}P, \log_{10} \dot{P})$ space. In order to ensure our distance correlator parameter $\lambda$ is not biased towards variations in $\log_{10}P$ or  $\log_{10} \dot{P}$ (which arises since we used a fixed number of bins applied over a different dynamical range in each of the dimensions), we standardize each direction of parameter space using the mean and standard deviation of the distributions inferred from the ATNF catalog. Once the best fit model is determined, we derive $p-$values which are normalized to the best-fit model by constructing a null distribution based on bootstrapped $\chi^2$ samples from the data set and the best-fit model set. This procedure is adopted as the high $\chi^2$ tail of the null distribution can be highly sensitivity to the samples used in the bootstrapping procedure, and one can incorrectly infer inflated $p-$values in scenarios which have significantly worse observed $\chi^2$ statistics.

Note that when comparing with the ATNF pulsar distribution, we remove pulsars for which $ 4 \times 10^{10} \, {\rm G} \, \leq B \leq B_q$, and $\dot{P} > 0$, \ie we attempt to remove pulsars which are not spinning down, magnetars, and millisecond pulsars, as these are not intended to be captured by the population synthesis models. Since we are comparing (at least in some cases\footnote{Even though some models do not impose a death line, we remove these pulsars in all population fits in order to ensure consistent between the analyses.}) with models that impose a sharp pulsar death line, we also remove pulsars in the ATNF catalog which fall below this threshold; in practice, this is only a small number of objects, and we have verified that including these pulsars has effectively no impact on any of the conclusions.

In practice, our analysis works directly on the PDFs, and thus don't carry any information on the total number of pulsars in the observable sample -- said another way, there is no information in this statistic as to whether the predicted pulsar birth rate is reasonably compatible with theoretical and observational expectations (if, say, 10 pulsars were produced in the mock observable pulsar sample, one would have to appeal for birth rates 2 orders of magnitude higher than what is expected theoretically). While the pulsar birth rate is not precisely known, it is expected to follow directly from the theoretically  inferred core collapse supernova rate of $\Gamma_{\rm cc} \sim 1.63 \pm 0.46 / {\rm century}$~\cite{rozwadowska2021rate}. In general, we perform two sets of analysis, one in which no constraint is imposed on the predicted birth rate, and one in which samples are rejected in the inferred birth rate (obtained by computing the population assuming 2 pulsars are born per century, and comparing the size of the predicted mock population with that of the ATNF catalog) exceeds 4 / century or is falls below 0.5 / century. Here, we have chosen sufficiently broad boundaries on the selection thresholds in order to avoid biasing any of our fits. 

\begin{figure*}
    \centering
    \includegraphics[width=\linewidth]{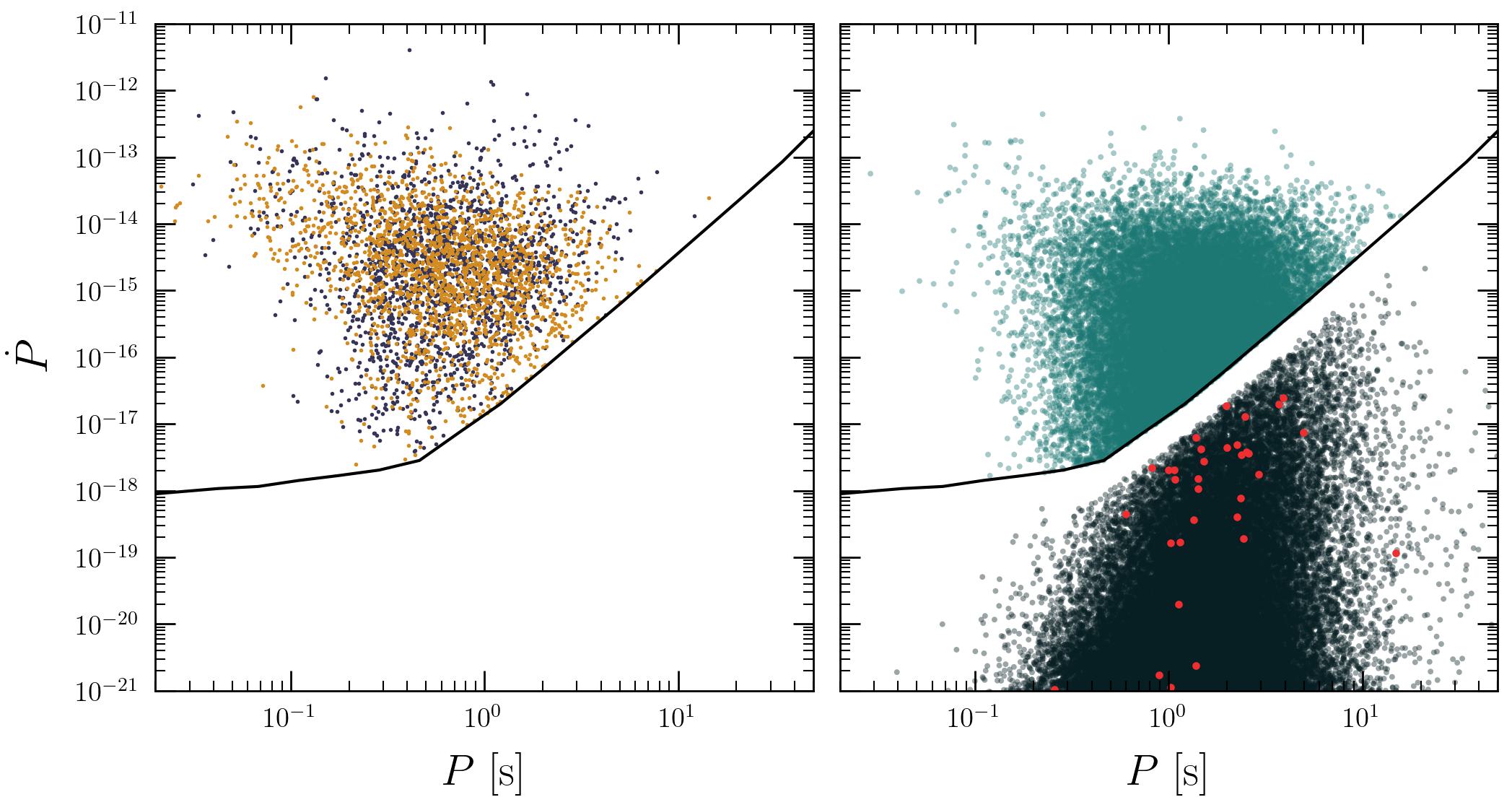}
    \caption{Comparison of observed and simulated pulsar populations. {\emph{Left:}} $P-\dot{P}$ diagram from sub-selection of the ATNF catalog (purple), compared with the observable pulsar population in Model 1. Black line shows the fiducial death line. {\emph{Right:}} Unobservable part of the mock pulsar population; green points correspond to pulsars which are designated as `active', but which are not observed due to selection cuts, black points are `dead' pulsars, and red points are active pulsars which would have been designated  as `observable' (passing all selection cuts) had one not enforced an abrupt notion of pulsar death. For the sake of simplicity, we 
    only show $5 \times 10^5$ samples of the unobservable pulsar population. Note that the gap appearing between the black points and the death line arises because we impose an abrupt shift of the spin down coefficient $\kappa_1$ at pulsar death (which arises from the fact that the spin down rate of a plasma filled magnetosphere is larger than that predicted in vacuum) -- this assumption is conservative in the sense that it suppresses the number of potentially observable dead pulsars.     }\label{fig:ppdot_m1}
\end{figure*}

We provide the full list of models  in Table~\ref{tab:popsyn}, highlighting the adopted distributions, evolutionary models, selection cuts, and inferred p-values. Here, we have normalized the p-values to the best-fit model, since an absolute p-value is inherently subject to the unknown level uncertainties, the binning scheme, the treatment of the covariance, etc. (note, however, that in our fiducial analysis, our best-fit models yield $p-$values $\sim 0.20$). A number of important trends which can be inferred directly from this table include: 
\begin{enumerate}
    \item Models which do not impose a pulsar death line (namely, Models 7, 10-17) do not fit the data well, leading to a significant degradation in the inferred $p-$value. This is not surprising, as all of the models predict $\mathcal{O}(100)$ pulsars sitting well below the death line, a feature which is clearly incompatible in a `by-eye' analysis in Figs.~\ref{fig:ppdot_m1}-\ref{fig:ppdot_m17}.
    \item The initial distribution of rotational periods of pulsars at birth does not strongly favor either the log-normal or the Weibull distribution.
    \item There is not a strong preference for either magnetic field decay model, however it is clear that magnetic fields decaying on $\mathcal{O}(Gyr)$ timescales appear to be strongly disfavored.
    \item In the case of the best-fitting model, imposing a constraint on the birth rate does not have an appreciable effect (we note that the inferred birth rate tends to larger than one might naively expect, and is near the upper threshold of the analysis, although given the uncertainties we hesitate to make any statement beyond this).
\end{enumerate}

The quality of the fit in all cases is highly sensitive to the fit parameters, and thus the posteriors of these parameters are expected to be highly peaked around their best-fit values (although there do exist a few degeneracies in the parameter space which allow for parameters to be shifted in particular directions without spoiling the fit).

Note that we have fit the data using well-known, although slightly outdated, pulsar surveys. The Meerkat telescope has come online in recent years and represents a significant improvement in the state-of-the-art radio instrument, with pulsar physics being one of the fundamental science goals\footnote{The Square Kilometer Array, coming online in the near future, also intends to perform a dedicated pulsar search to improve the understanding of the broader pulsar population~\cite{SKAOPulsarScienceWorkingGroup:2025zsy}. }. The galactic plane survey will lead to significant improvement in the flux density sensitivity, and has already released early science results, including the discover of new pulsars~\cite{padmanabh2023mpifr}. We expect this survey to further strengthen the case for a need for a pulsar death line. In order to illustrate this point, we re-simulate the population corresponding to the Model 10, which corresponds to one of the best-fitting models within the class of models which do not impose a death line. Here, we impose a fixed flux sensitivity threshold of $20 \mu$Jy in order to simulate the sensitivity of Meerkat. The observed population in shown in Fig.~\ref{fig:meerkat}, and is found to produce a noticeably worse fit to the observed pulsar population than what is listed in Table~\ref{tab:popsyn} (specifically, we find the $p-$value degrades by a factor of a few), providing further strength for our conclusions that a death line is a crucial aspect of the pulsar population. 

For the sake of completeness, we include figures showing the observed, and unobserved, pulsar populations for each of the models studied in this work. In the left panels of Figs.~\ref{fig:ppdot_m1}-\ref{fig:ppdot_m17}, we compare the observed pulsar population of the ATNF catalog (purple) used in the fits to the mock population of observed pulsars for each Model in Table~\ref{tab:popsyn} (gold). The fiducial death line (black) is shown for reference. In the right panel of theses figures we instead display a random subset of the unobserved part of the simulated population (the fully population is sufficiently large to be problematic for both storage and figure rendering, and thus this is just down for practical reasons), where the color coding corresponds to: active pulsars (green), dead pulsars that {\bf{ would}} have been unobservable even if abrupt pulsar death had not been imposed (black), and dead pulsars which are only unobservable because we have assumed they crossed the pulsar death line. Note that for models which do not impose pulsar death, all points (by definition) are colored green. Instead, for models imposing pulsar death, the red points highlight potential pulsars that could be revived by the presence of axion hair. 

Given that the precise quality of each fit is difficult to judge by eye, we also include in Figs.~\ref{fig:hist_m1}-\ref{fig:hist_m17} two one-dimensional histograms comparing the predicted best fit pulsar population for each model with the observed ATNF population. One-sigma statistical error bars are included on the ATNF population for reference. Again, one can notice a general trend among those populations which do not adopt a death line to over predict the abundance of low $\dot{P}$ and high $P$ pulsars.

\begin{figure*}
    \centering
    \includegraphics[width=\linewidth]{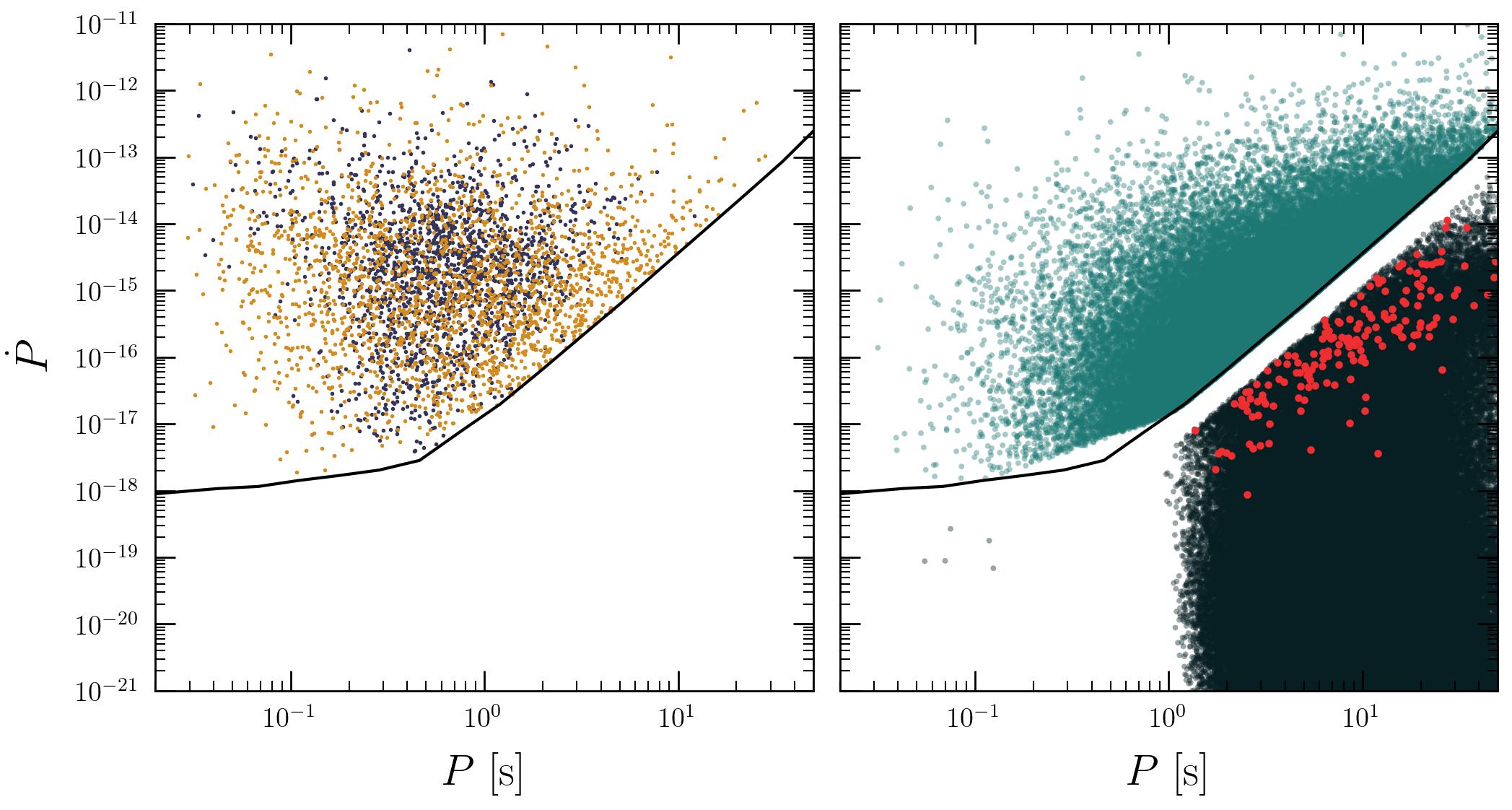}
    \caption{Same as Fig.~\ref{fig:ppdot_m1}, but for Model 2.}
\end{figure*}

\begin{figure*}
    \centering
    \includegraphics[width=\linewidth]{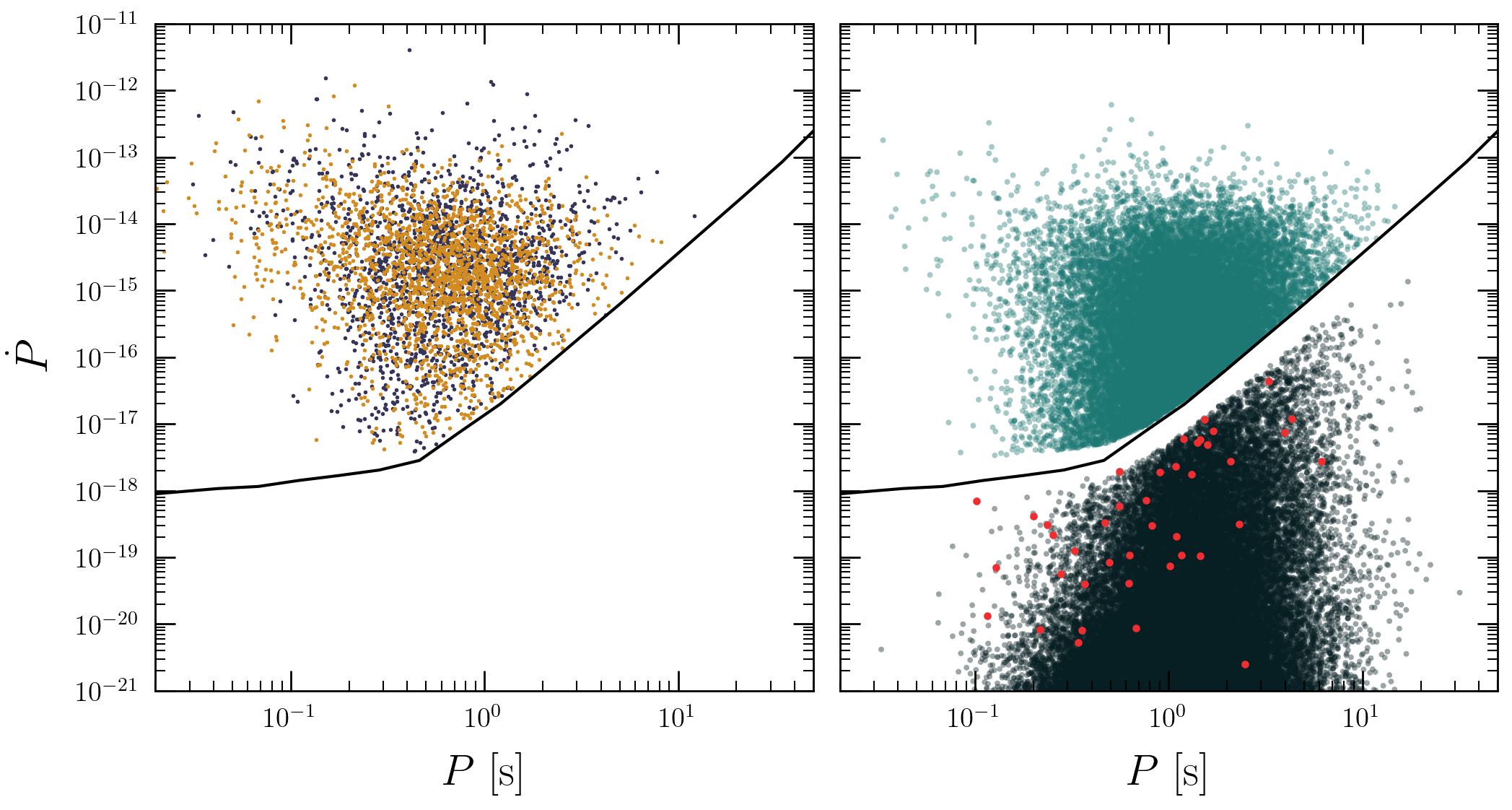}
    \caption{Same as Fig.~\ref{fig:ppdot_m1}, but for Model 3.}
\end{figure*}

\begin{figure*}
    \centering
    \includegraphics[width=\linewidth]{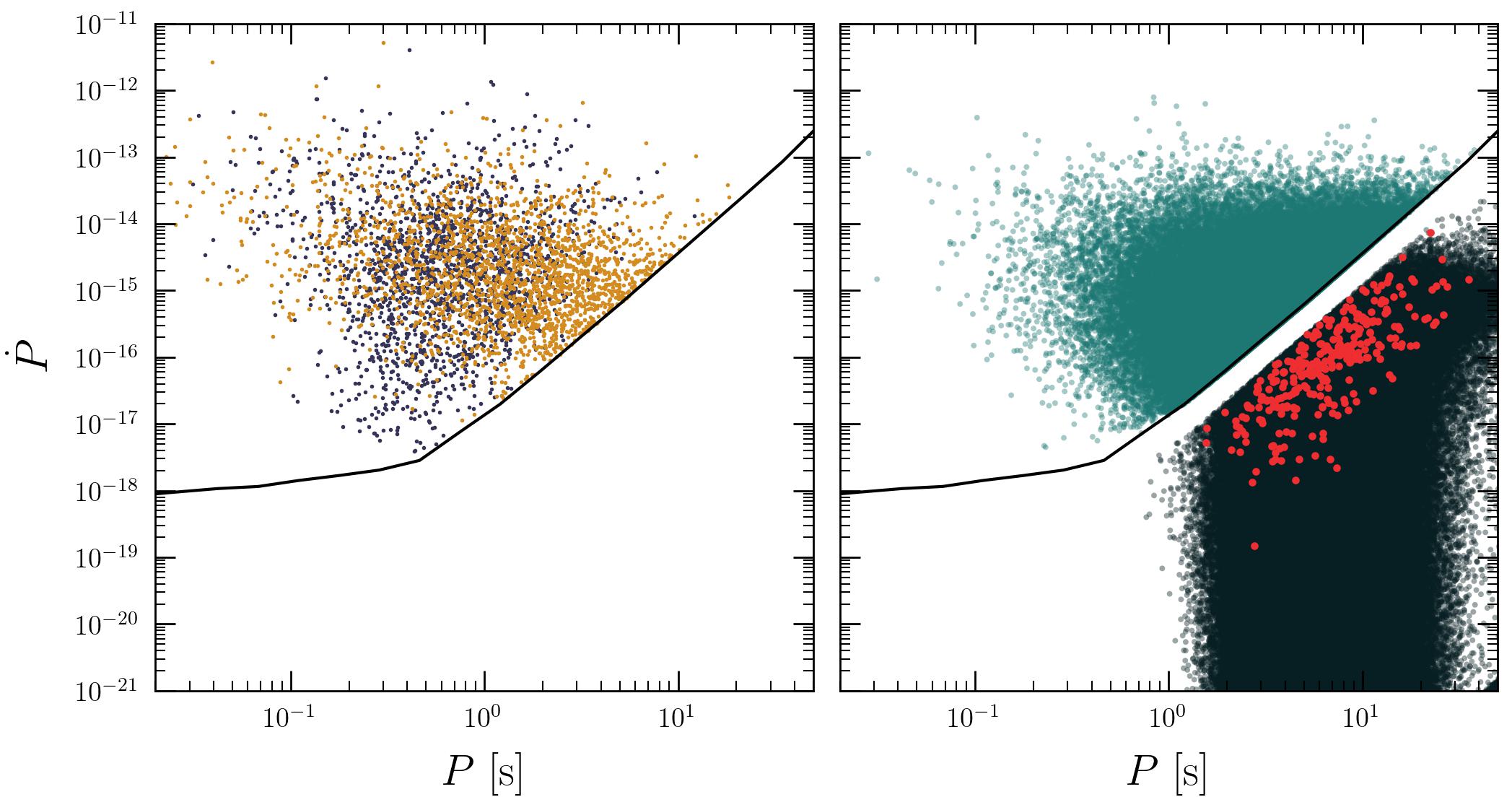}
    \caption{Same as Fig.~\ref{fig:ppdot_m1}, but for Model 4.}
\end{figure*}

\begin{figure*}
    \centering
    \includegraphics[width=\linewidth]{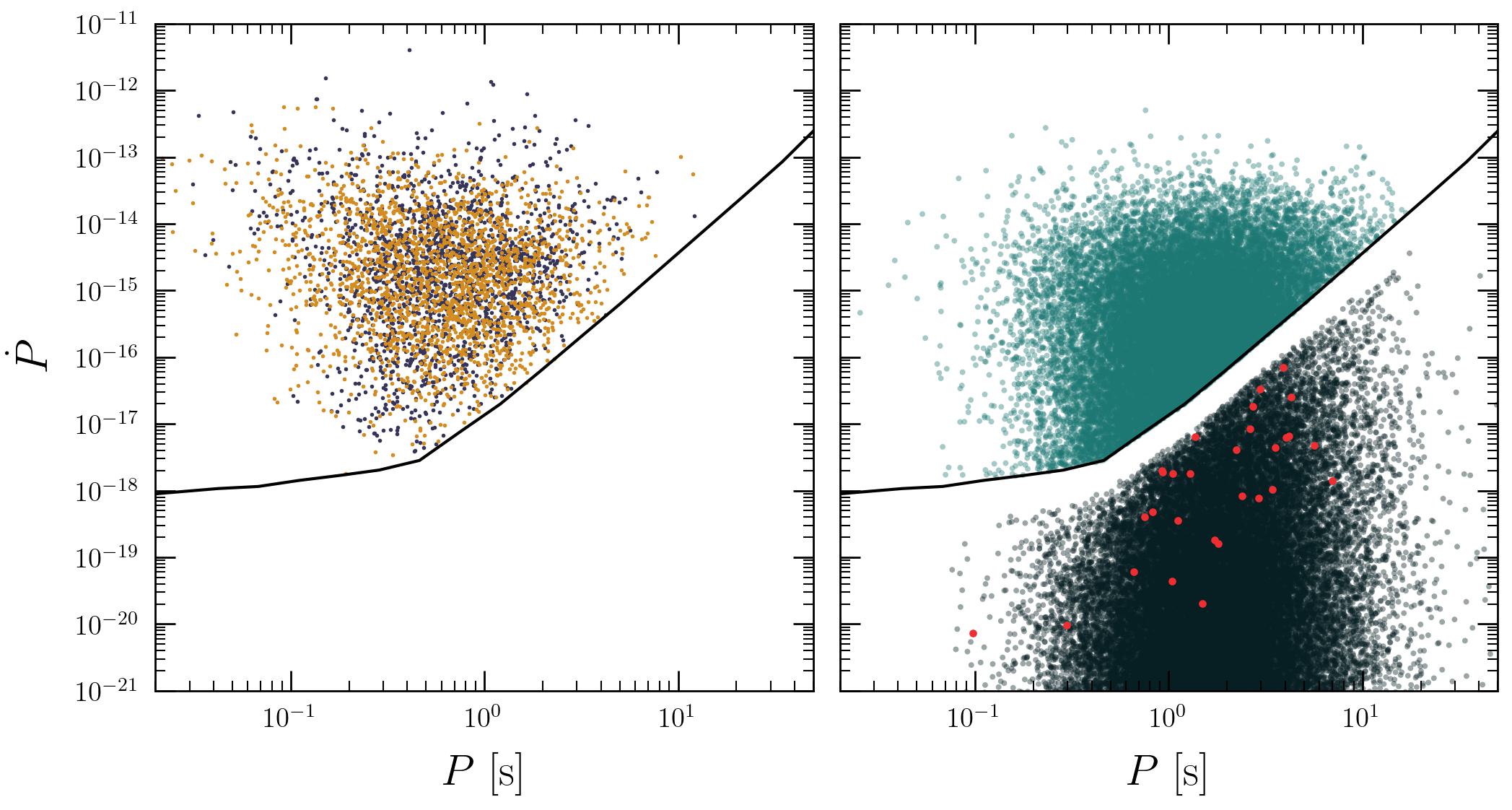}
    \caption{Same as Fig.~\ref{fig:ppdot_m1}, but for Model 5.}
\end{figure*}

\begin{figure*}
    \centering
    \includegraphics[width=\linewidth]{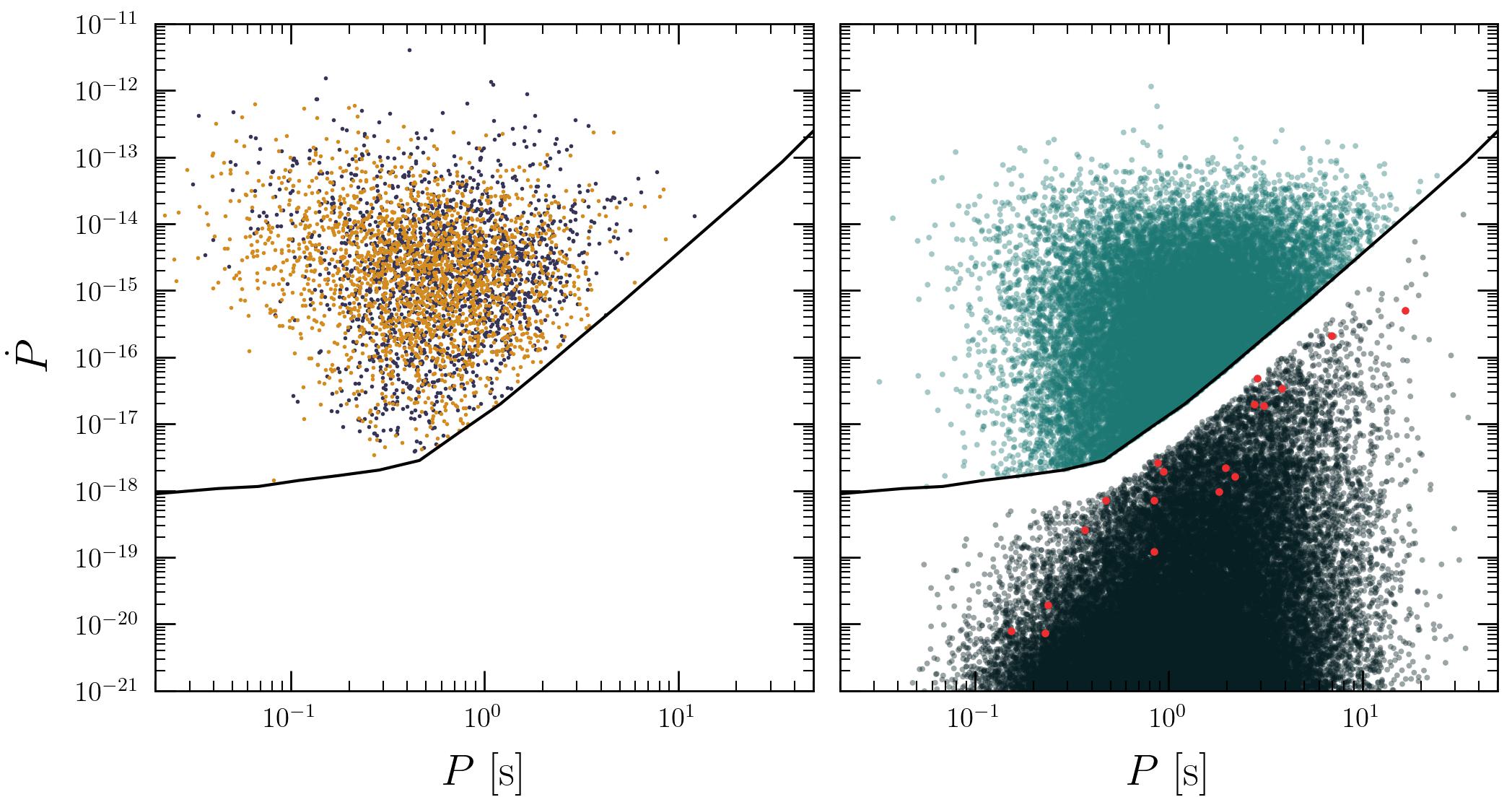}
    \caption{Same as Fig.~\ref{fig:ppdot_m1}, but for Model 6.}
\end{figure*}

\begin{figure*}
    \centering
    \includegraphics[width=\linewidth]{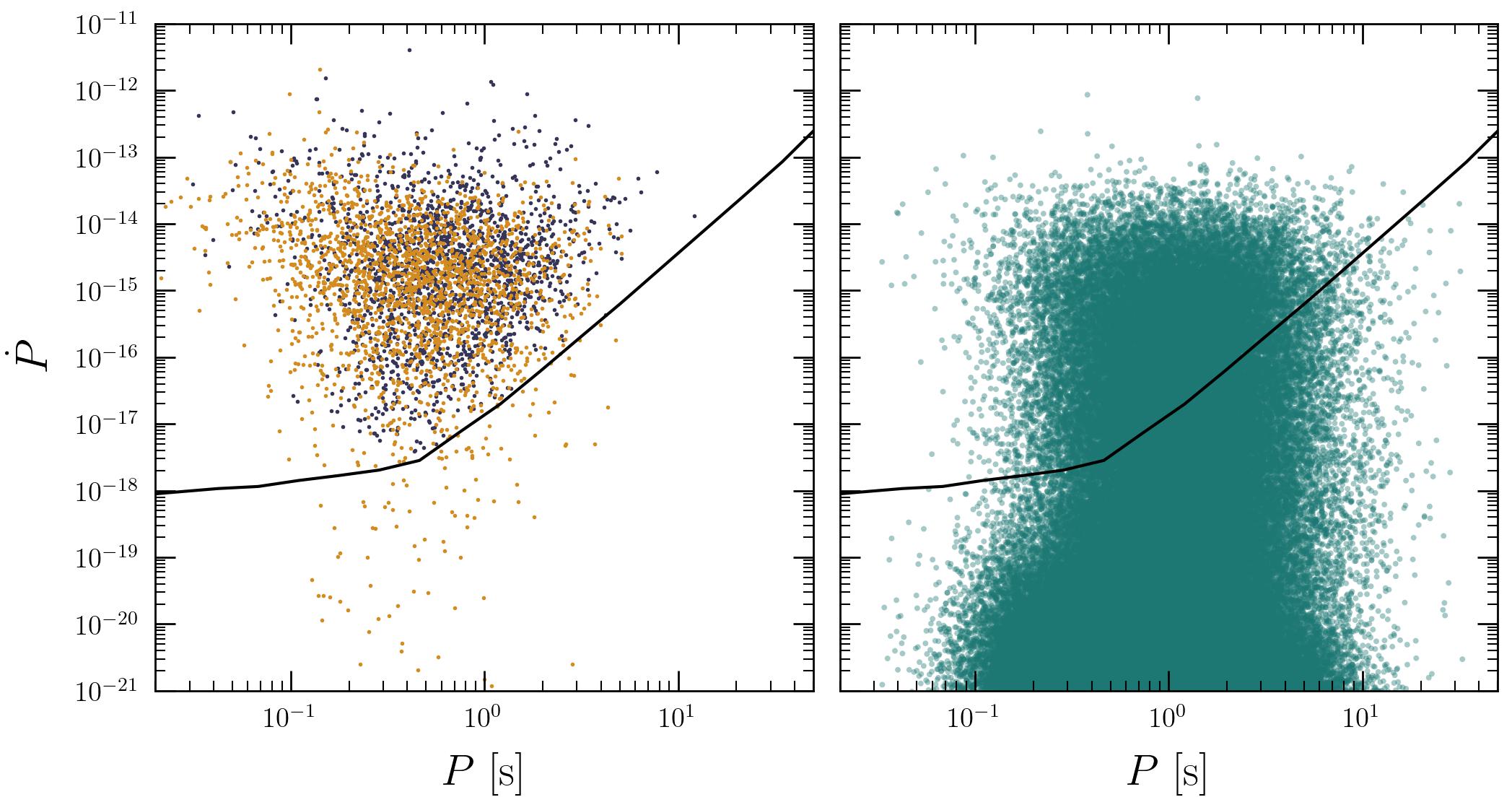}
    \caption{Same as Fig.~\ref{fig:ppdot_m1}, but for Model 7.}
    \label{fig:ppdot_m7}
\end{figure*}

\begin{figure*}
    \centering
    \includegraphics[width=\linewidth]{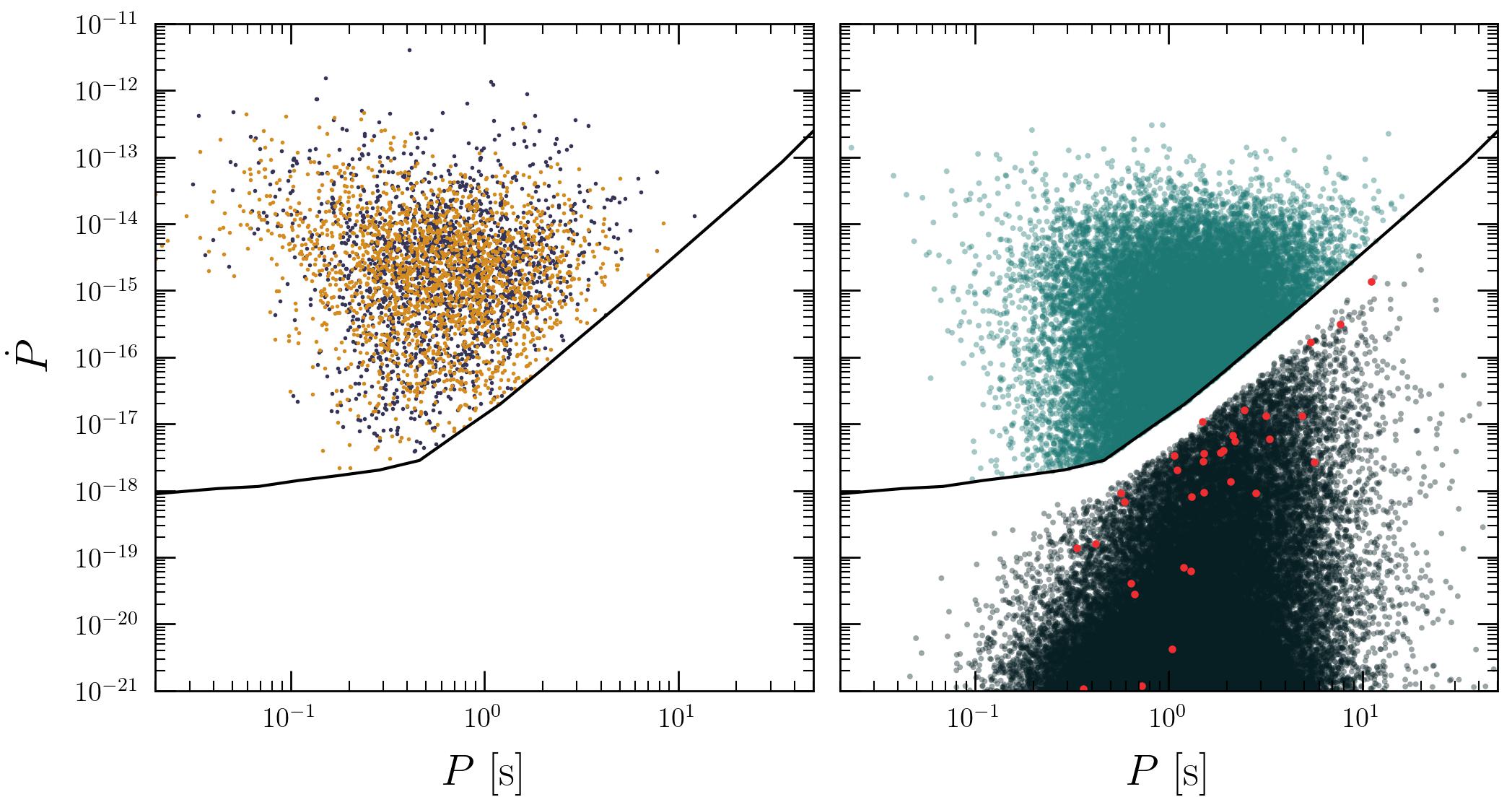}
    \caption{Same as Fig.~\ref{fig:ppdot_m1}, but for Model 8.}
\end{figure*}
\begin{figure*}
    \centering
    \includegraphics[width=\linewidth]{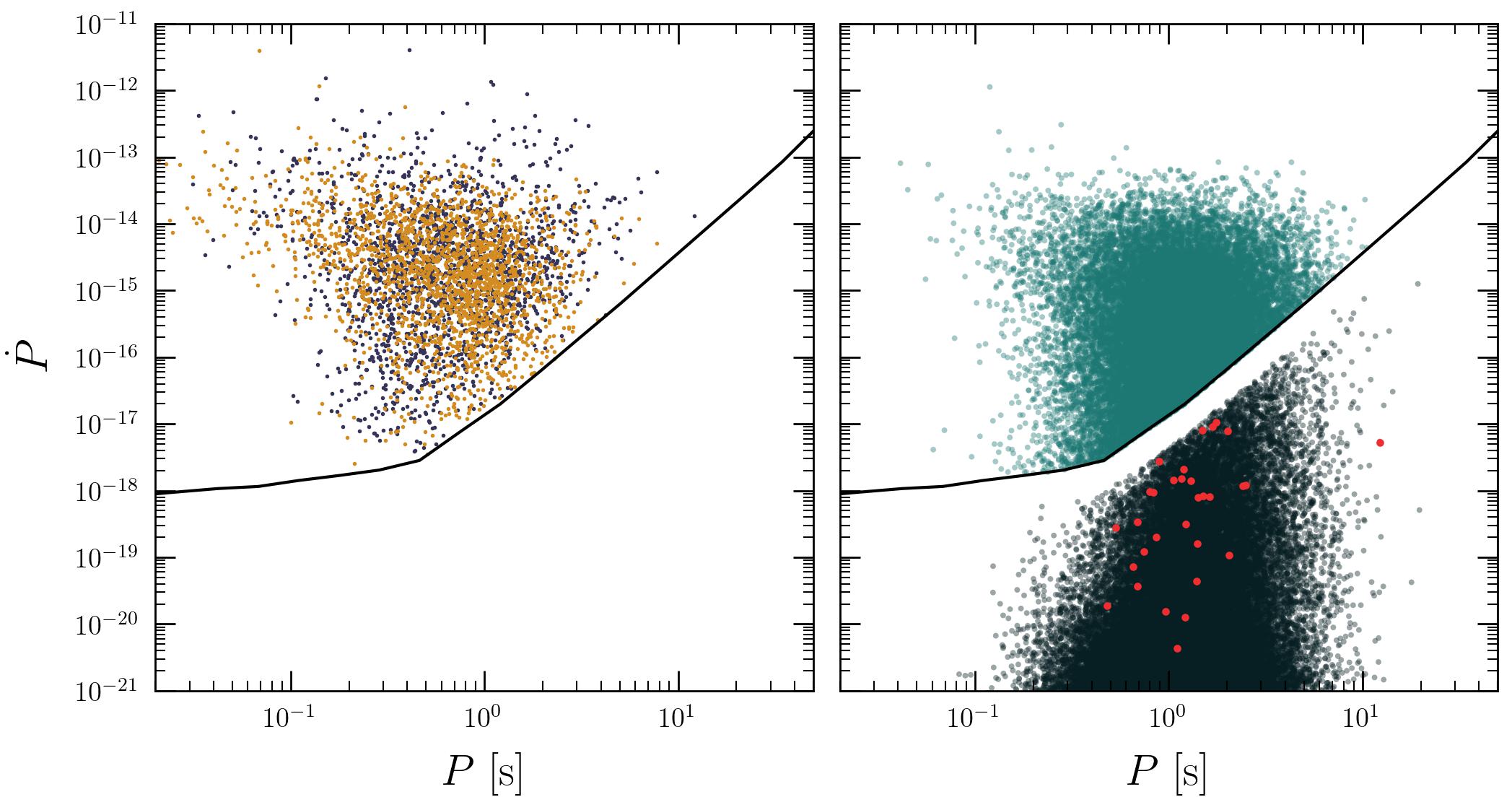}
    \caption{Same as Fig.~\ref{fig:ppdot_m1}, but for Model 9.}
    \label{fig:ppdot_m9}
\end{figure*}

\begin{figure*}
    \centering
    \includegraphics[width=\linewidth]{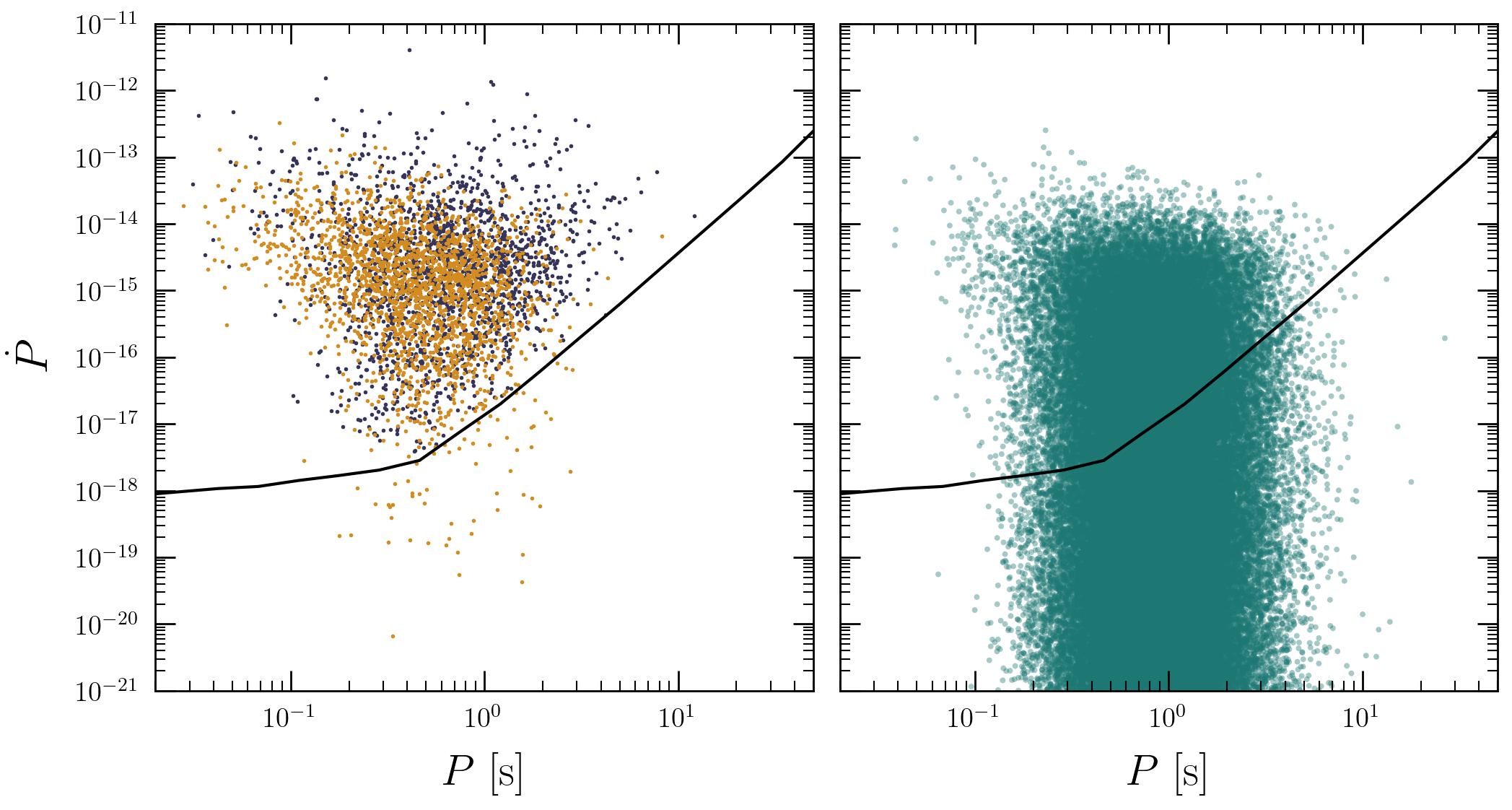}
    \caption{Same as Fig.~\ref{fig:ppdot_m1}, but for Model 10.}\label{fig:ppdot_m10}
\end{figure*}

\begin{figure*}
    \centering
    \includegraphics[width=\linewidth]{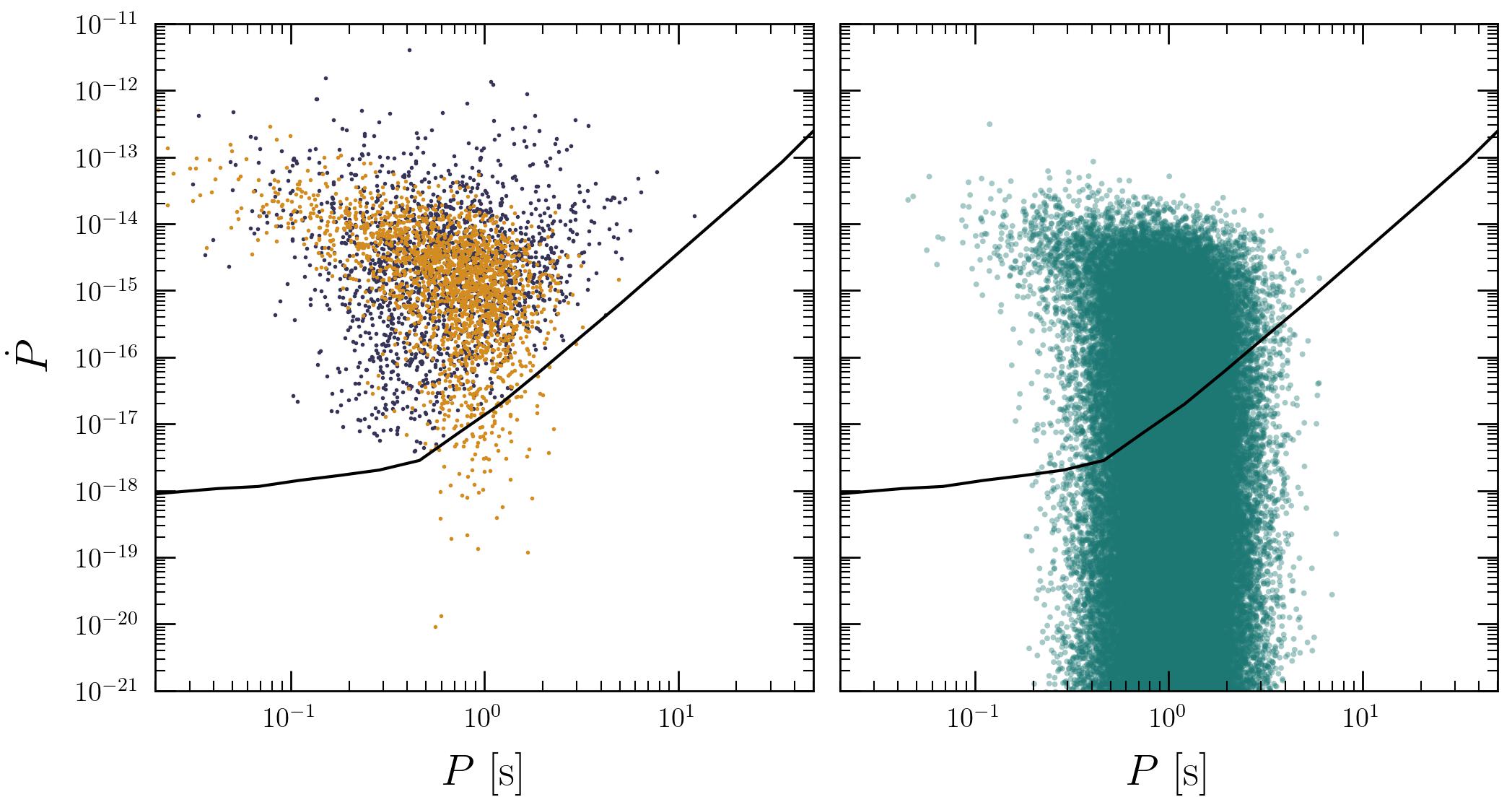}
    \caption{Same as Fig.~\ref{fig:ppdot_m1}, but for Model 11.}
    \label{fig:ppdot_m11}
\end{figure*}

\begin{figure*}
    \centering
    \includegraphics[width=\linewidth]{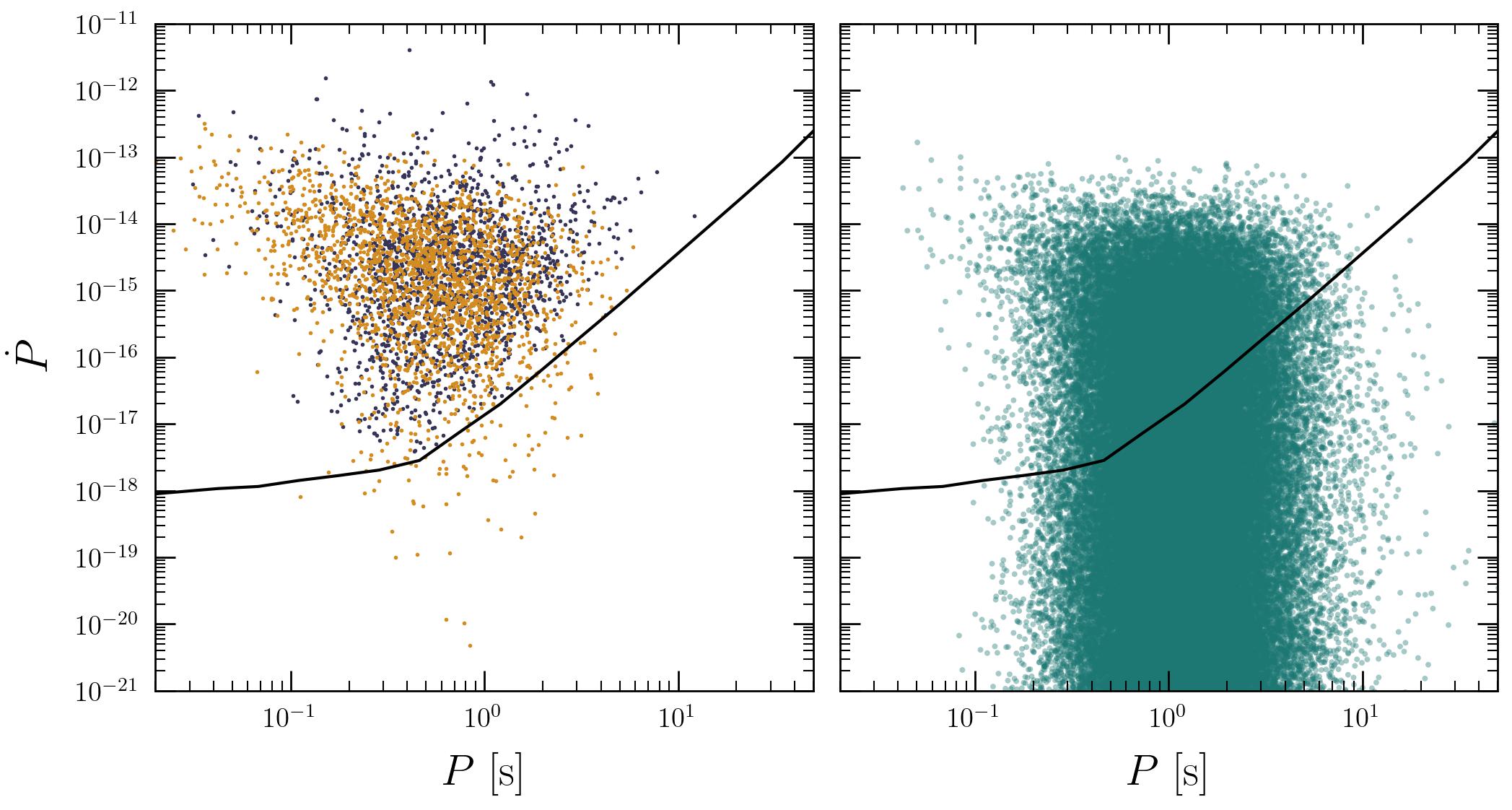}
    \caption{Same as Fig.~\ref{fig:ppdot_m1}, but for Model 12.}
    \label{fig:ppdot_m12}
\end{figure*}

\begin{figure*}
    \centering
    \includegraphics[width=\linewidth]{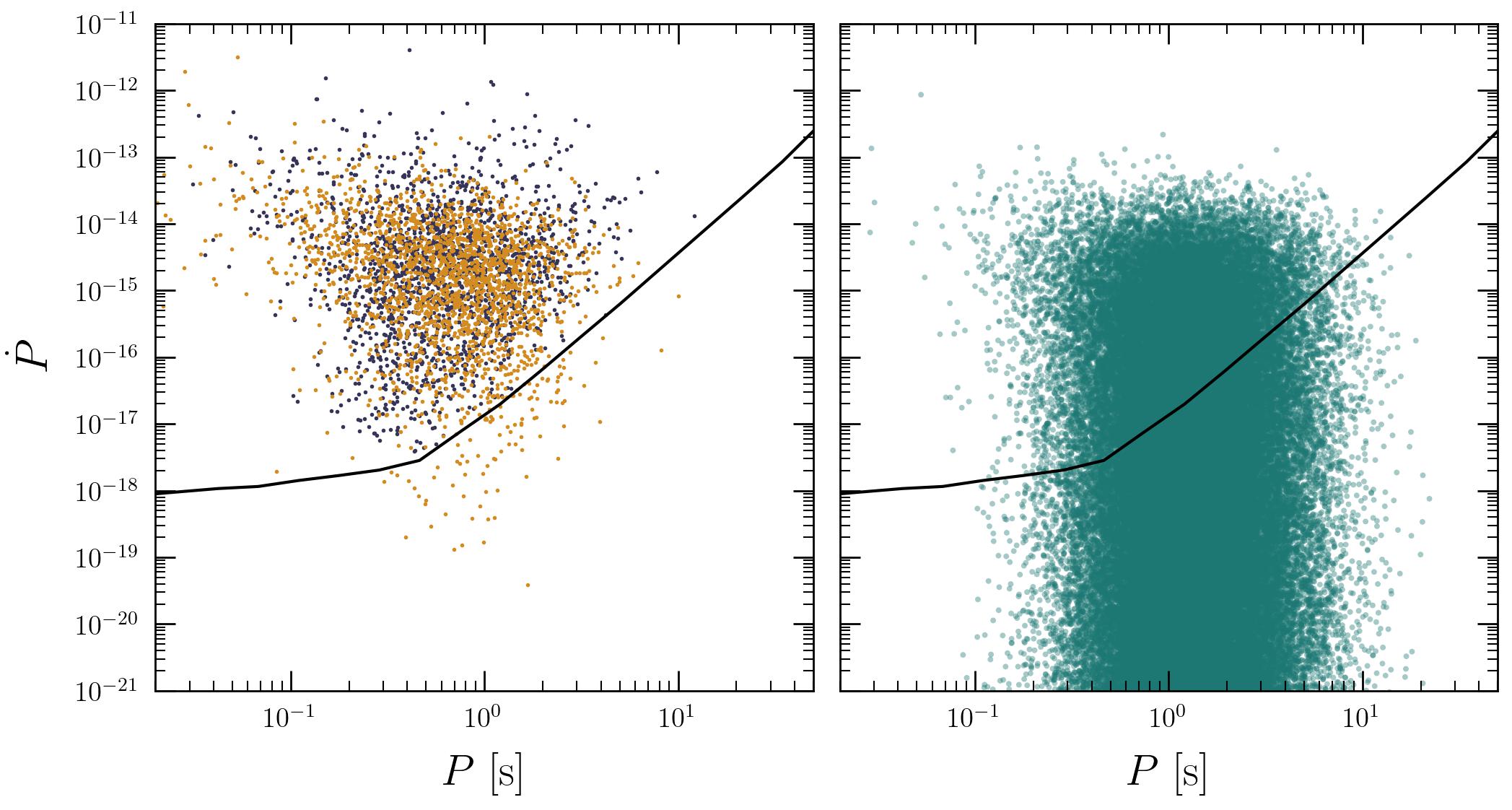}
    \caption{Same as Fig.~\ref{fig:ppdot_m1}, but for Model 13.}
    \label{fig:ppdot_m13}
\end{figure*}

\begin{figure*}
    \centering
    \includegraphics[width=\linewidth]{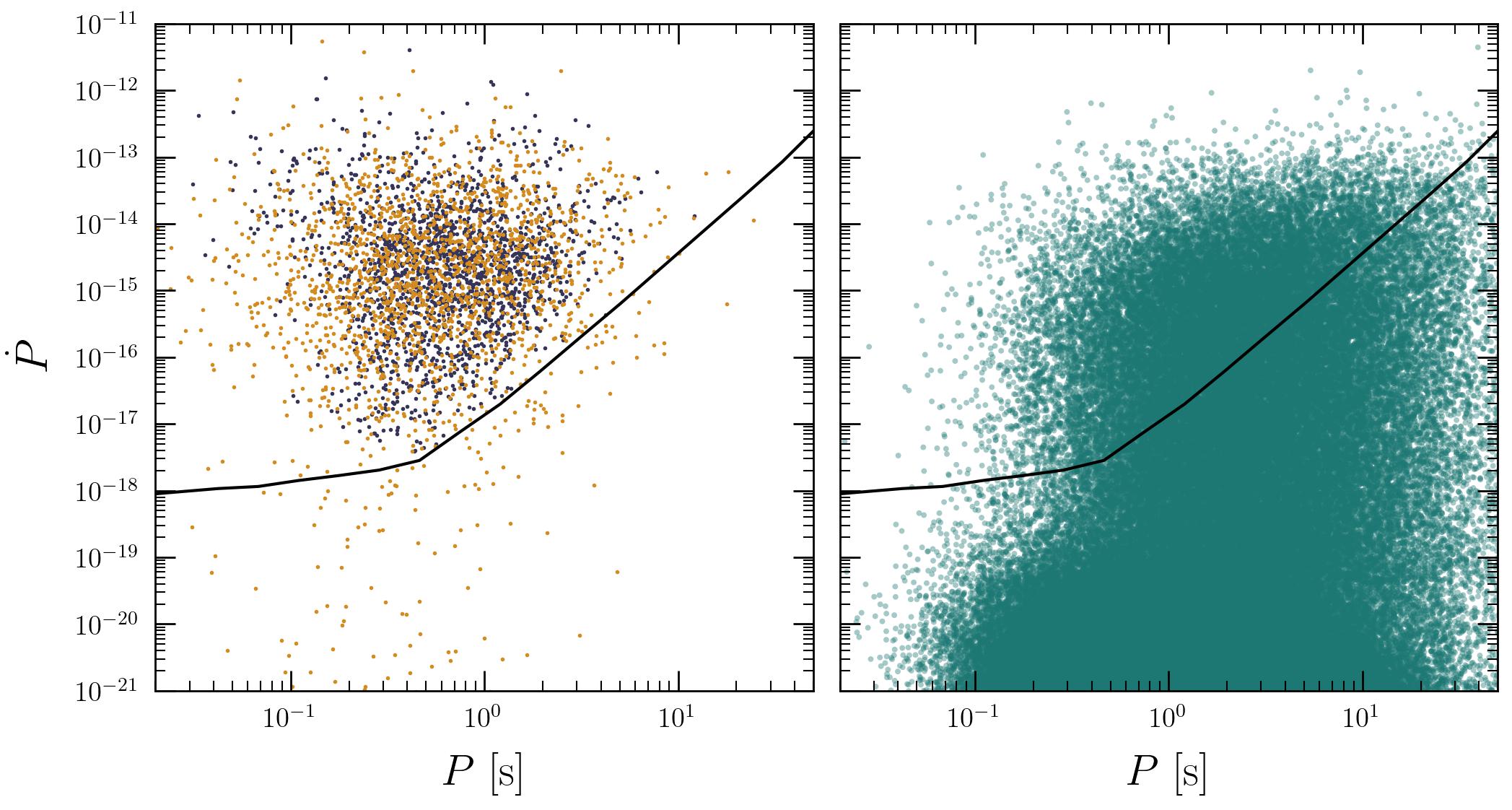}
    \caption{Same as Fig.~\ref{fig:ppdot_m1}, but for Model 14.}
    \label{fig:ppdot_m14}
\end{figure*}

\begin{figure*}
    \centering
    \includegraphics[width=\linewidth]{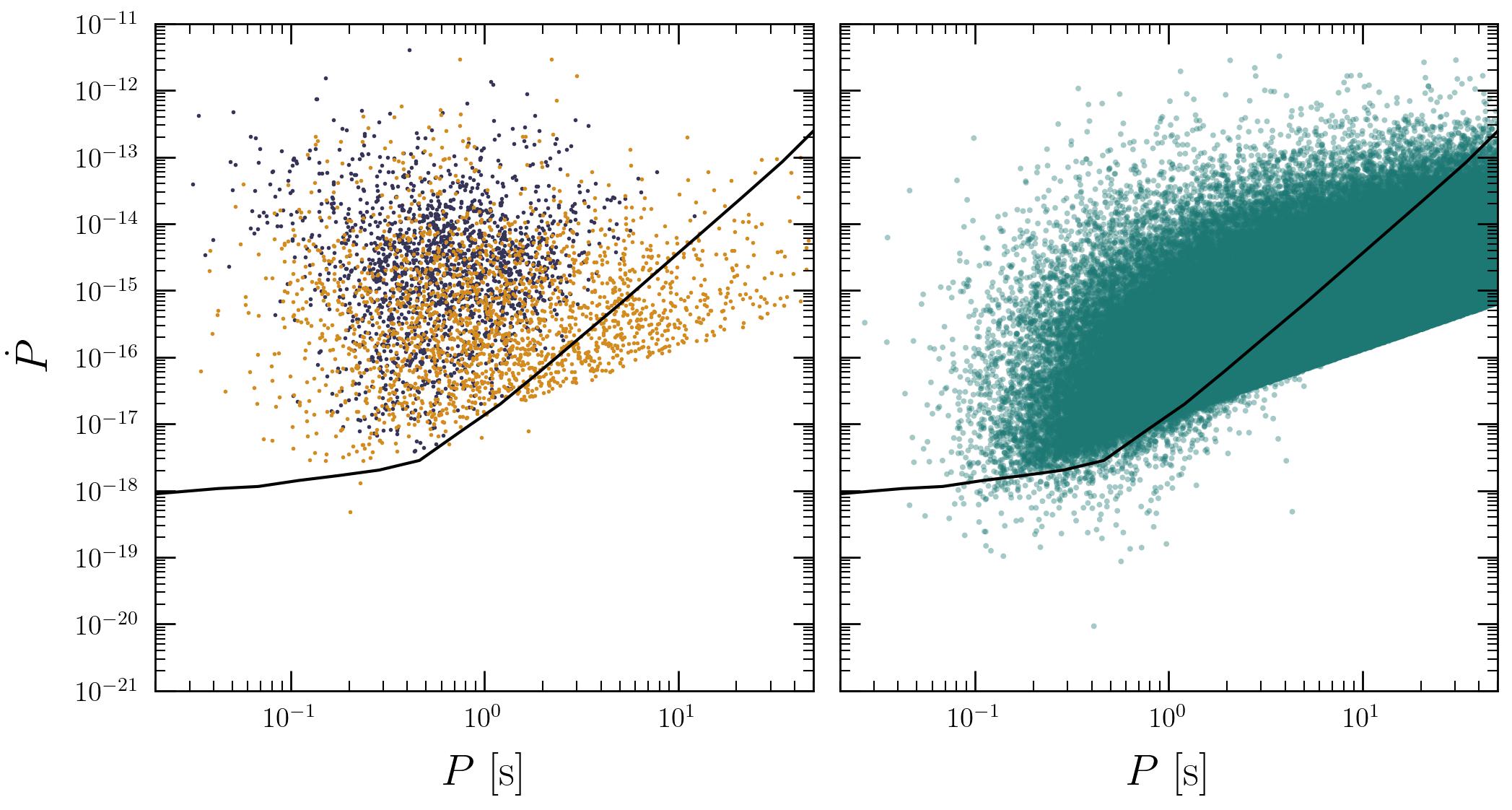}
    \caption{Same as Fig.~\ref{fig:ppdot_m1}, but for Model 15.}
    \label{fig:ppdot_m15}
\end{figure*}

\begin{figure*}
    \centering
    \includegraphics[width=\linewidth]{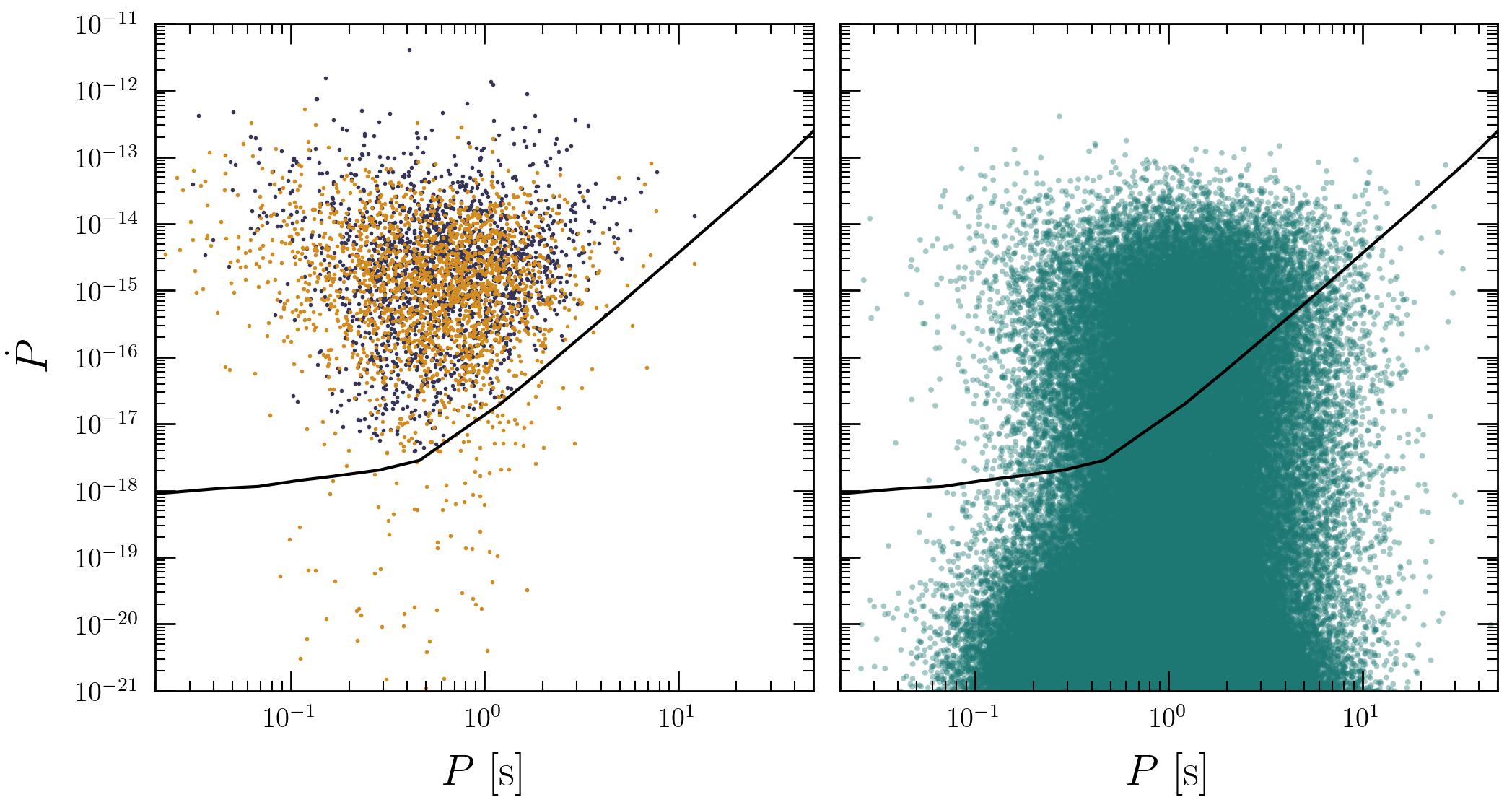}
    \caption{Same as Fig.~\ref{fig:ppdot_m1}, but for Model 16.}
    \label{fig:ppdot_m16}
\end{figure*}

\begin{figure*}
    \centering
    \includegraphics[width=\linewidth]{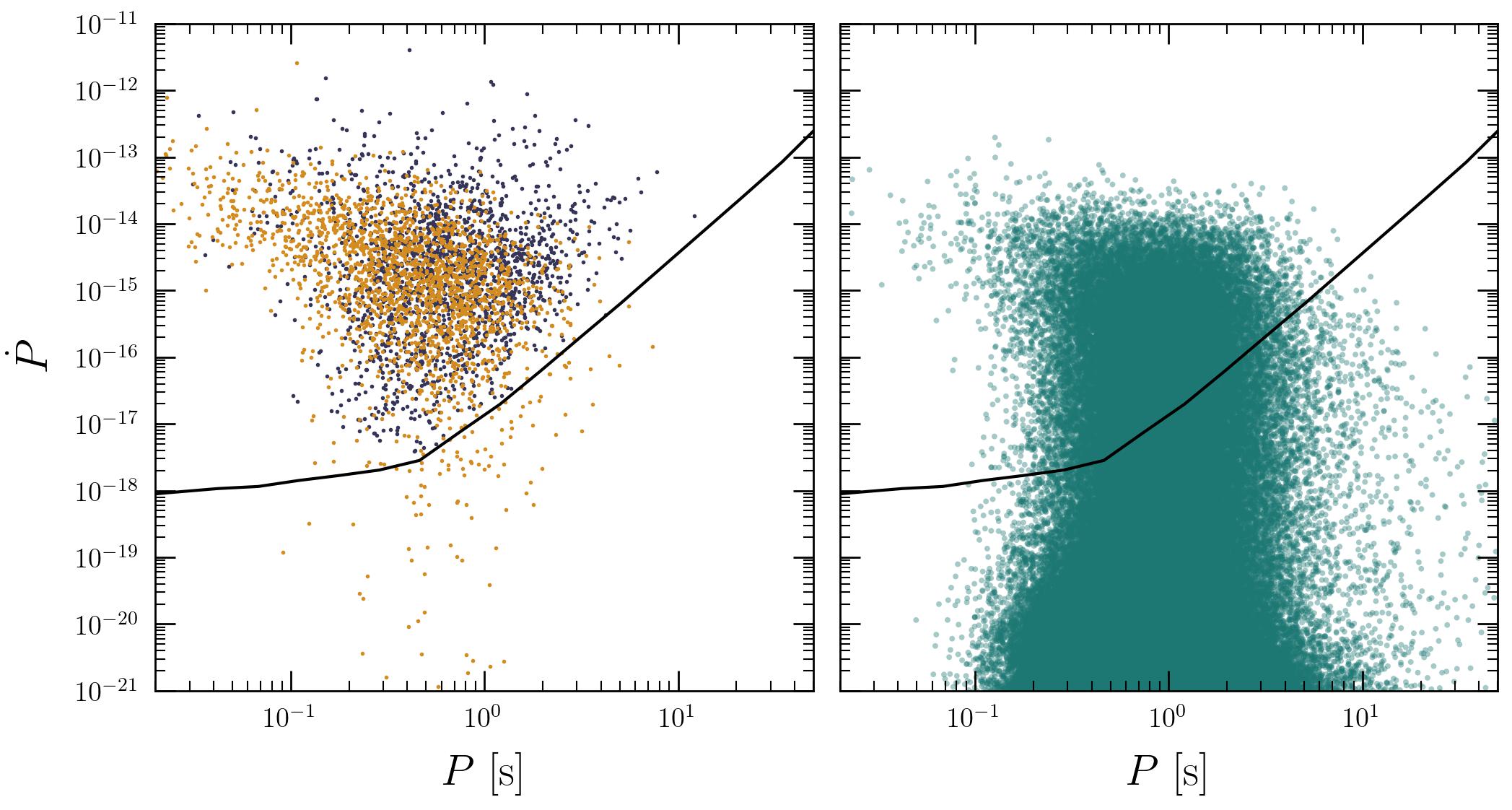}
    \caption{Same as Fig.~\ref{fig:ppdot_m1}, but for Model 17.}
    \label{fig:ppdot_m17}
\end{figure*}

\begin{figure*}
    \centering
    \includegraphics[width=.8\textwidth]{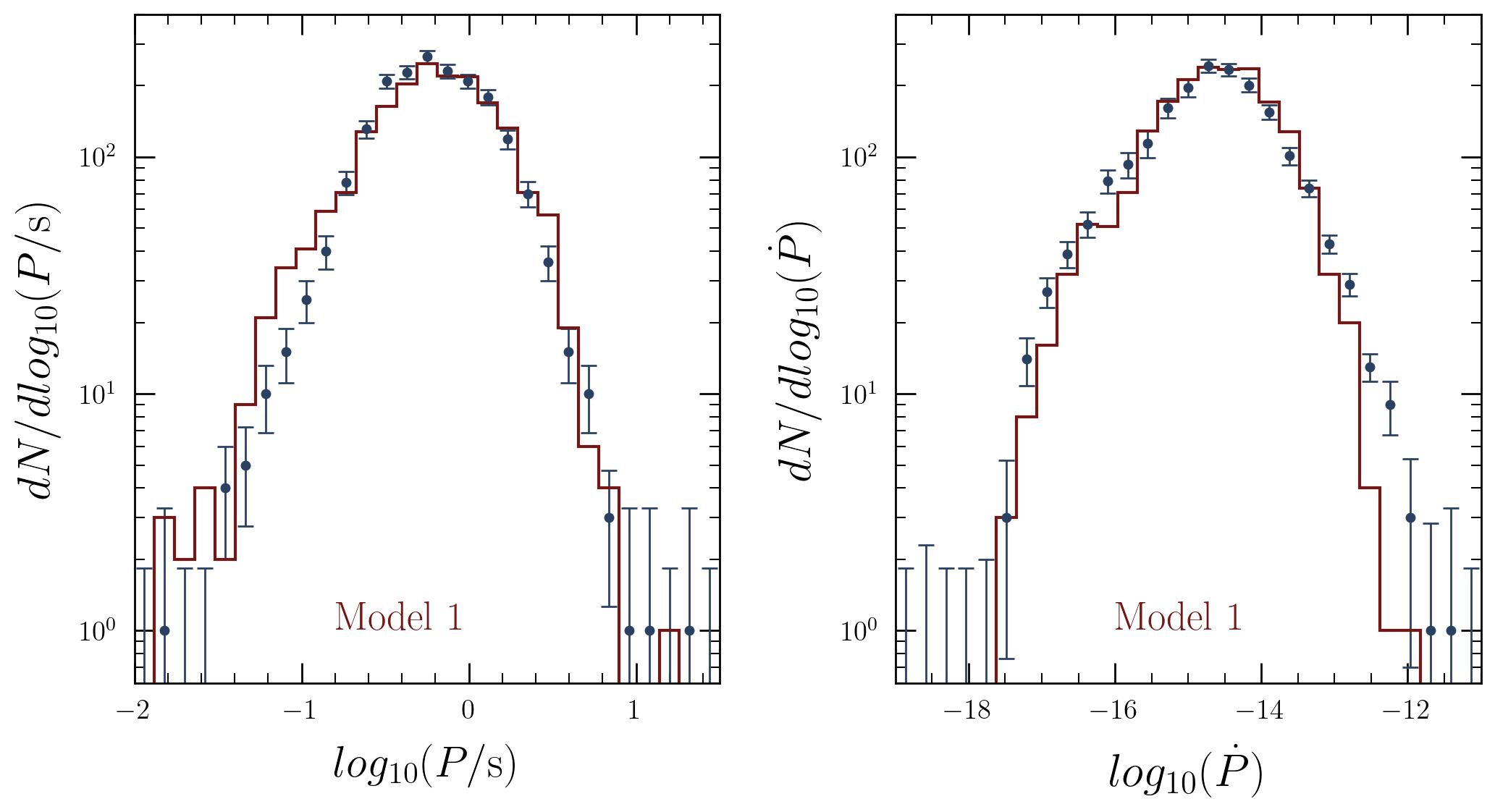}
    \caption{One-dimensional histograms in $P$ and $\dot{P}$ comparing the observed pulsar population in the ATNF catalogue (blue points), as well as the distributions obtained from the best-fit mock population using Model 1. }
    \label{fig:hist_m1}
\end{figure*}

\begin{figure*}
    \centering
    \includegraphics[width=.8\textwidth]{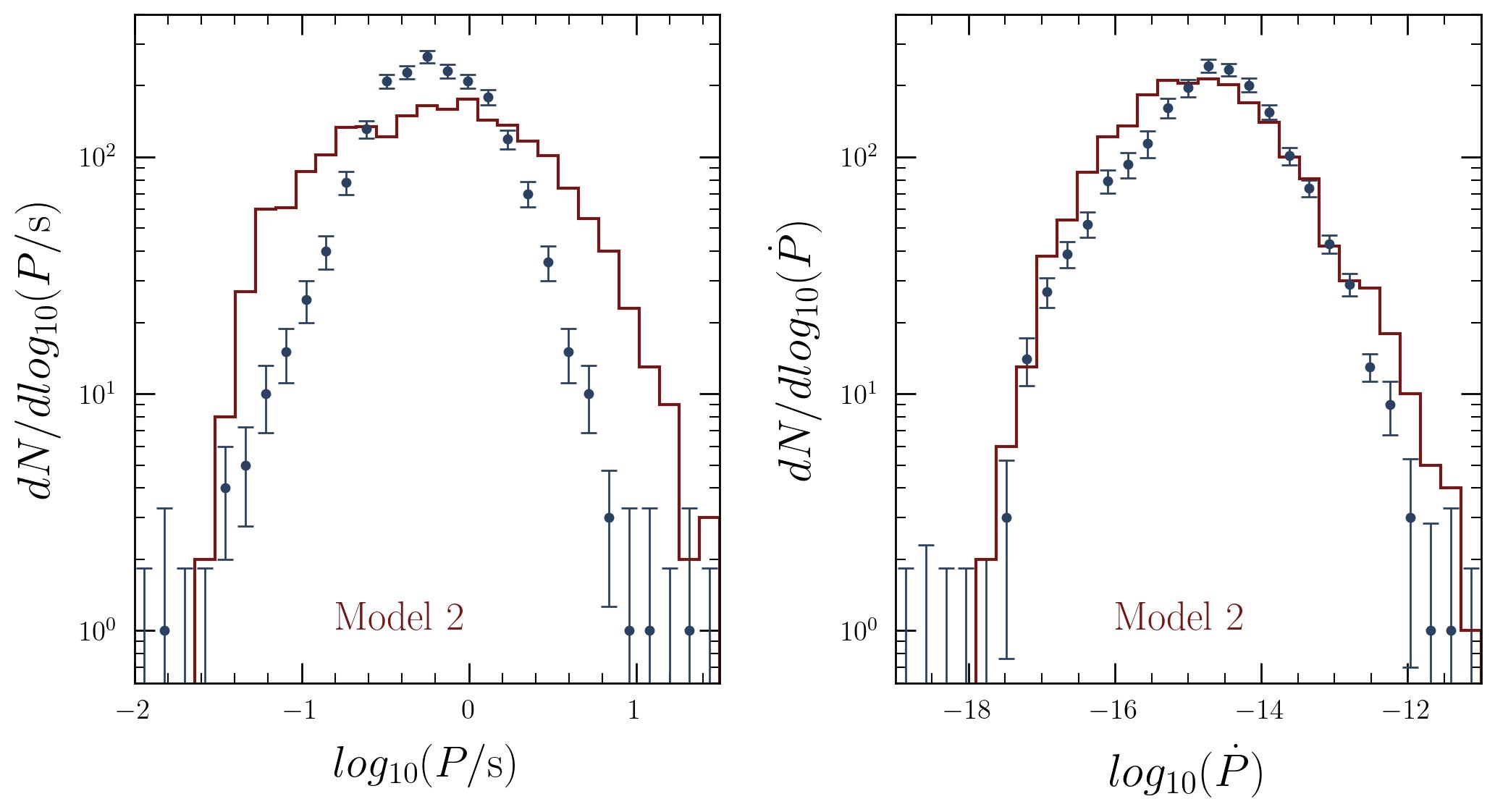}
    \caption{Same as Fig.~\ref{fig:hist_m1} but for Model 2. }
\end{figure*}

\begin{figure*}
    \centering
    \includegraphics[width=.8\textwidth]{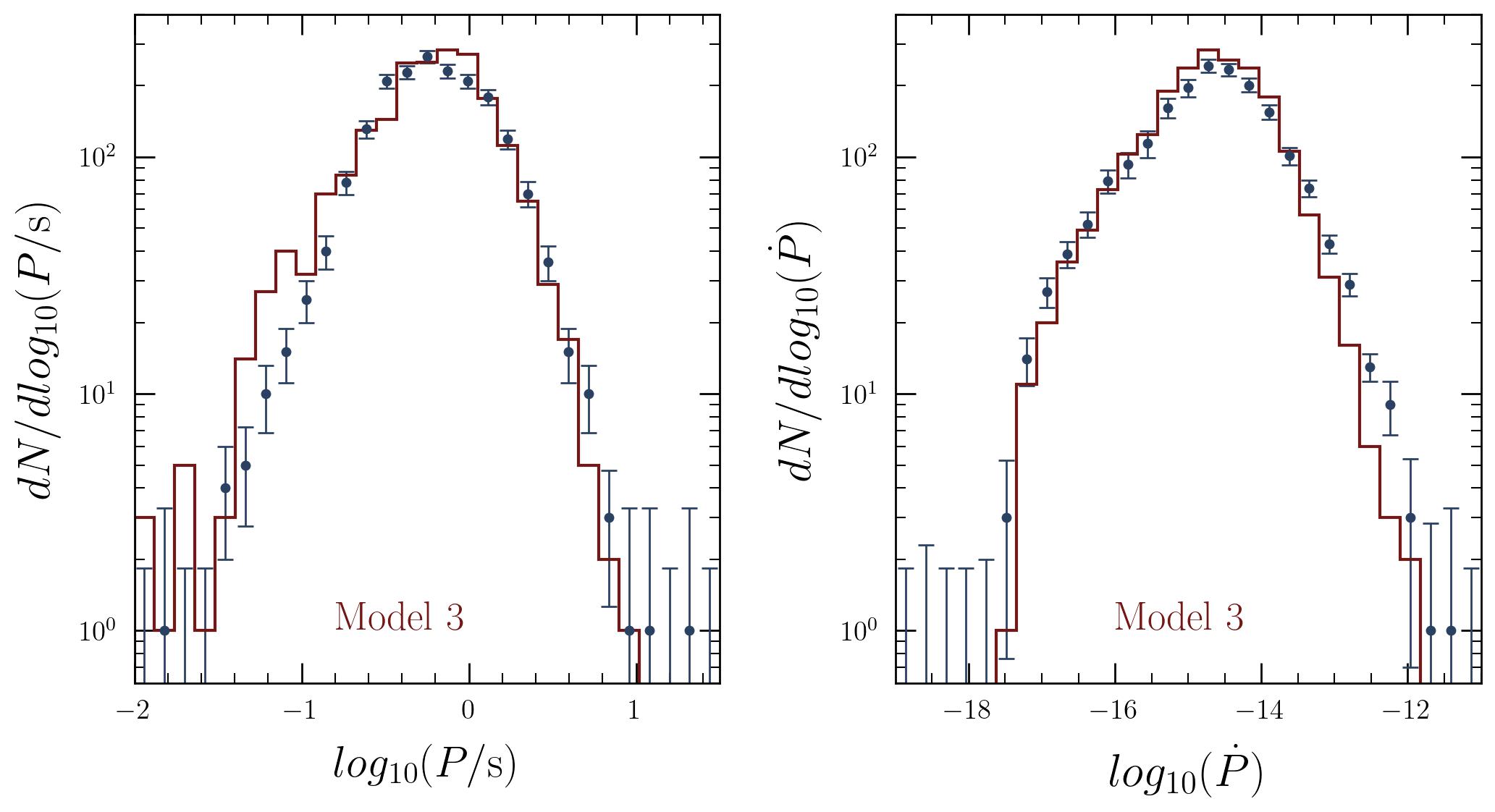}
    \caption{Same as Fig.~\ref{fig:hist_m1} but for Model 3. }
\end{figure*}

\begin{figure*}
    \centering
    \includegraphics[width=.8\textwidth]{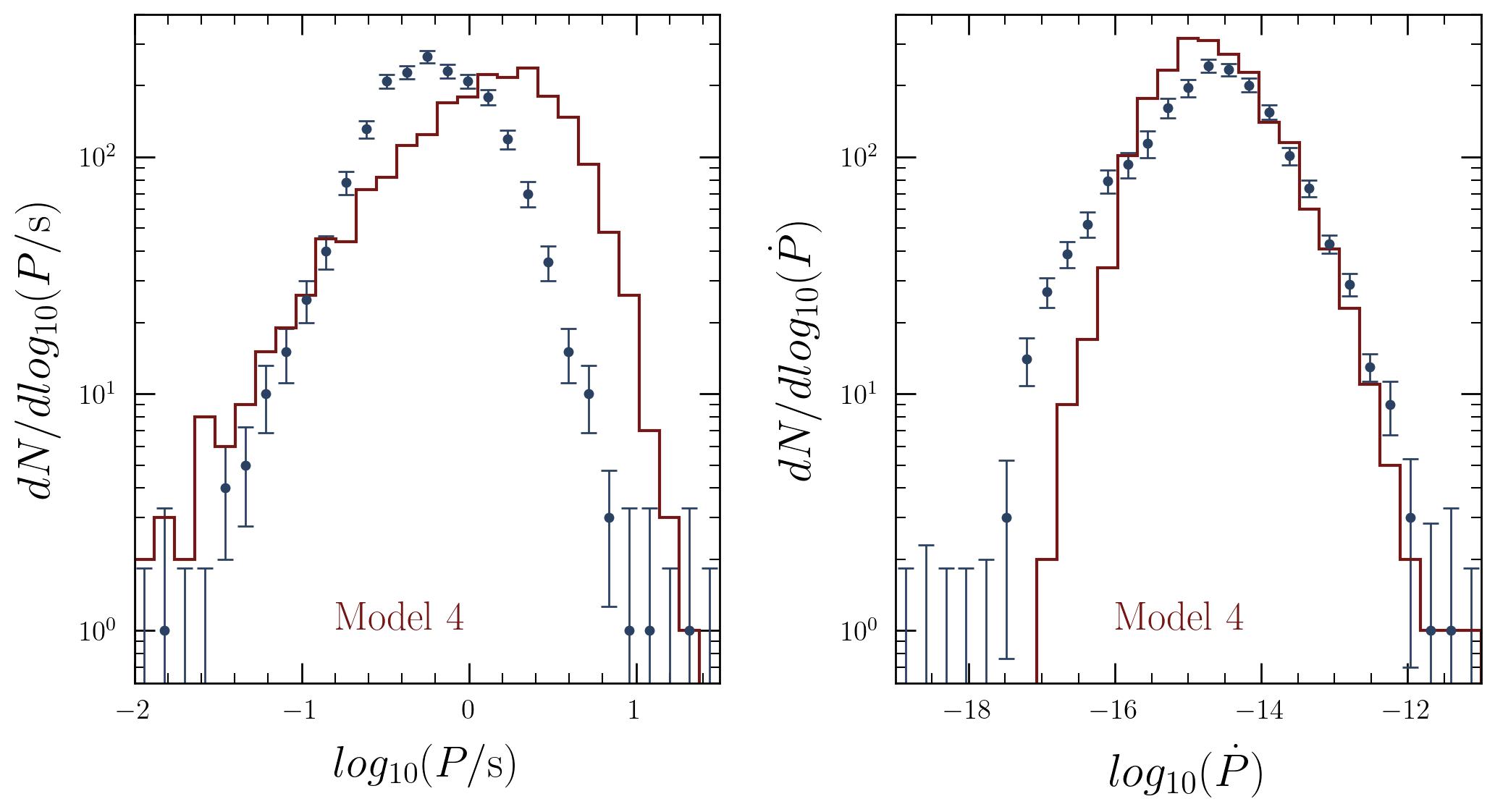}
    \caption{Same as Fig.~\ref{fig:hist_m1} but for Model 4. }
\end{figure*}

\begin{figure*}
    \centering
    \includegraphics[width=.8\textwidth]{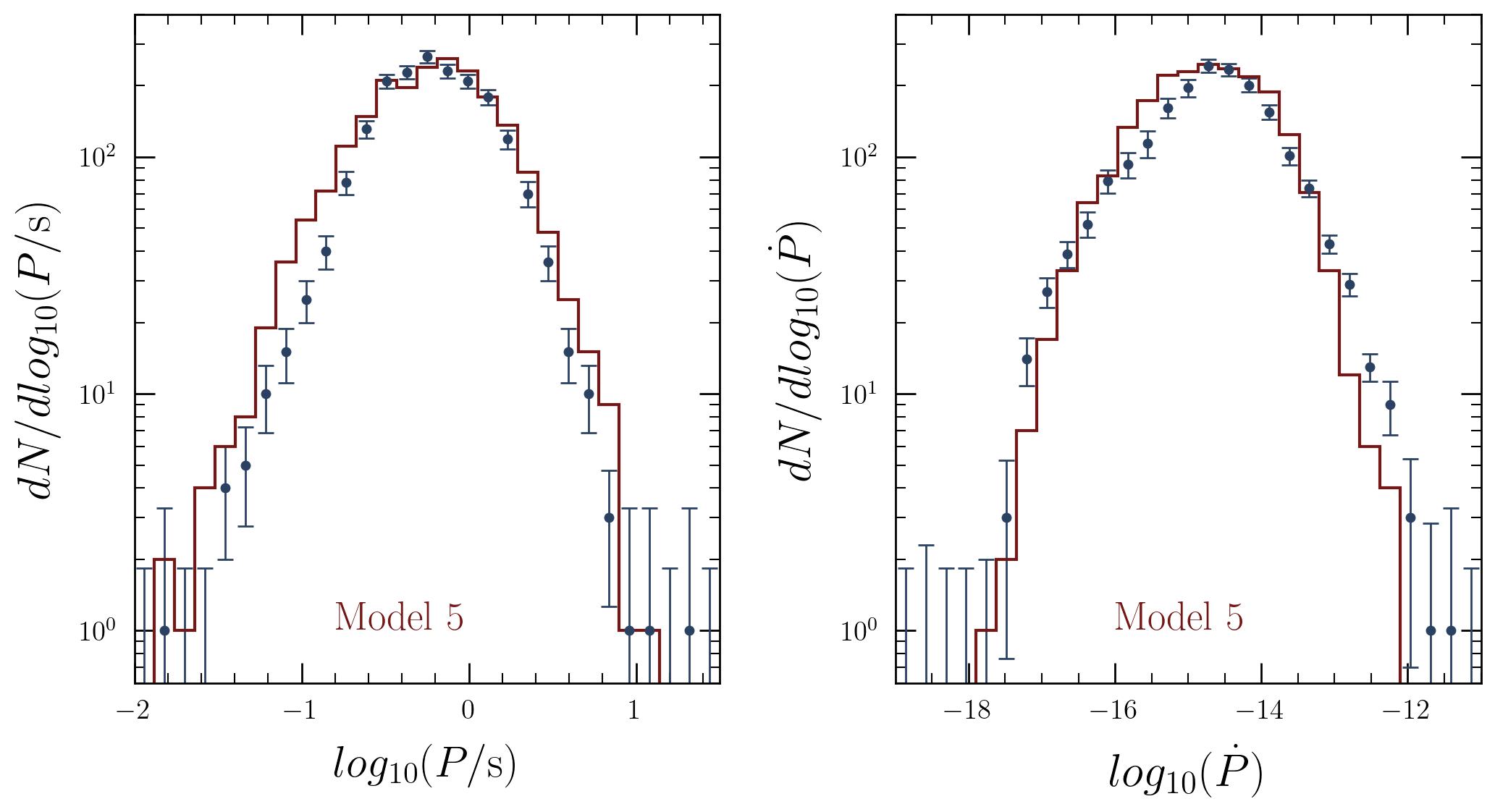}
    \caption{Same as Fig.~\ref{fig:hist_m1} but for Model 5. }
\end{figure*}

\begin{figure*}
    \centering
    \includegraphics[width=.8\textwidth]{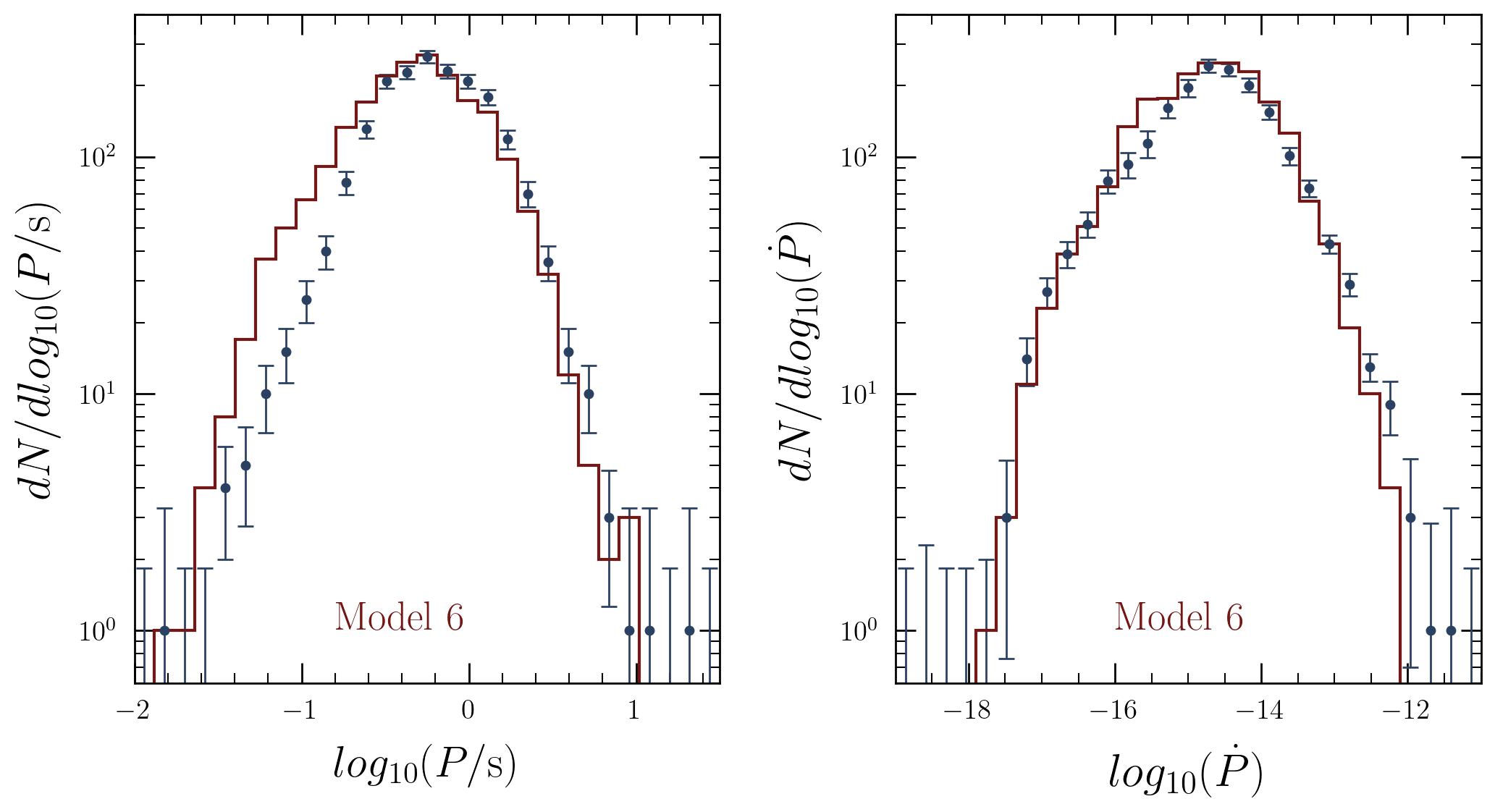}
    \caption{Same as Fig.~\ref{fig:hist_m1} but for Model 6. }
\end{figure*}

\begin{figure*}
    \centering
    \includegraphics[width=.8\textwidth]{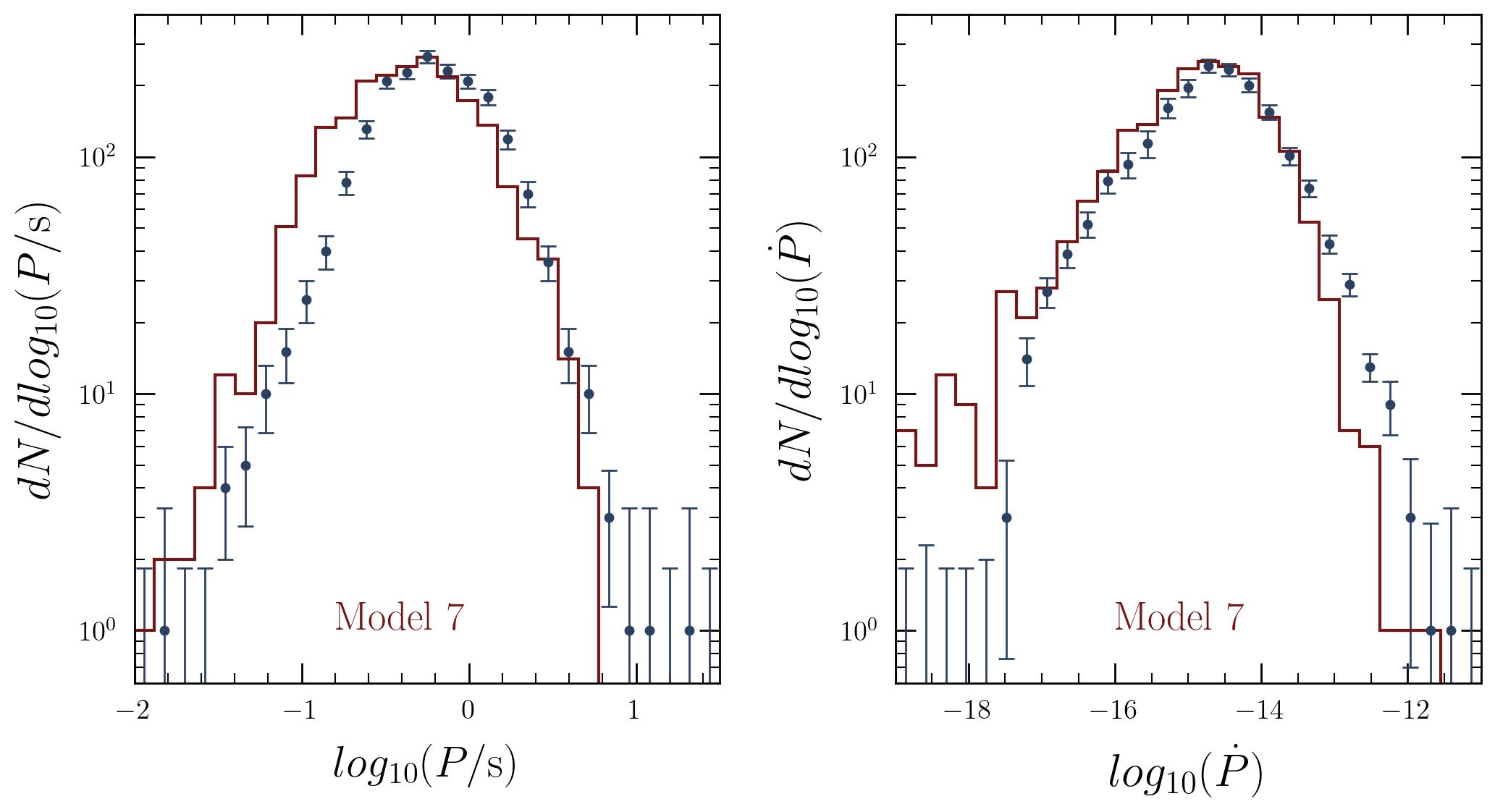}
    \caption{Same as Fig.~\ref{fig:hist_m1} but for Model 7. }
\end{figure*}

\begin{figure*}
    \centering
    \includegraphics[width=.8\textwidth]{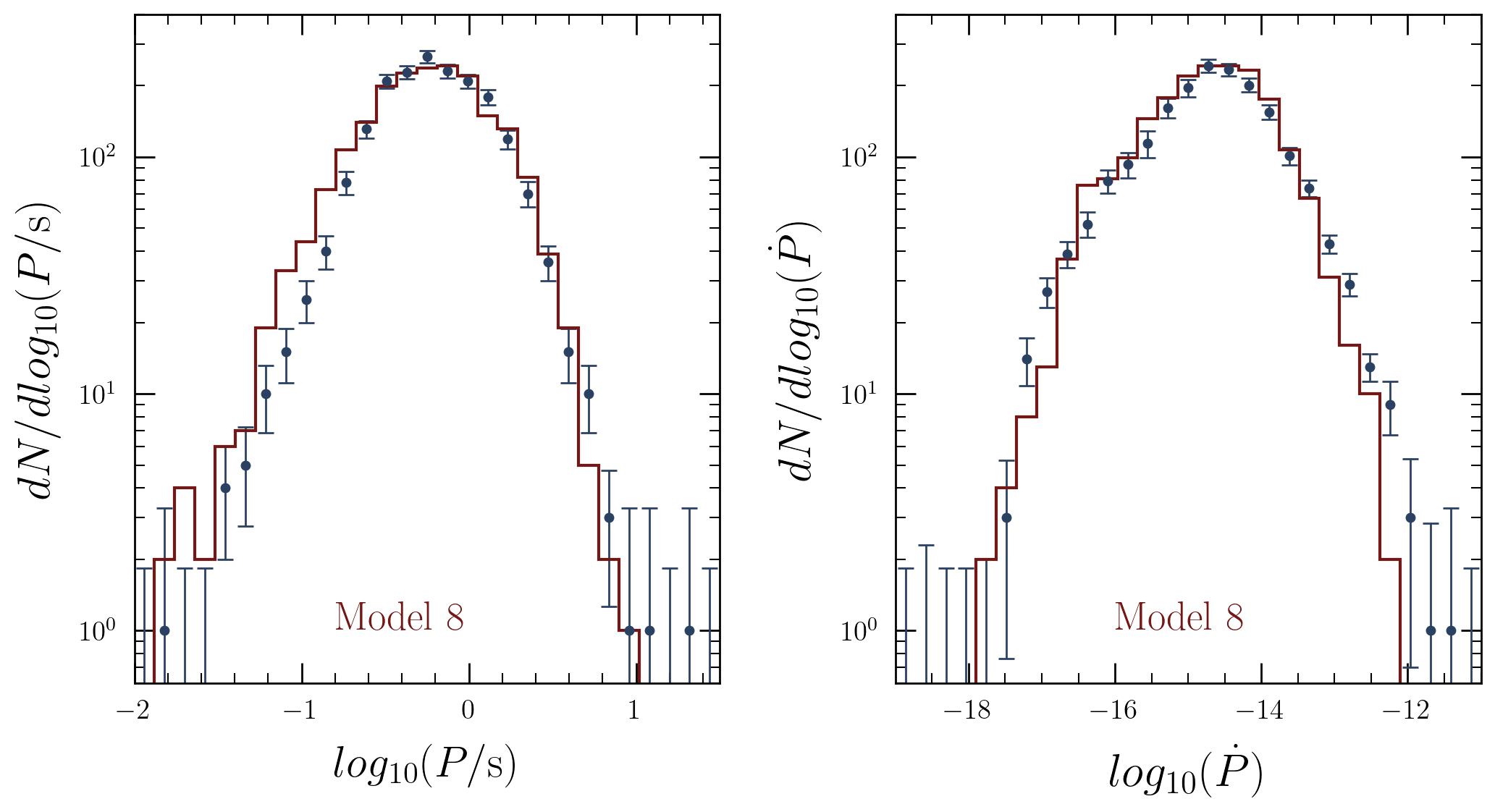}
    \caption{Same as Fig.~\ref{fig:hist_m1} but for Model 8. }
\end{figure*}

\begin{figure*}
    \centering
    \includegraphics[width=.8\textwidth]{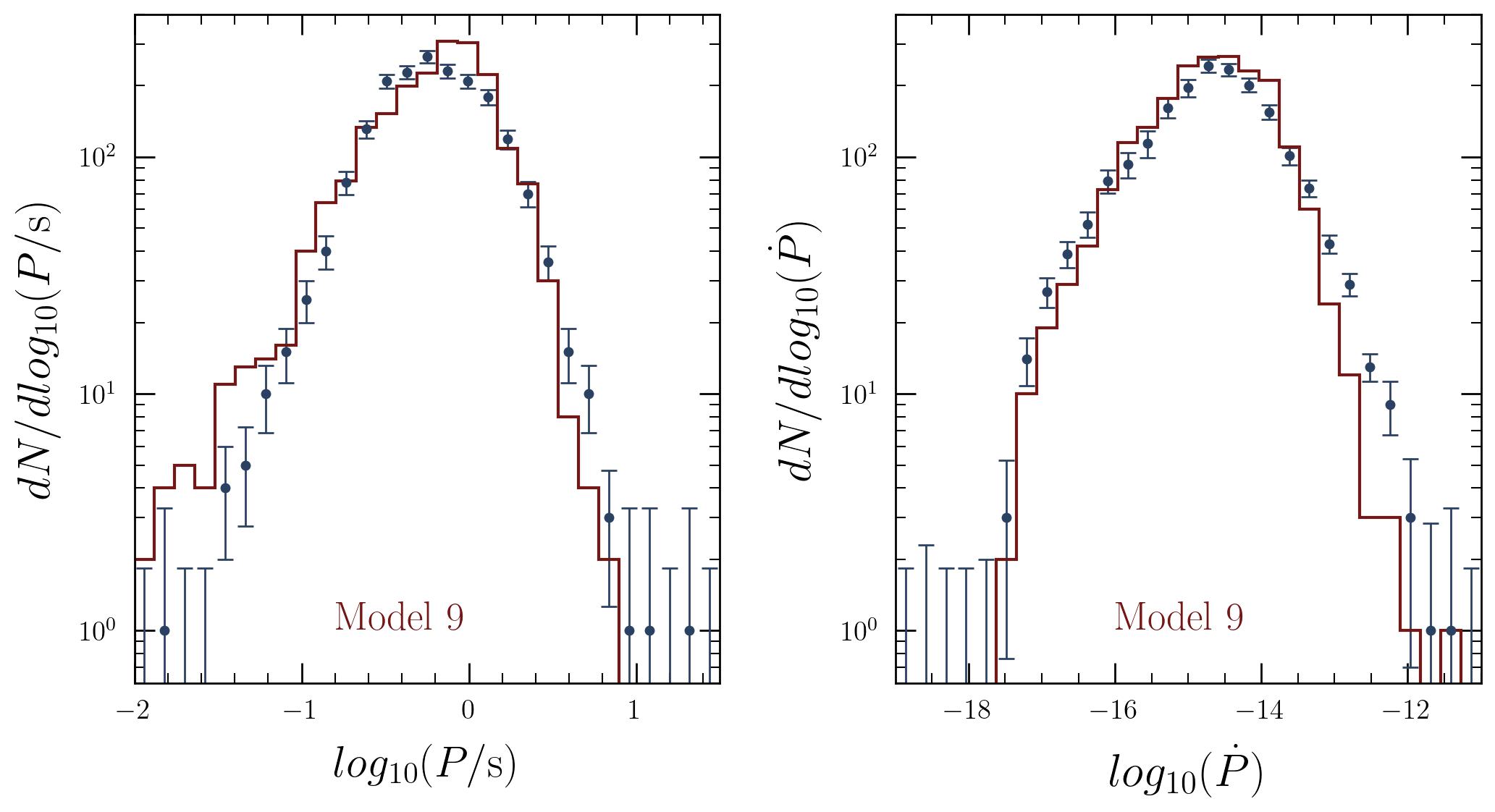}
    \caption{Same as Fig.~\ref{fig:hist_m1} but for Model 9. }
\end{figure*}

\begin{figure*}
    \centering
    \includegraphics[width=.8\textwidth]{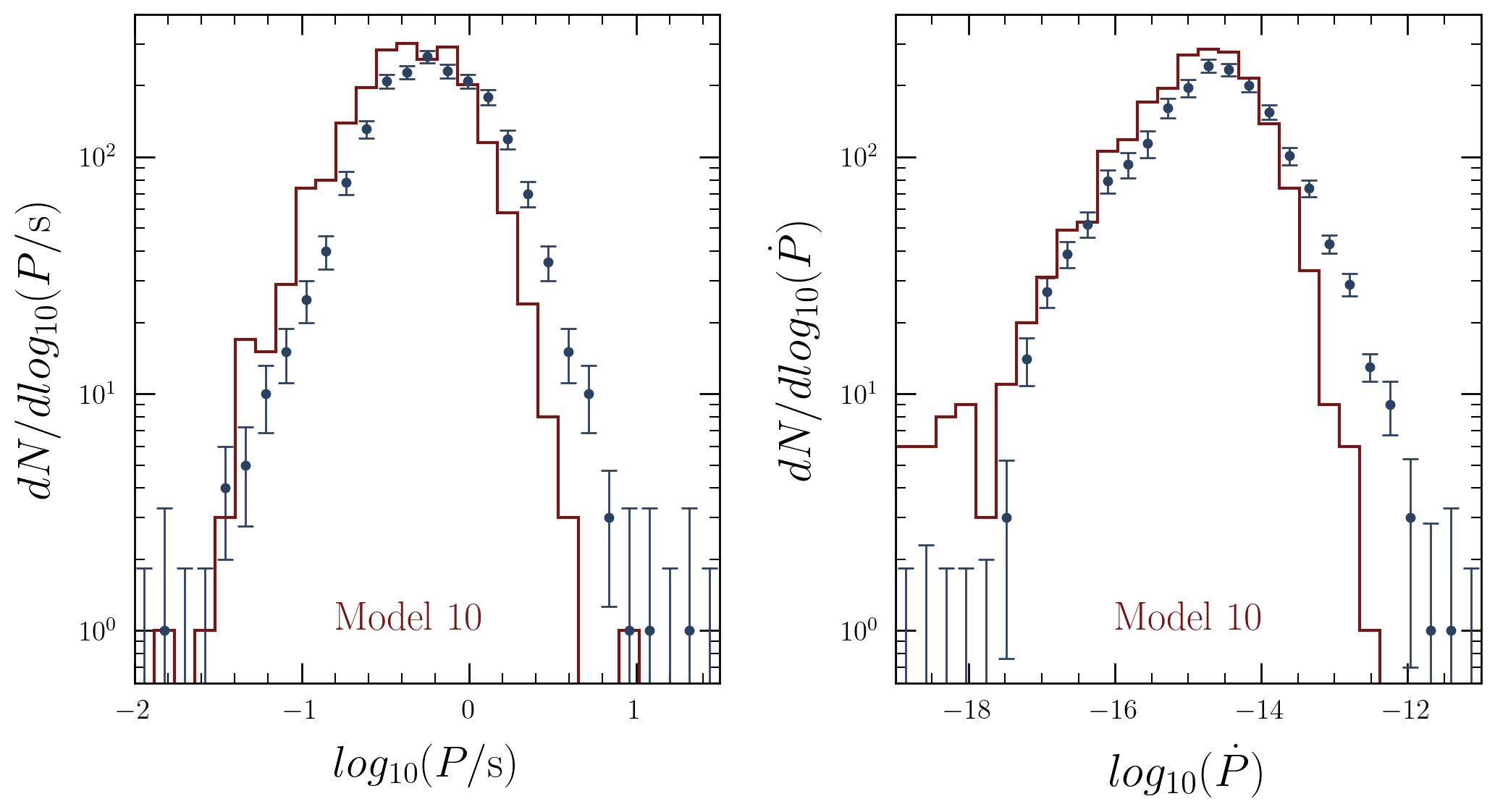}
    \caption{Same as Fig.~\ref{fig:hist_m1} but for Model 10. }
\end{figure*}

\begin{figure*}
    \centering
    \includegraphics[width=.8\textwidth]{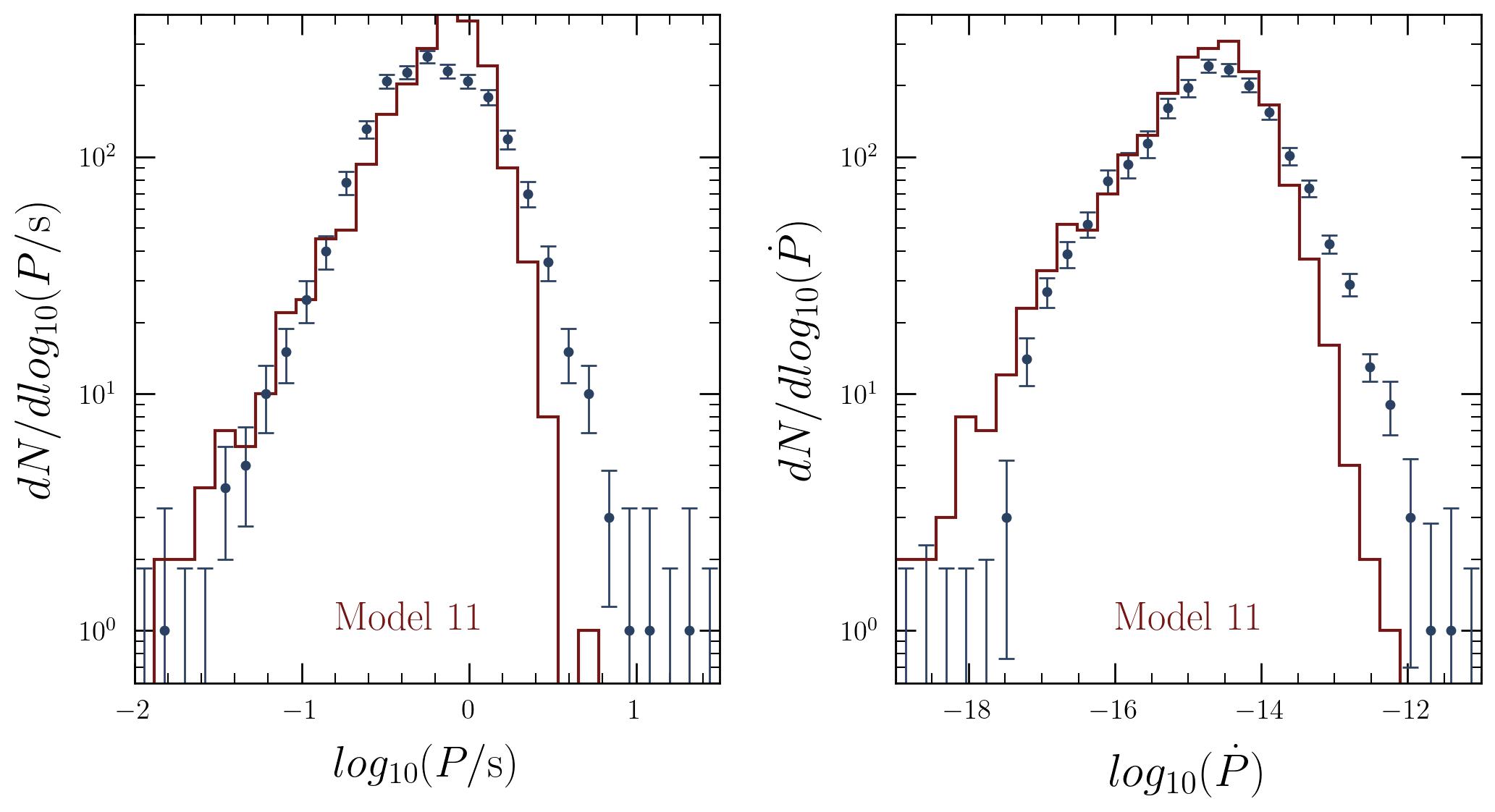}
    \caption{Same as Fig.~\ref{fig:hist_m1} but for Model 11 }
\end{figure*}

\begin{figure*}
    \centering
    \includegraphics[width=.8\textwidth]{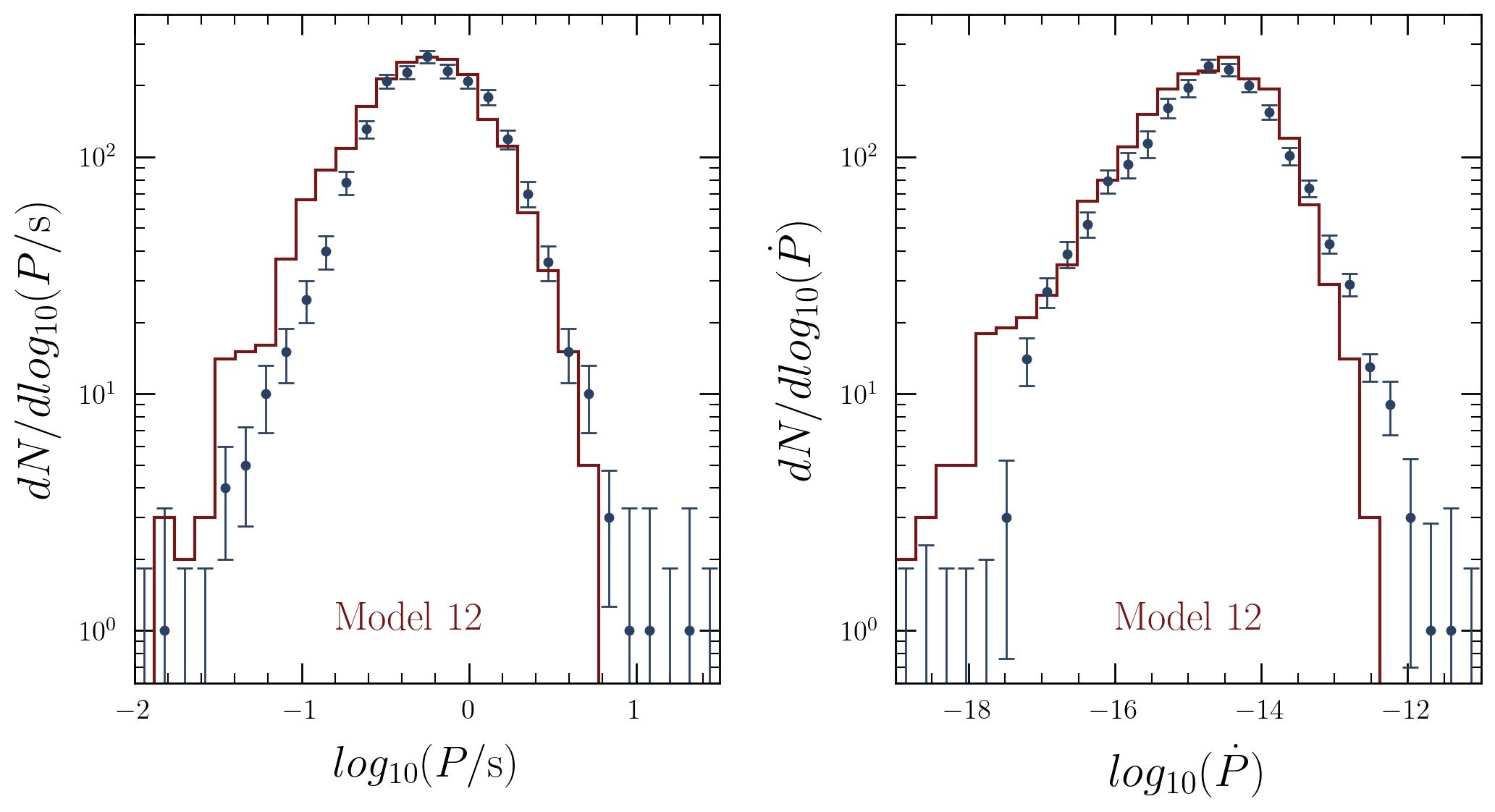}
    \caption{Same as Fig.~\ref{fig:hist_m1} but for Model 12 }

\end{figure*}

\begin{figure*}
    \centering
    \includegraphics[width=.8\textwidth]{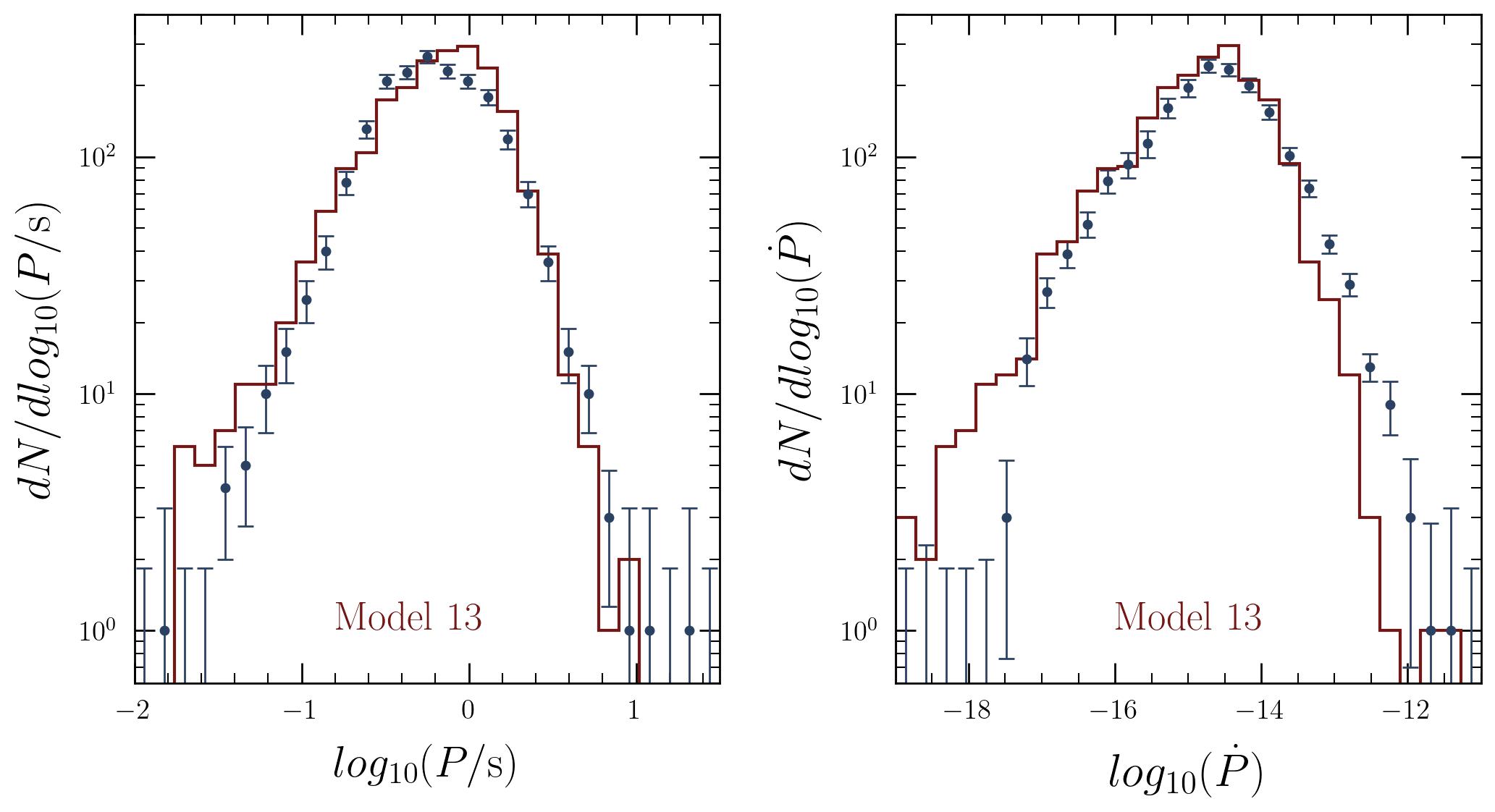}
    \caption{Same as Fig.~\ref{fig:hist_m1} but for Model 13 }

\end{figure*}

\begin{figure*}
    \centering
    \includegraphics[width=.8\textwidth]{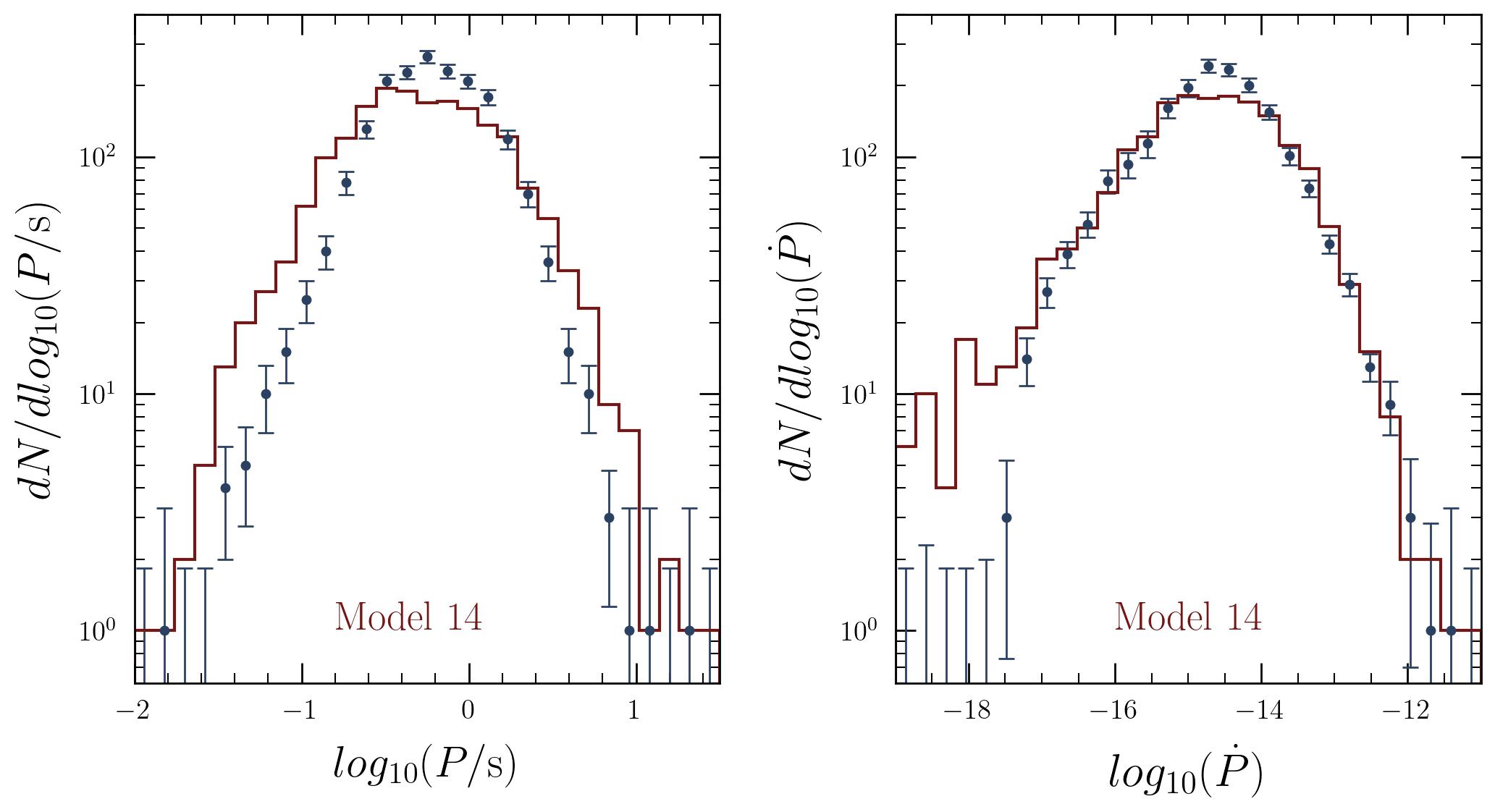}
    \caption{Same as Fig.~\ref{fig:hist_m1} but for Model 14 }
\end{figure*}

\begin{figure*}
    \centering
    \includegraphics[width=.8\textwidth]{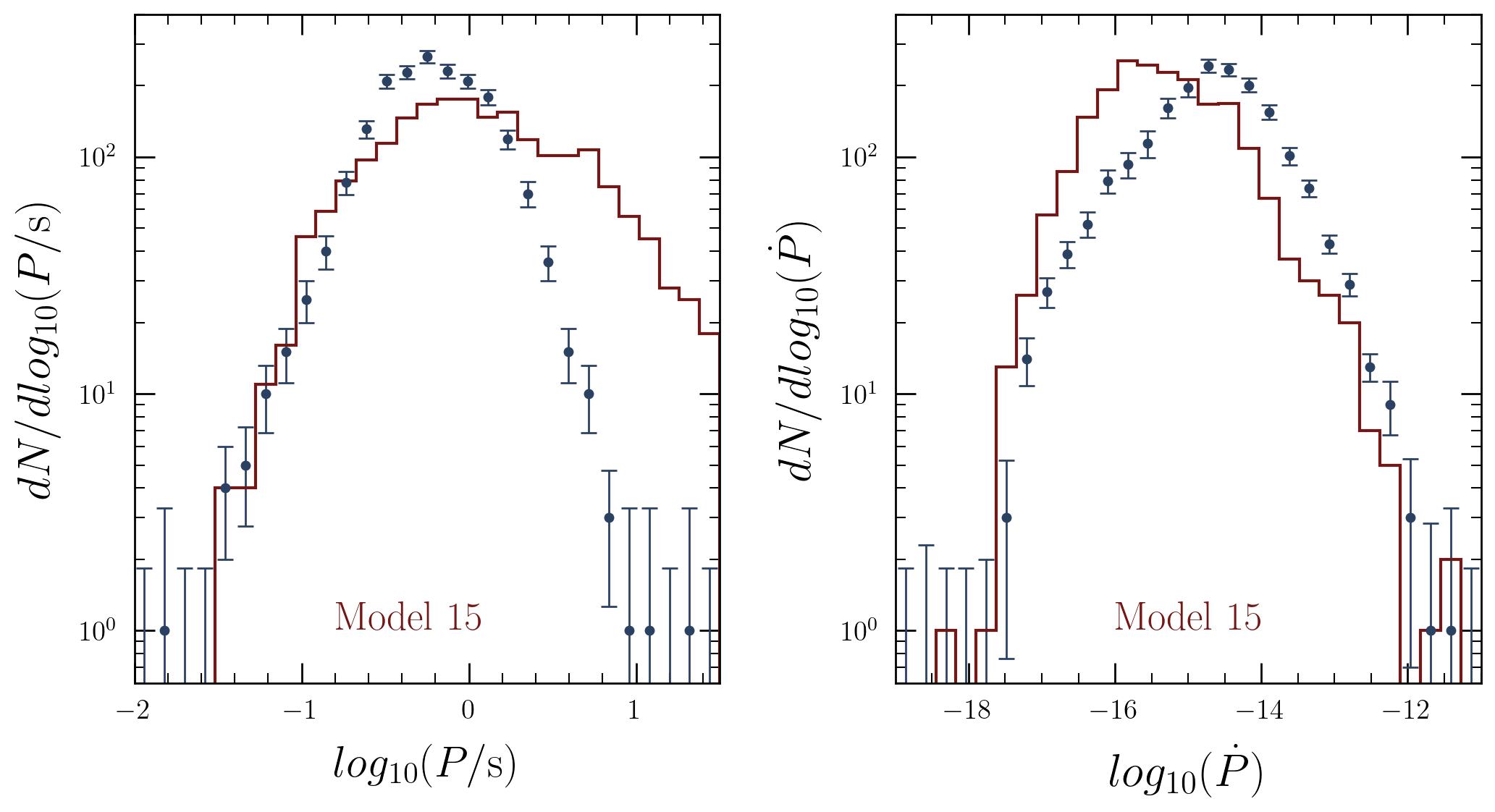}
    \caption{Same as Fig.~\ref{fig:hist_m1} but for Model 15 }
\end{figure*}

\begin{figure*}
    \centering
    \includegraphics[width=.8\textwidth]{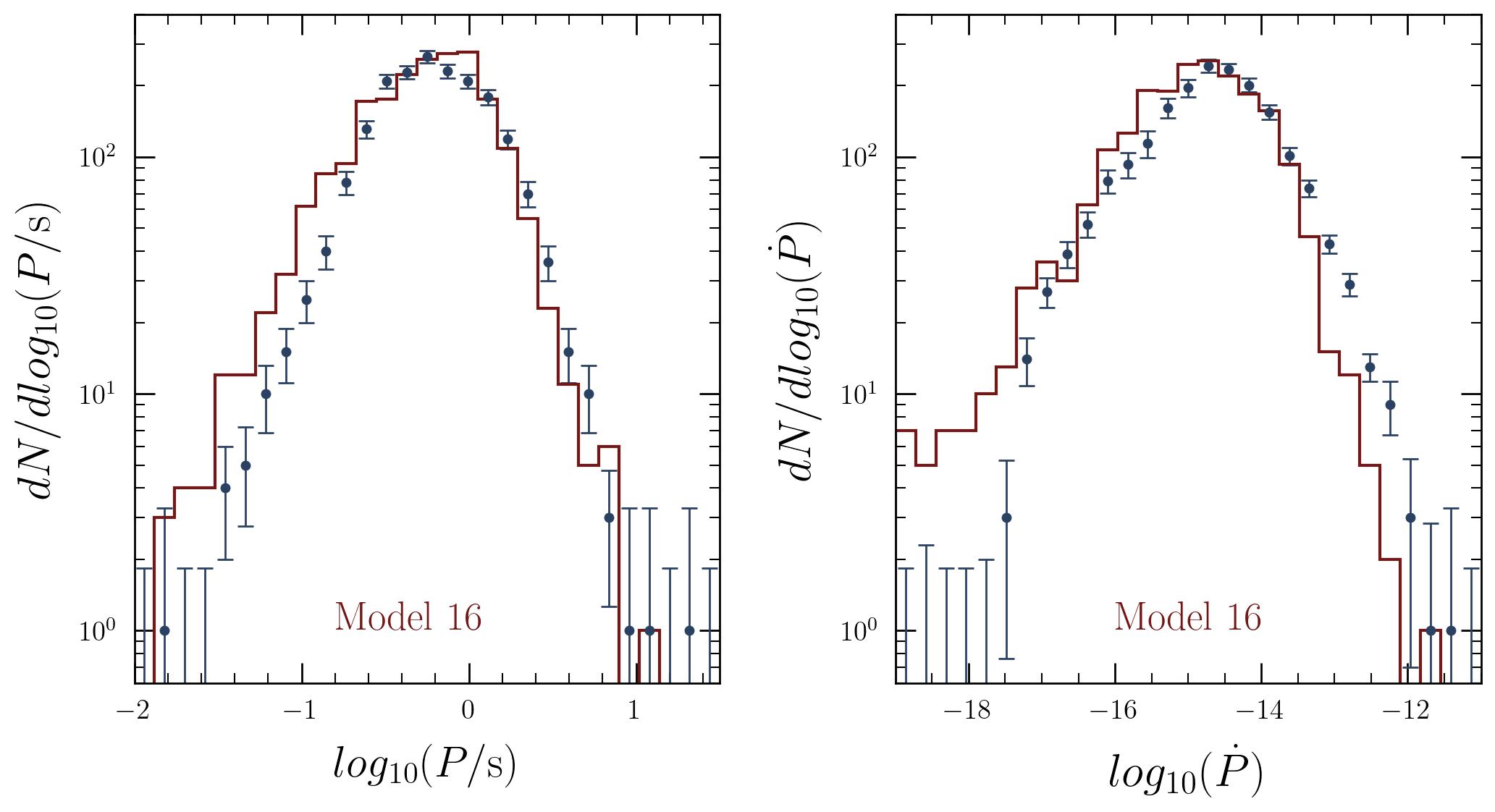}
    \caption{Same as Fig.~\ref{fig:hist_m1} but for Model 16 }
\end{figure*}

\begin{figure*}
    \centering
    \includegraphics[width=.8\textwidth]{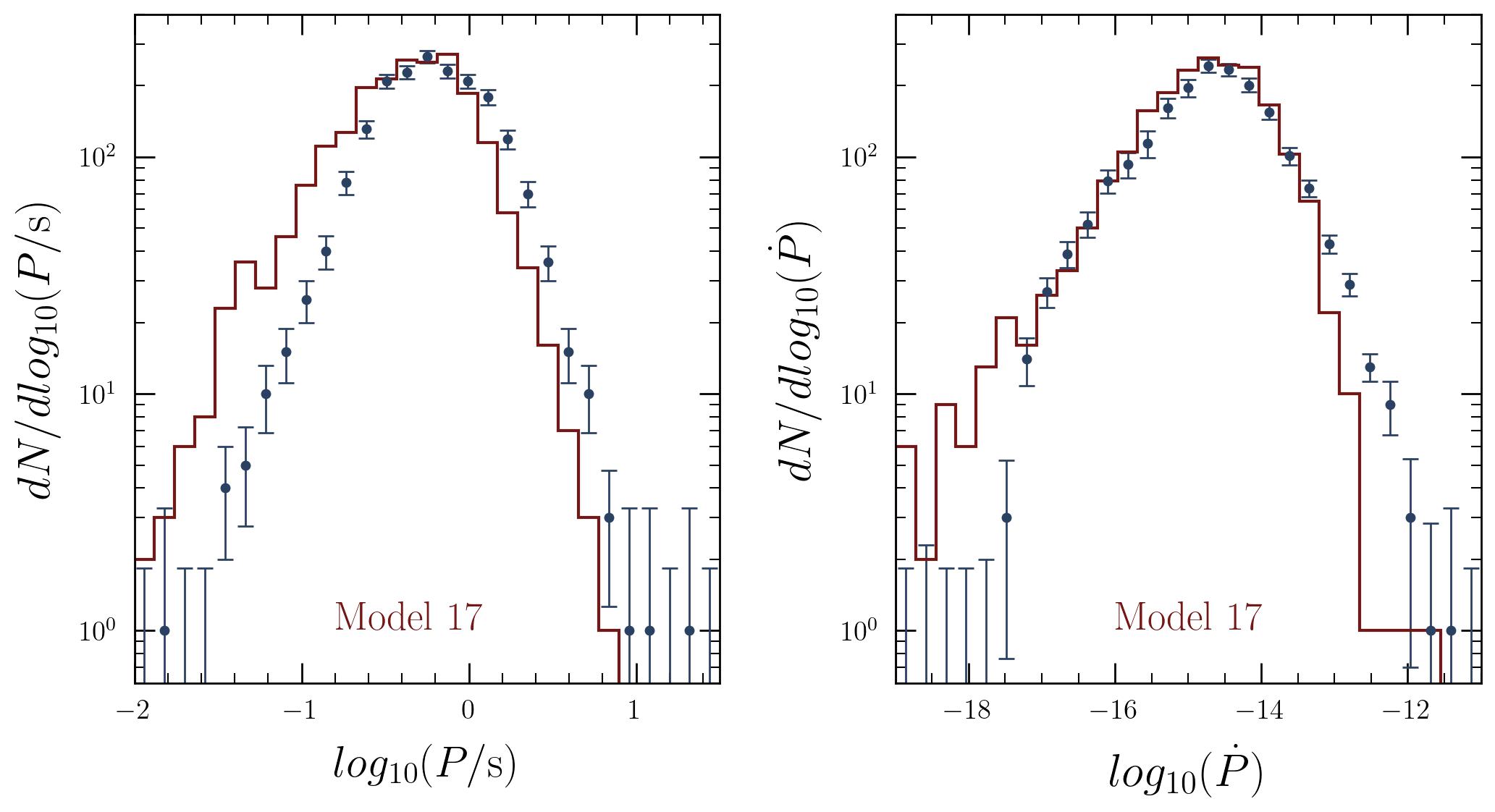}
    \caption{Same as Fig.~\ref{fig:hist_m1} but for Model 17 }
    \label{fig:hist_m17}
\end{figure*}

\end{document}